\documentclass[apj]{emulateapj}

\usepackage{verbatim}
\usepackage{amsmath}

\def\W{{\cal W}}

\def\H{{\cal H}}
\def\I{{\cal I}}
\def\C{{\cal C}}
\def\M{{\cal M}}
\def\N{{\cal N}}

\newcommand{\simgt}{\,\hbox{\lower0.6ex\hbox{$\sim$}\llap{\raise0.6ex\hbox{$>$}}}\,}
\newcommand{\simlt}{\,\hbox{\lower0.6ex\hbox{$\sim$}\llap{\raise0.6ex\hbox{$<$}}}\,}

\begin{document}

\title{Turbulence and Radio Mini-halos in the Sloshing Cores of Galaxy Clusters}

\author{J.A. ZuHone\altaffilmark{1}, M. Markevitch\altaffilmark{1}, G. Brunetti\altaffilmark{2}, S. Giacintucci\altaffilmark{3}}

\altaffiltext{1}{Astrophysics Science Division, Laboratory for High Energy Astrophysics, Code 662, NASA/Goddard Space Flight Center, Greenbelt, MD 20771, USA}
\altaffiltext{2}{INAF – Istituto di Radioastronomia, via Gobetti 101, 40129 Bologna, Italy}
\altaffiltext{3}{Department of Astronomy, University of Maryland, College Park, MD, 20742-2421, USA}

\begin{abstract}
A number of relaxed, cool-core galaxy clusters exhibit diffuse, steep-spectrum radio sources in their central regions, known as radio mini-halos. It has been proposed that the relativistic electrons responsible for the emission have been reaccelerated by turbulence generated by the sloshing of the cool core gas. We present a high-resolution MHD simulation of gas sloshing in a galaxy cluster coupled with subgrid simulations of relativistic electron acceleration to test this hypothesis. Our simulation shows that the sloshing motions generate turbulence on the order of $\delta{v} \sim$~50-200~km~s$^{-1}$ on spatial scales of $\sim$50-100~kpc and below in the cool core region within the envelope of the sloshing cold fronts, whereas outside the cold fronts, there is negligible turbulence. This turbulence is potentially strong enough to reaccelerate relativistic electron seeds (with initial $\gamma \sim 100-500$) to $\gamma \sim 10^4$ via damping of magnetosonic waves and non-resonant compression. The seed electrons could remain in the cluster from, e.g., past AGN activity. In combination with the magnetic field amplification in the core, these electrons then produce diffuse radio synchrotron emission that is coincident with the region bounded by the sloshing cold fronts, as indeed observed in X-rays and the radio. The result holds for different initial spatial distributions of preexisting relativistic electrons. The power and the steep spectral index ($\alpha \approx 1-2$) of the resulting radio emission are consistent with observations of minihalos, though the theoretical uncertainties of the acceleration mechanisms are high. We also produce simulated maps of inverse-Compton hard X-ray emission from the same population of relativistic electrons.
\end{abstract}


\section{Introduction\label{sec:intro}}

A number of relaxed, cool-core clusters are hosts to faint, diffuse radio emission with a radius comparable to the size of the cooling region ($r \simlt$~100-300~kpc) and a steep spectrum ($\alpha > 1$; $S_{\nu} \propto \nu^{-\alpha}$). These sources, called mini-halos, are relatively rare, with currently only around 10 clusters with confirmed detections. Examples include Perseus \citep{bur92,sij93}, A2029 \citep{gov09}, Ophiuchus \citep{gov09,mur10}, RXC J1504.1-0248 \citep{gia11}, and RXJ 1347-1145 \citep{git07}, to name a few. Questions still remain about the physical properties and the origin of these sources. 

Though clusters hosting mini-halos have central active galaxies, they are not sufficient by themselves to power the diffuse radio emission. The radiative timescale of the electrons at the required energies for the observed emission ($\sim{10^8}$~years) is much shorter than the time required for these electrons to diffuse across the cooling region \citep{bru03}. Two physical mechanisms have been identified as possibly responsible for the radio emission: reacceleration of pre-existing, low-energy electrons in the intracluster medium (ICM) by turbulence in the core region \citep{git02,git04}, and the generation of secondary particles via inelastic collisions between relativistic cosmic-ray protons and thermal protons \citep{pfr04,kes10b,kes10c}. 

\begin{figure*}
\begin{center}
\plottwo{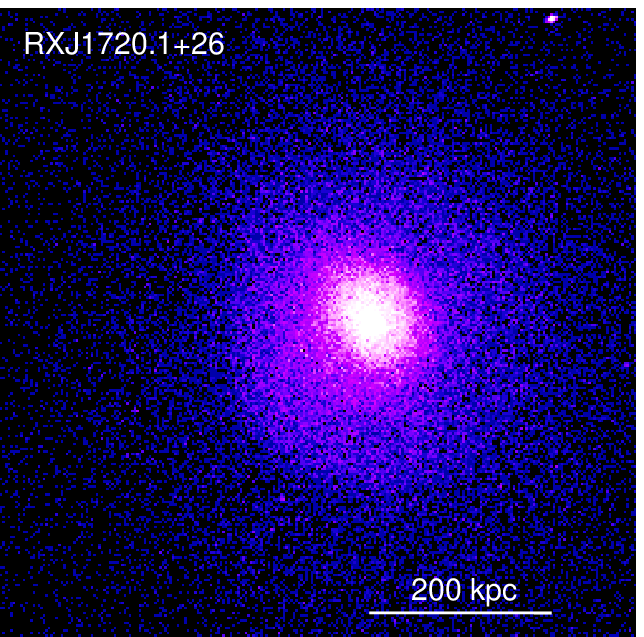}{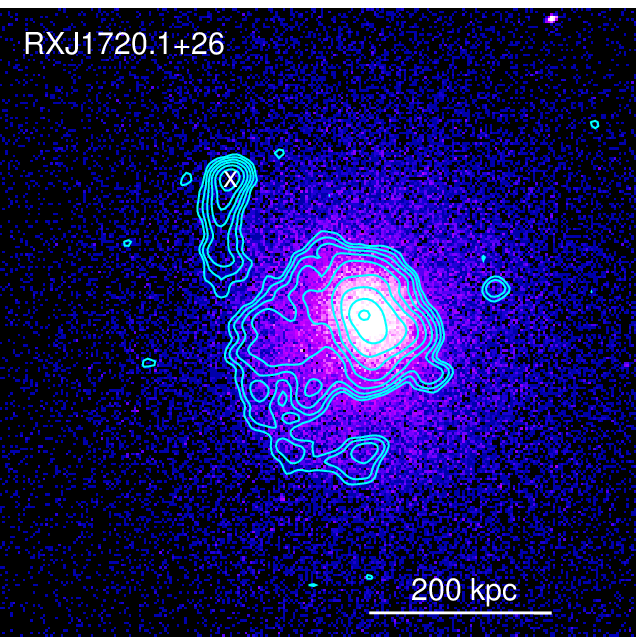}
\caption{(a) Chandra X-ray image of RXJ\,1720.1+26, one of the clusters
exhibiting a sloshing core and a minihalo (Mazzotta \& Giacintucci
2008). (b) Same image with the 610 MHz radio contours overlaid (from
Giacintucci et al.\ 2012, in preparation). Contours start from $+3\sigma$ and
are spaced by factor 2, where $1\sigma=30\mu$Jy/beam and the beam size is $8''\times6''$. The white
cross marks an unrelated head-tail radio source. The minihalo radio emission is confined within and
traces the cold fronts that are visible as brightness edges in the
X-ray image NW and SE of the center.}
\label{fig:RXJ1720}
\end{center}
\end{figure*}

In the reacceleration model, the seed electrons may be provided by buoyant bubbles inflated by the central AGN and disrupted by gas motions in the core. In the absence of a reacceleration mechanism, electrons in such disrupted bubbles cool rapidly and emit at radio frequencies well below those currently observable. However, a key question is the origin of the turbulence responsible for reaccelerating the electrons. \citet{git02} originally proposed that the cooling flow of gas inward in the core may generate turbulence. However, recent X-ray observations indicate that, while ``cooling flows'' as envisioned in \citet{fab94} probably do not materialize, even relatively relaxed clusters have large-scale gas motions in their cores. The observational signature of these gas motions are the spiral-shaped ``cold fronts'' seen in the majority of cool-core clusters \citep[for a review see][]{mar07}. These cold fronts are believed to be produced by the cold gas of the core ``sloshing'' in the cluster's deep potential well. \citet{fuj04} showed that sloshing motions can produce significant turbulence in the cluster core. \citet{maz08} discovered spatial correlations between radio mini-halo emission and cold fronts in the X-ray images of two clusters--the minihalos apparently contained within the region delineated by the cold fronts. A similar correlation is seen in the Perseus Cluster (Markevitch \& Churazov 2012, in preparation). These authors suggested the correlation arises from turbulence generated by the sloshing motions, a hypothesis that we test in this work. Figure \ref{fig:RXJ1720} shows a particularly striking example of this in the cluster RXJ1720.1+26, clearly showing that the radio emission of the minihalo is bounded by the cold fronts as seen in X-rays. 

In the last few years, a suite of hydrodynamic and magnetohydrodynamic simulations of idealized cluster mergers have been carried out to test the sloshing scenario for the origin of the cold fronts in cool cores \citep[][hereafter AM06]{AM06}, determine the physical effects of sloshing on the cluster thermal gas and magnetic field \citep[][hereafter ZMJ10 and ZML11]{zuh10,zuh11}, and make comparisons to observations of specific clusters with sloshing cold fronts \citep[][]{rod11a}. For the first time, we use a high-resolution numerical simulation to examine the connection between gas sloshing and radio minihalos in clusters. Our simulation models the cool, magnetized core gas sloshing in the gravitational potential well of a massive, initially relaxed galaxy cluster, similar to the previous works mentioned above. Using a filtering procedure similar to that employed in studies of turbulence in a cosmological context \citep[][]{dol05,vaz06,vaz09,vaz10,vaz11,vaz12}, we have determined that the sloshing motions generate turbulence in the cluster cool core that is spatially contained within the volume of the cold fronts. 

To test the predictions of the reacceleration model for minihalos, we have included in the simulation a population of passive tracer particles to provide a Lagrangian description of the sloshing gas. These particles provide us with the ability to follow relativistic electrons associated with a parcel of magnetized fluid, integrating the change in their energies determined by the fluid properties (in particular, the local turbulent velocity) along each tracer particle's trajectory. The connection between turbulent velocity and the acceleration of electrons is treated in a ``subgrid'' way--we use the prescription from \citet{bru07}, which is detailed in Section \ref{sec:elec_accel}. Given the spatial and spectral distribution of relativistic electrons along with the magnetic field of the cluster, we produce maps of simulated synchrotron emission that may be compared to observations of actual mini-halos. Our simulation can also be exploited for evaluating the relative importance of particle reacceleration and cosmic ray-thermal proton collisions for the origin of minihalos, which will be discussed in a future paper (ZuHone et al 2012, in preparation).

This paper is organized as follows: in Section 2, we describe the MHD simulation and the treatment of turbulence. In Section 3, we describe the treatment of relativistic particles. In Section 4, we describe the characteristics of the turbulence that develops in the cluster core, and based on them, estimate the reacceleration coefficients. In Section 5, we detail the evolution of the relativistic electrons and the resulting synchrotron and inverse-Compton emission. Finally, in Sections 6 and 7, we discuss and summarize our results and future developments of this work. We have assumed a flat $\Lambda$CDM cosmology with $h$ = 0.7, $\Omega_{m} = 0.3$, and $\Omega_b = 0.02h^{-2}$.

\section{Methods: Treatment of the Magnetized Cluster Gas\label{sec:gas_model}}

\subsection{MHD Simulations\label{sec:MHD_sims}}

In our simulation we solve the ideal MHD equations. Written in conservation form in Gaussian units, they are:
\begin{eqnarray}
\frac{\partial{\rho}}{\partial{t}} + \nabla \cdot (\rho{\bf v})= 0 \\
\frac{\partial{(\rho{\bf v})}}{\partial{t}} + \nabla \cdot \left(\rho{\bf vv} - \frac{\bf BB}{4\pi}\right) + \nabla{p} = \rho{\bf g} \\
\frac{\partial{E}}{\partial{t}} + \nabla \cdot \left[{\bf v}(E+p) - \frac{{\bf B}({\bf v \cdot B})}{4\pi}\right] = \rho{\bf g \cdot v} \\
\frac{\partial{\bf B}}{\partial{t}} + \nabla \cdot ({\bf vB} - {\bf Bv}) = 0
\end{eqnarray}
where
\begin{eqnarray}
p = p_{\rm th} + \frac{B^2}{8\pi} \\
E = \frac{\rho{v^2}}{2} + \epsilon + \frac{B^2}{8\pi}
\end{eqnarray}
where $p_{\rm th}$ is the gas pressure, and $\epsilon$ is the gas internal energy per unit volume. We assume an ideal equation of state with $\gamma = 5/3$. 

We performed our simulation using FLASH 3, a parallel hydrodynamics/$N$-body astrophysical simulation code developed at the Center for Astrophysical Thermonuclear Flashes at the University of Chicago \citep{fry00,dub09}. FLASH uses adaptive mesh refinement (AMR), a technique that places higher resolution elements of the grid only where they are needed. We are interested in capturing sharp ICM features like shocks and cold fronts accurately, as well as resolving the inner cores of the cluster dark matter halos. It is particularly important to be able to resolve the grid adequately in these regions. AMR allows us to do so without needing to have the whole grid at the same resolution. 

FLASH 3 solves the equations of magnetohydrodynamics using a directionally unsplit staggered mesh algorithm \citep[USM;][]{lee09}. The USM algorithm used in FLASH 3 is based on a finite-volume, high-order Godunov scheme combined with a constrained transport method (CT), which guarantees that the evolved magnetic field satisfies the divergence-free condition \citep{eva88}. In our simulations, the order of the USM algorithm corresponds to the Piecewise-Parabolic Method (PPM) of \citet{col84}, which is ideally suited for capturing shocks and contact discontinuties (such as the cold fronts that appear in our simulations). 

The gravitational potential on the grid is set up as the sum of two ``rigid bodies'' corresponding to the contributions to the potential from both clusters. This approach to the modeling the potential is used for simplicity and speed over solving the Poisson equation for the matter distribution, and is an adequate approximation for our purposes. The details of this setup may be found in \citet{zuh11}. A comparison between the rigid-potential and fully-modeled self-gravitating setups for the dark matter components of merging clusters (and a justification of using the former method for this work) may be found in \citet{rod11b}.

We refer the reader to \citet{zuh11} for the details of our initial setup of the ICM model and the magnetic field. Here we sketch only briefly the details of the simulation performed in this work. Our initial conditions consist of a massive ($M \sim 10^{15} M_\odot, T \sim 10$~keV) cool-core galaxy cluster, and a smaller ($M \sim 2 \times 10^{14} M_\odot$) gasless subcluster, separated at an initial distance $d$ = 3~Mpc, an initial impact parameter $b$ = 500~kpc, and placed on a bound mutual orbit (the core passage occurs at $t \approx 1.8$~Gyr in our adopted time frame). The size of the computational domain is $L$ = 2~Mpc, with a finest cell size of $\Delta{x}$ = 1~kpc. The maximum resolution covers a spherical region of $r \sim 300$~kpc, centered on the DM peak of the main cluster, which encompasses all of the phenomena of interest in this study. 

The ICM of the main cluster is magnetized, which is set up initially with a tangled magnetic field configuration with an average magnetic pressure proportional to the gas pressure, with $\beta_{\rm pl} = p/p_B = 100$. This choice is consistent with constraints put on the radial profiles of the magnetic field by Faraday rotation measurements \citep{bon10} and simulations \citep{dol99,dub08}. This initial magnetic field configuration is not relaxed, and the differences in magnetic pressure and tension will generate spurious turbulent velocities $|\delta{\bf v}| \simlt 100$~km~s$^{-1}$ in the gas. To examine exclusively the turbulence generated by the sloshing motions, we modeled our main cluster in isolation for $t \sim 5$~Gyr, increasing the numerical viscosity during this period to damp out the motions until they are relatively small ($|\delta{\bf v}| \simlt 10$~km~s$^{-1}$) and the magnetic field has reached a relaxed configuration. The gas density and temperature profiles of the main cluster are largely unaffected, with the largest deviations from the initial profiles of $\sim 10\%$ in the central few zones, immaterial for our purposes. Figure 1 of \citet{zuh11} shows the initial and final profiles of the main cluster.  

Finally, our MHD simulation includes passive tracer particles flowing with the ICM. These particles will be used to model the effects of reacceleration and energy losses on relativistic electrons traveling along these trajectories (see \ref{sec:elec_accel}). Tracer particles are stored at intervals of 10~Myr and from these snapshots the individual particle trajectories are extracted. The entire simulation contains approximately 10 million passive tracer particles initially distributed with their number density proportional to the local gas density. 

For the details of the process of gas sloshing in galaxy cluster cores, e.g. how it is initiated, the formation of sloshing cold fronts, and its effects on the magnetic field, we refer the reader to past simulation works on sloshing \citep{AM06,rod11a,zuh10,zuh11}. We assume these results in the following discussions. 

\subsection{Separating Turbulence from Bulk Motions\label{ref:turb_model}}

We are assuming in this work that relativistic electrons are reaccelerated by turbulence. Therefore, in order to determine the acceleration rate in our simulations, we must determine the magnitude and location of turbulence. Since not all motions in the ICM are turbulent (and not all turbulent motions efficiently accelerate particles), a distinction must be made between ``bulk'' or ``laminar'' flows and turbulent motions. Substructures and mergers will drive motions in the gas on the scale of the merging subclusters that are not turbulent in nature. This applies to our isolated binary cluster merger simulations, where the most significant non-turbulent motion is the sloshing of the cool core, with sharp cold fronts and tangential velocity discontinuities. Its contribution to the overall velocity power spectrum needs to be properly filtered out to examine the turbulent motions alone. This typically involves some process of spatial averaging, where the bulk motions are filtered out and only the turbulent (small-scale, random) component remains.\footnote{A more sophisticated approach involves the implementation of a subgrid model for the unresolved turbulent motions, such as in \citet{sca08,iap11}} However, the choice of a region to average over is often a judgment call. The simplest approaches typically take the average velocity of the cluster as a whole \citep{sun03,hal11}, or compute the average in concentric radial shells \citep{iap08,lau09}, none of which is sufficient for us in the presence of sloshing motions. 

\begin{figure}
\resizebox{\hsize}{!}{\includegraphics{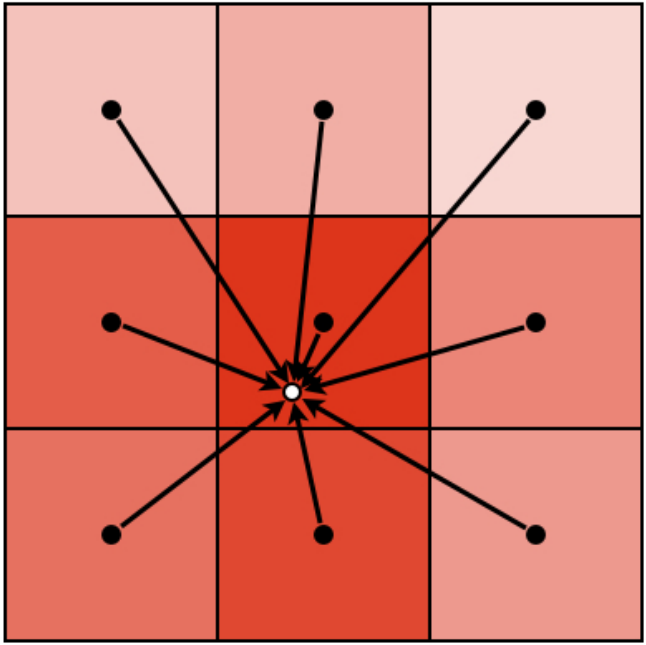}}
\caption{Schematic diagram to illustrate the method for interpolating mean velocities to a point in space. Boxes represent the volumes over which the mean velocity is determined. The velocity is interpolated to any point using a triangle-shaped cloud filter. The shading of each box represents the approximate contribution of each box to the local mean.\label{fig:interpolation}}
\end{figure} 

One successful choice of averaging, known as the ``local mean velocity field'' technique, has been applied in both SPH and AMR hydrodynamics simulations of galaxy clusters \citep[e.g.,][]{dol05,vaz06,vaz09,vaz10,vaz11,vaz12}, with success in separating the turbulent and bulk velocity fields in clusters where both kinds of motions are significant. This technique involves dividing the computational domain into ``boxes'' over which the average velocity is obtained, and the mean velocity local to a point in space is interpolated from these boxes and subtracted from the total velocity field to yield the turbulent component of the velocity field. 

Following these previous works, we use the following simple prescription to estimate the turbulence in our simulations. We begin by setting up a coarse 3D uniform grid of cubical boxes of width $\ell$ that spans the entire simulation domain. Within each box, a mean velocity is computed from the cells on the FLASH adaptive grid contained in that box. At any point in the simulation, a local mean velocity field $\overline{\bf v}({\bf x})$ may be constructed by interpolation from these boxes via a ``Triangle-Shaped Cloud'' (TSC) window function \citep[][a schematic illustration of this procedure is given in Figure \ref{fig:interpolation}]{hoc88}. From this local mean field, we may compute a local ``filtered'' turbulent field $\delta{\bf v} = {\bf v} - \overline{\bf v}$. 

The resulting turbulent velocity field is a conservative estimate of the field that may arise in a real cluster merger such as the one we are simulating. The TSC interpolation will filter out turbulent modes with scales larger than the box size $\ell$. Additionally, the dissipation scale where the energy in the turbulent cascade is dissipated into heat is set not by the actual viscosity of the ICM but by the resolution of our simulation, which results in a dissipation scale of several times the cell size. Thus, the power of turbulence that we measure in our simulations is a lower bound to that which may be expected in a real cluster merger. We will address this point in detail in Section \ref{sec:turb_char}. We will also discuss a separation of turbulent motions into compressive and solenoidal components in Section \ref{sec:turb_coeff}. 

\section{Methods: Treatment of Relativistic Electrons\label{sec:methods_rel}}

The reacceleration model for radio halos assumes a population of relativistic electrons already present in the ICM, which are reaccelerated by turbulence, e.g. that generated by sloshing motions. In our simulation model, we assume that relativistic electrons are advected along with the fluid, so we may use our tracer particle trajectories to determine the effect of the properties of the fluid along each trajectory on the energies of the relativistic electrons. We do this by performing a set of post-processing ``simulations'' on the tracer particle trajectories. There is a sufficient number of tracer particles to sample the cool core very well. A similar post-processing approach was used to study simulated gas mixing, metal distribution, and iron line profiles in a set of clusters drawn from a cosmological simulation in \citet{vaz10}. 

\subsection{Model for Relativistic Electron Spectra\label{sec:elec_model}}

In order to compute the synchrotron emission for each tracer particle as a function of time, it is necessary to model the energy distribution of relativistic electrons $N(\gamma)$ that may be initialized and evolved for each tracer particle along its trajectory. A rigorous approach to this problem would be to integrate a Fokker-Planck equation for $N(\gamma)$ along each trajectory, with $N(\gamma)$ sampled discretely in bins of electron energy. This approach would provide a complete description of the relativistic particle diffusion in momentum space as well as systematic energy gain and losses. While feasible in the near future, this approach is beyond the current state of software development for our code, and we chose to adopt a simpler method. Instead, the distribution of relativistic electrons $N(\gamma)$ is discretely sampled in a ``Monte Carlo'' fashion with a large number of ``relativistic particles''  for each tracer particle. The distinction between these two methods is similar to the distinction between Eulerian grid-based methods versus Lagrangian particle-based methods (smooth-particle hydrodynamics) for solving the equations of hydrodynamics. In the latter case, the individual gas particles do not represent any real concentrations of gas {\it per se}, but are simply discrete Lagrangian samples of the distribution function of the gas; similarly, the relativistic particles are not electrons but merely sample the underlying distribution.

Each relativistic particle has a Lorentz factor $\gamma_{i,j}$ that represents the average relativistic energy $\gamma_{i,j}m_ec^2$ for the sample $i$ for the tracer particle $j$. The energy disribution function $N_j(\gamma)$ for each tracer particle $j$ may be written as:
\begin{equation}
N_j(\gamma) = K_j\displaystyle\sum_i\delta(\gamma-\gamma_{i,j}),
\end{equation}
where the sum is taken over the number of samples for the tracer particle and $K_j$ is a normalization constant. To get a proper scaling for the total energy in relativistic electrons, as our default case we assume the initial energy density in relativistic electrons of each tracer particle is proportional to the tracer particle's associated thermal energy (that is, assuming the uniform initial ratio of relativistic electron to thermal pressures): 
\begin{equation}
\eta = \frac{m_ec^2\displaystyle\int\gamma{N_j(\gamma)}d{\gamma}}{m_j\varepsilon_j} = \frac{K_jm_ec^2\displaystyle\sum_i(\gamma_{i,j}-1)}{m_j\varepsilon_j},
\end{equation}
where $\varepsilon_j$ is the internal energy per unit mass of the thermal gas and each tracer particle has a gas mass $m_j$ associated with it, such that the total gas mass of particles equals the total mass in gas on the grid within the region under consideration. This condition sets the normalization constant $K_j$. The results in this work only depend on $\eta$ as a normalization constant, so we will express electron energies and radiative quantities in terms of $\eta$, but we assume a default value of $\eta = 10^{-3}$ to compare the results of our numerical simulations to observations. 

We generate a physically plausible initial spectrum of relativistic electrons by taking a power-law spectrum with index $s = 2.5$ (defined as $N(\gamma) \propto \gamma^{-s}$) and passively evolving it for 1~Gyr under the constant conditions of $n_{\rm th} = 0.01$~cm$^{-3}$ and $B = 5~\mu$G, representative values for the cluster core regions in this model. The result is a spectrum that has significantly cooled, with essentially no electrons at energies higher than $\gamma \approx 500$. This initial distribution of electrons has no significant emission at observable radio wavelengths. We assign this spectrum to each tracer particle at $t = 2.55$~Gyr. At this epoch, the characteristic size of the sloshing region is approximately 4$\ell = 120$~kpc, and by beginning the relativistic electron evolution at this point, we hope to minimize contamination to our derived turbulent spectrum from bulk motions. This is also roughly the epoch at which turbulence begins to be significant (Section \ref{sec:turb_char}).

As a starting point, we also assume that the number density of relativistic electrons is proportional to the density of thermal electrons (however, in Section \ref{sec:different_space}, we will discuss alternative initial spatial distributions of particles). Since our tracer particles are advected with the thermal gas and represent parcels of constant gas mass, this simply implies that the number of relativistic electrons per tracer particle remains constant. For the results presented here, we generated 10$^{4}$ relativistic particles for each tracer particle. 

We select only certain tracer particles for consideration instead of using all of them contained within the simulation volume. Our default case begins with the tracer particles filling a large region of radius $R$ = 300~kpc in the initial undisturbed cluster, covering the region that will eventually contain sloshing and cold fronts. 

\subsection{Two Mechanisms of Reacceleration by MHD turbulence\label{sec:elec_accel}}

The critical component of the model for radio mini-halos is the reacceleration of relativistic particles by MHD turbulence. Relativistic particles interacting with MHD turbulence are subject to several mechanisms that lead to their acceleration, both through resonant and non-resonant coupling with turbulent waves. In the case of compressible MHD turbulence generated at large scales in the ICM, \citet{bru07} have shown that reacceleration is accomplished primarily by the damping of fast magnetosonic waves through the Transit-Time Damping (TTD) resonance \citep[e.g.,][]{mel68}, the condition for which is
\begin{equation}
\omega - k_{\parallel}v_{\parallel} = 0,
\end{equation}
where $\omega$ is the frequency of the magnetosonic wave, and $k_{\parallel}$ and $v_{\parallel}$ are the parallel (projected along ${\bf B}$) wavenumber and particle speed, respectively. To compute the momentum-diffusion coefficient of relativistic electrons due to TTD resonance, we use Equation 40 from \citet{bru07} (assuming the mode phase velocity $\langle{V_{\rm ph}}\rangle \approx c_s$), where the spectrum of compressive modes is calculated self-consistently by taking into account the effect of dampings due to TTD with thermal and relativistic particles:
\begin{eqnarray}\nonumber
D_{\rm pp,TTD} &=& \frac{\pi^2}{2c}\frac{p^2}{B_0^2}\displaystyle\int_0^{\pi/2}d\theta{c_s^2}\frac{\sin^3{\theta}}{|\cos{\theta}|}\H\left(1-\frac{c_s/c}{\cos{\theta}}\right) \\
&& \left(1-(\frac{c_s/c}{\cos{\theta}})^2\right)^2\displaystyle\int_{k_{\rm min}}^{k_{\rm cut}}{dk}\W_B(k)k.
\label{eqn:Dpp1}
\end{eqnarray}
In the integral over $k$, $k_{\rm min}$ specifies the wavenumber corresponding to the scale of the largest turbulent eddies and $k_{\rm cut}$ specifies the wavenumber corresponding to the minimum scale of turbulence where the timescale of dampings equals that of the turbulent cascade. $p$ is the electron momentum, $B_0$ is the local magnetic field, $\H$ is the Heaviside step function, and the spectrum of the magnetic field fluctuations associated with the compressive modes $\W_B(k)$ is
\begin{equation}
\W_B(k) = \frac{1}{\beta_{\rm pl}}\left\langle\frac{\beta_{\rm pl}|B_k^2|}{16\pi\W(k)}\right\rangle\W(k),
\label{eqn:Dpp2}
\end{equation}
where the quantity in brackets $\langle{...}\rangle$ is averaged with respect to $\theta$ and is of order unity. Noting that $\beta_{\rm pl} = p_{\rm th}/p_B$, Equation \ref{eqn:Dpp1} may be simplified to
\begin{equation}
D_{\rm pp,TTD} \approx \frac{\pi}{16\rho{c}}p^2\langle{...}\rangle\I\displaystyle\int_{k_{\rm min}}^{k_{\rm cut}}{dk}\W(k)k,
\label{eqn:Dpp3}
\end{equation}
where $\I$ is the integral over $\theta$ and is of order a few. The integral over $k$ may be rearranged to give 
\begin{equation}
\displaystyle\int_{k_{\rm min}}^{k_{\rm cut}}{dk}\W(k)k = {\langle{k}\rangle}\displaystyle\int_{k_{\rm min}}^{k_{\rm cut}}{dk}\W(k) \approx {\langle{k}\rangle}\rho{v_t}^2R^c,
\label{eqn:Dpp4}
\end{equation}
where ${\langle{k}\rangle}$ is the average wavenumber of the turbulent cascade as measured in the simulation, $v_t$ is the turbulent velocity of the tracer particle, and $R^c$ is the fraction of the turbulent energy in the form of compressible motions. The last equality follows from the definition of the power spectrum. The scaling of fast modes with wavenumber $k$ is of Kraichnan form, with $\W(k) \propto k^{-3/2}$ \citep{cho03}, as opposed to the purely hydrodynamic Kolmogorov scaling of $k^{-5/3}$. We will assume this form for the power spectrum of the turbulent cascade throughout this paper. The mean wavenumber ${\langle{k}\rangle}$ is then expressed in terms of $k_{\rm min}$ and $k_{\rm cut}$ as:
\begin{equation}
\langle{k}\rangle = \frac{\displaystyle\int_{k_{\rm min}}^{k_{\rm cut}}\W(k)kdk}{\displaystyle\int_{k_{\rm min}}^{k_{\rm cut}}\W(k)dk} \approx \frac{\displaystyle\int_{k_{\rm min}}^{k_{\rm cut}}{k^{-3/2}k}dk}{\displaystyle\int_{k_{\rm min}}^{k_{\rm cut}}{k^{-3/2}}dk} = \sqrt{k_{\rm min}k_{\rm cut}}
\label{eqn:avg_k}
\end{equation}

In cgs units, the resulting momentum-diffusion coefficient for the relativistic electrons via TTD is given by:
\begin{equation}
D_{\rm pp,TTD} \approx 4 \times 10^{-11}{\langle{k}\rangle}{v_t}^2R^cp^2.
\label{eqn:Dpp5}
\end{equation}
Fast particles diffusing in compressive MHD turbulence also experience statistical compression and expansion that leads to additional stochastic non-resonant acceleration. In the fast diffusion limit, the momentum-diffusion coefficient is \citep[][Equation 55]{bru07}:
\begin{equation}
D_{\rm pp,C} \sim \frac{2}{9}p^2\frac{V_l^2}{D},
\label{eqn:Dpp6}  
\end{equation}
where the spatial diffusion coefficient $D$ under our physical conditions is is \citep[][Equation 56]{bru07}:
\begin{equation}
D \sim \frac{cl_{\rm mfp}}{3}.
\label{eqn:Dpp7}
\end{equation}
The two mechanisms, TTD resonance and non-resonant compression, are essentially driven by the same turbulent modes and involve independent particle-mode couplings. Consequently, as a first approximation, the acceleration process may be thought of as the combination of the two mechanisms \citep{cho06,bru07}, with a total momentum-diffusion coefficient $D_{\rm pp} = D_{\rm pp,TTD} + D_{\rm pp,C}$.

\subsection{Gains and Losses on the Relativistic Particles\label{sec:gains_and_losses}} 
Each relativistic particle has a reacceleration efficiency of 
\begin{equation}
\chi = \frac{4D_{\rm pp}}{p^2},
\label{eqn:Dpp8}
\end{equation}
so that the rate of momentum increase for each relativistic particle is 
\begin{equation}
\left(\frac{dp}{dt}\right)_{\rm acc} = \chi{p}
\label{eqn:Dpp9}
\end{equation}
We will detail the assumptions that go into the computation of the reacceleration coefficient in our simulations in Section \ref{sec:turb_coeff}, in particular, the evaluation of ${\langle{k}\rangle}$, $R^c$, and $v_t$. 

The rate of energy losses due to synchrotron and Inverse-Compton scattering off of cosmic microwave background (CMB) photons for individual electrons is (in cgs units):
\begin{eqnarray}\nonumber
\left(\frac{dp}{dt}\right)_{\rm rad} &=& -{4.8 \times 10^{-4}}p^2\left[\left(\frac{B_{{\mu}G}}{B_{\rm CMB}}\right)^2\frac{\sin^2\theta}{2/3}+(1+z)^4\right]\\
&=& -\frac{\beta{p}^2}{m_ec} 
\label{eqn:radloss}
\end{eqnarray}
Since our relativistic particles are not individual electrons but samples of the electron distribution function (and hence represent many electrons), we assume for simplicity that each relativistic particle represents an isotropic distribution of pitch angles, with $\langle\sin^2\theta\rangle = 2/3$. $B_{\rm CMB} = 3.2(1+z)^2 \mu$G is the equivalent magnetic field strength for the CMB at present, where $z$ is the cosmological redshift. 

In our simulation the core passage of the disturbing subcluster occurs at $t \sim$~1.8~Gyr from the beginning of the simulation, and our particle trajectories begin at $t$ = 2.55~Gyr (after the onset of turbulence due to sloshing, see Section \ref{sec:electron_spectra}), and we assign the redshift $z = 0$ to the epoch $t$ = 5~Gyr of the simulation, in order to reproduce some of the observed nearby clusters exhibiting cold fronts in their cores. Under these conditions, the redshift at each epoch is computed from the simulation time assuming a $h = 0.7$, $\Omega_M = 0.3$, and $\Omega_\Lambda = 0.7$ $\Lambda$CDM cosmology.  

The Coulomb losses are given by (in cgs units):
\begin{equation}
\left(\frac{dp}{dt}\right)_{\rm coll} = -{3.3 \times 10^{-29}}n_{\rm th}\left[1+\frac{\ln{(\gamma/n_{\rm th})}}{75}\right]
\end{equation}
where $n_{\rm th}$ is the number density of thermal particles.

\subsection{Solving for the Evolution of the Relativistic Particles\label{sec:evolution_model}}

If spatial diffusion is not important, formally the time evolution of the relativistic electron momentum distribution $N(p,t)$ is a solution to the Fokker-Planck equation \citep{bru07}:
\begin{eqnarray}\nonumber
\frac{\partial{N(p,t)}}{\partial{t}} &=& \frac{\partial}{\partial{p}}\left[N(p,t)\left(\left|\frac{dp}{dt}\right|_{\rm rad} + \left|\frac{dp}{dt}\right|_{\rm coll}-\frac{4D_{pp}}{p}\right)\right] \\ 
&+& \frac{\partial^2}{\partial{p}^2}\left[D_{pp}N(p,t)\right]
\label{eqn:fokker-planck}
\end{eqnarray}
Solving this equation numerically can be expensive, particularly for the case of many individual tracer particle trajectories as in our simulation. However, as previously described, we have chosen to evolve relativistic ``sample'' particles instead of $N(p,t)$ explicitly, which can be thought of as the probability density for the random variable $P_t$, which corresponds to the momenta of the sample particles. For a given Fokker-Planck equation and distribution function $N(p,t)$ there is a corresponding stochastic differential equation (SDE) for the evolution of $P_t$ for an ensemble of sample particles. By following the momentum trajectories of many sample particles, we may reliably approximate the behavior of $N(p,t)$ and the observable quantities that depend on it, such as the resultant synchrotron and IC emission. SDEs have been used extensively in other astrophysical contexts, in particular, for the integration of cosmic ray trajectories in the heliospheric and galactic magnetic fields \citep[][]{zha99,flo09,pei10,str11,kop12}. The main differences between our approach and many of these works is that a) we are only integrating the momentum of each particle as the relativistic particles are assumed to follow the tracer particle trajectories in space, and b) we are integrating the equations forward in time instead of backwards to the original source of particles. 

The SDE that corresponds to the above Fokker-Planck equation is given (in the It${\rm \bar{o}}$ formulation) by:
\begin{equation}
dP_t = a(p,t)dt + b(p,t)dW_t
\label{eqn:energy_evo}
\end{equation}
where the ``drift'' term is 
\begin{equation}
a(p,t) = \left|\frac{dp}{dt}\right|_{\rm rad}+\left|\frac{dp}{dt}\right|_{\rm coll}-\frac{4D_{pp}}{p}
\label{eqn:drift}
\end{equation}
and the "stochastic" term is 
\begin{equation}
b(p,t)dW_t = \sqrt{2D_{pp}}dW_t \sim \sqrt{2D_{pp}dt}\N(0,1)
\label{eqn:stochastic}
\end{equation}
where $dW_t$ is a standard Wiener (or "Brownian motion") process, and $\N(0,1)$ is a normal distribution with zero mean and unit variance (the "$\sim$" symbol here indicates "is distributed as"). The effects of each of these terms on the relativistic electron spectrum will be shown in the Appendix. 

We have integrated the drift term of this equation using a 4th-order Runge-Kutta method. To integrate the stochastic term, we use the Milstein method \citep{klo11}. This results in the following discretization for Equation \ref{eqn:stochastic}:
\begin{equation}
b(p,t)dW_t \approx b(p,t)\Delta{W_n} + \frac{1}{2}b(p,t)\frac{\partial{b(p,t)}}{\partial{p}}[(\Delta{W_n})^2-\Delta{t}]
\end{equation}
where 
\begin{equation}
\Delta{W_n} \sim \sqrt{\Delta{t}}\N(0,1)
\end{equation}
This equation is integrated for each relativistic particle along each tracer particle trajectory with a variable timestep for each tracer particle $\Delta{t}_j = \min\{0.1(p_{i,j}/\dot{p}_{i,j}), 0.1(p_{i,j}^2/D_{\rm pp,i,j})\}$ to ensure stability. To determine the fluid quantities ($\rho, T, \delta{v}, B$) at any point on the trajectory, they are linearly interpolated between the saved instances of each particle along that trajectory, which have a time resolution of 10~Myr. In the Appendix we provide tests of our method when compared to analytical solutions and the results of a Fokker-Planck calculation. A detailed description of the method and the code will be the subject of a future paper (ZuHone et al 2012, in preparation).

\section{Results: Turbulence\label{sec:turbulence}}

\subsection{Characteristics of Turbulence Generated in the Sloshing Region\label{sec:turb_char}}
\begin{figure}
\begin{center}
\plotone{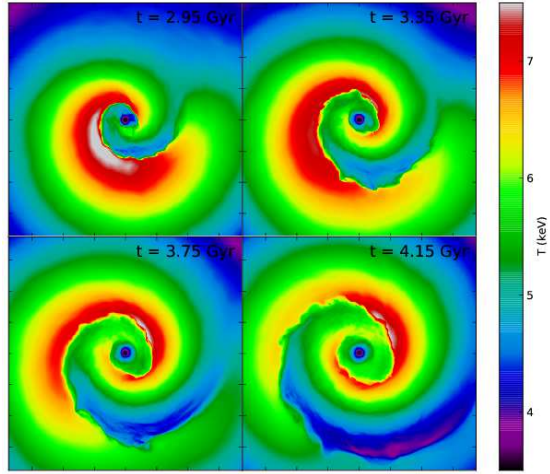}
\caption{Projected temperature maps in the $z$-projection for several different epochs of the simulation. The colorbar is temperature in keV. Each panel is 750~kpc on a side. Tick marks indicate a distance of 100~kpc. Core passage occurs at $t \ approx 1.8$~Gyr.\label{fig:temp_map}}
\end{center}
\end{figure}

In the sloshing/reacceleration hypothesis for radio mini-halos that we are testing in this work, the radio emission coincident with the envelope of the cold fronts as seen in X-rays is due to the turbulence that is associated with the sloshing motions. Figure \ref{fig:temp_map} shows the spiral shape in temperature that results from the sloshing motions for a few different epochs of the simulation. We now determine the location and spectrum of turbulence that results from the encounter with the subcluster. We note at the outset that a very recent work \citep{vaz12}, using an improved filtering technique, produced turbulent velocity maps and power spectra for gas sloshing in the core of a galaxy cluster and achieved results in general agreement with those we present here. 

Separating out the turbulent energy component from the bulk component in a simulation of gas sloshing is a challenge, mainly due to the presence of the sloshing motions themselves. Strong velocity shears and discontinuities exist at the cold front surfaces, and tracer particles situated near these surfaces will have a mean velocity interpolated to them that is a weighted sum of the velocities across the front, resulting in some of the bulk component of the particle's velocity remaining after the spatial filtering procedure. Therefore, tracer particles that travel near these surfaces may have their turbulent kinetic energy overestimated, which we should find a way to avoid. 

In our merger setup, where the subcluster's trajectory is in the $x-y$ plane, most of the kinetic energy associated with the sloshing motions is in the $v_x$ and $v_y$ components of the velocity field. It is in these directions in which it will be most difficult to filter out the contribution to the total velocity at a particle position from the sloshing motions. At the same time, we expect that the vast majority of the kinetic energy associated with the $v_z$ velocity component is due to the turbulent cascade generated by the sloshing motions, since there is only a slow outward expansion of the cold fronts in the $z$-direction. We will use this simple observation to separate the turbulent component in our simulation. 

\begin{figure*}
\begin{center}
\plottwo{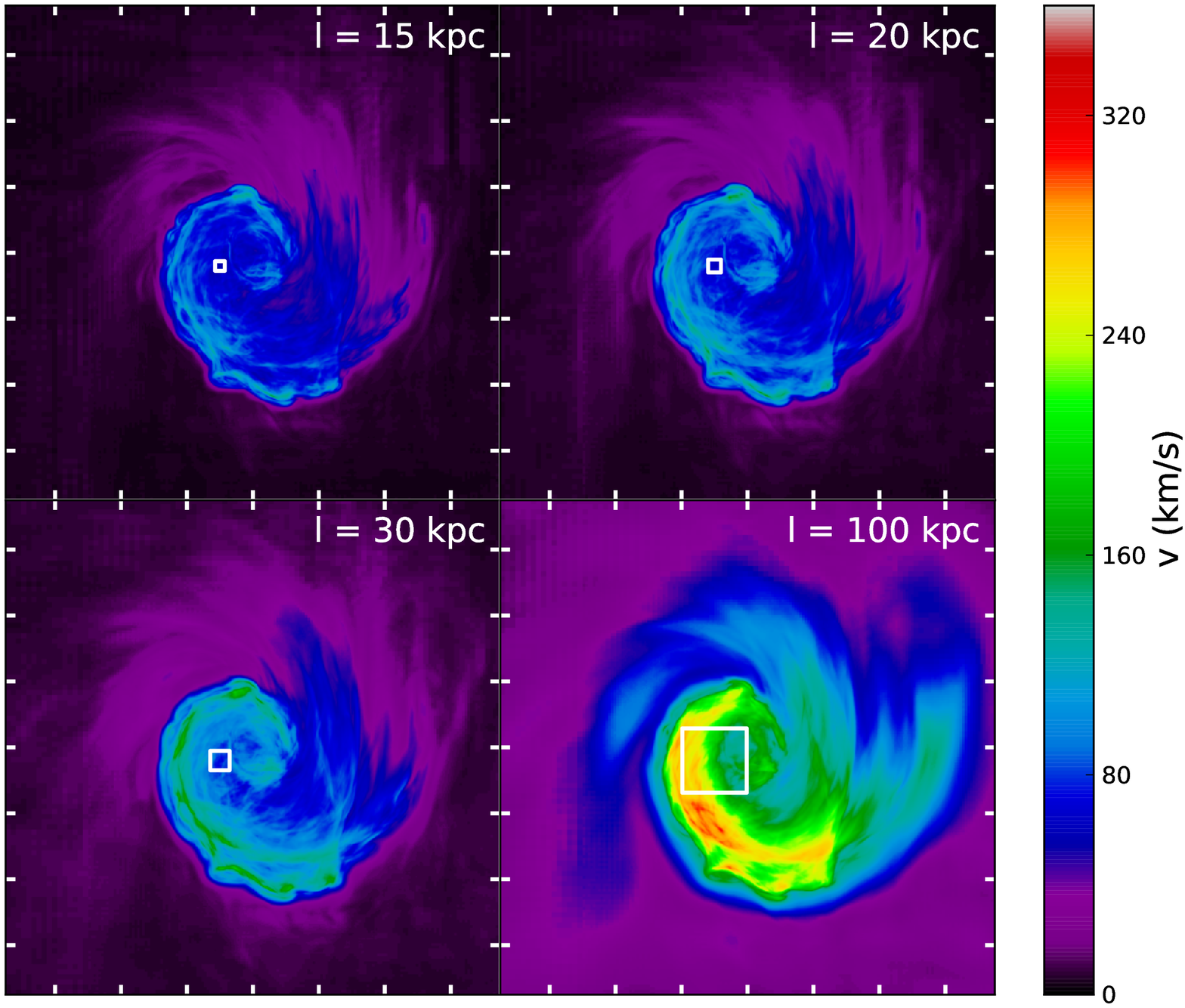}{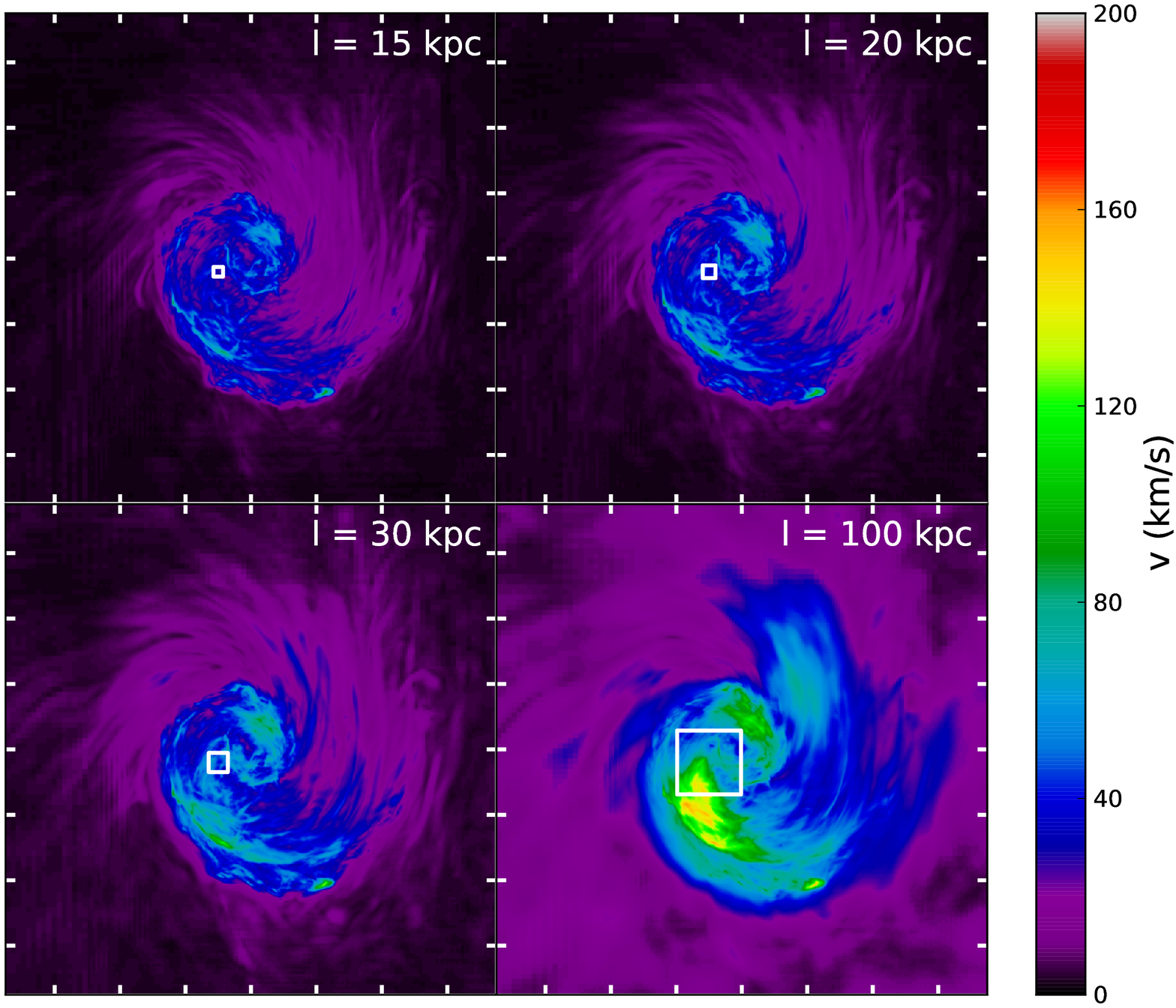}
\caption{Projected (mass-weighted) turbulent velocity map using all velocity components in km s$^{-1}$ at the epoch $t$ = 3.35~Gyr for varying filtering box size $\ell$. Left panels: Projection of all three velocity components. Right panels: Projection of the $v_z$ component only. In each panel a sample filtering box is shown. Each panel is 750~kpc on a side. Tick marks indicate 100~kpc distances.\label{fig:vels}}
\end{center}
\end{figure*}

To get a sense of the location of the tubulence in our simulations, we create projected maps of the mass-weighted turbulent velocity:
\begin{equation}
\delta{v}_{\rm rms} = \langle|\delta{\bf v}|^2\rangle^{\frac{1}{2}} = \left[\frac{\displaystyle\int\rho_g|\delta{\bf v}|^2dz}{\displaystyle\int\rho_g{dz}}\right]^{\frac{1}{2}}
\end{equation}
where the integrals are taken along the line of sight. The left panel of Figure \ref{fig:vels} shows the projected mass-weighted total velocity at the epoch $t = 3.45$~Gyr, $\sim$1.7~Gyr after the subcluster core passage, for a few different choices of the filtering box size $\ell$. There is a clear distinction between the core region bounded by the sloshing cold fronts and the region outside; inside this region, $\delta{v}_{\rm rms} \sim 50-200$~km~s$^{-1}$, while outside the cold fronts, $\delta{v}_{\rm rms} \sim 10-20$~km~s$^{-1}$. However, the highest values of projected $\delta{v}_{\rm rms}$ appear to be near the cold front surfaces, which is to be expected from the preceding discussion. The speed of these motions increases with increasing box size $\ell$, as the larger box size permits larger-scale flows to be interpreted as ``turbulent'' by the filtering scheme. 

The right panel of Figure \ref{fig:vels} shows the projected mass-weighted $z$-component of velocity at the same epoch $t = 3.45$~Gyr, for the same choices of the filtering box size $\ell$. In this case, there is still a clear distinction between the sloshing core region and the outside region of the simulation domain. However, the $v_z$ field does not in general have anomalously large values along the cold front surfaces. This indicates that (given our filtering procedure) the $v_z$ component of the filtered velocity is a better indicator of the turbulent pattern in our simulated ICM. As expected, the turbulent energy within the sloshing region is smaller than in the case where the $v_x$ and $v_y$ components of velocity were included. In our simulations, we will use $\delta{v_z}$ to estimate the turbulent velocity, scaling $\delta{v_z}$ by an appropriate factor to reflect the fact that the total kinetic energy in turbulence is larger than just the contribution from this component alone. Note that because it is the sloshing and shear in the $x-y$ plane that drives turbulence, the true $v_x$ and $v_y$ turbulent components will be relatively higher than they would be in isotropic turbulence, and this should be taken into account. We also expect, on the basis of Figure \ref{fig:vels}, that the overall spatial pattern of turbulence in all three directions will be similar, so by taking the $v_z$ component only, there is no significant loss of spatial information. 

\begin{figure}
\begin{center}
\plotone{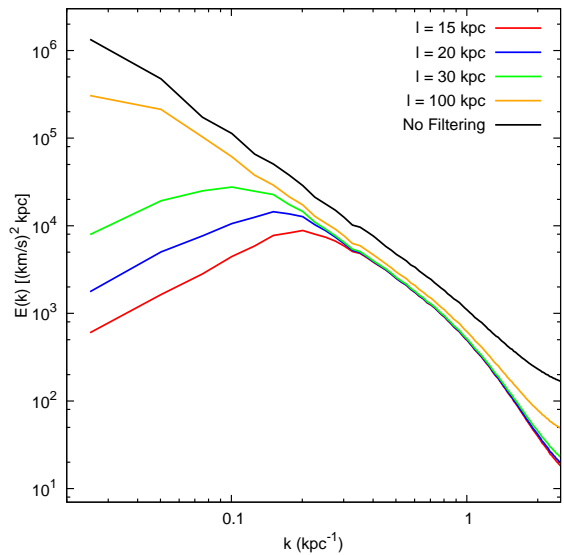}
\caption{The total velocity power spectrum for differing filtering box sizes $\ell$. The epoch is $t$ = 3.45~Gyr.\label{fig:Pk_diff_scales}}
\end{center}
\end{figure}

\begin{figure}
\begin{center}
\plotone{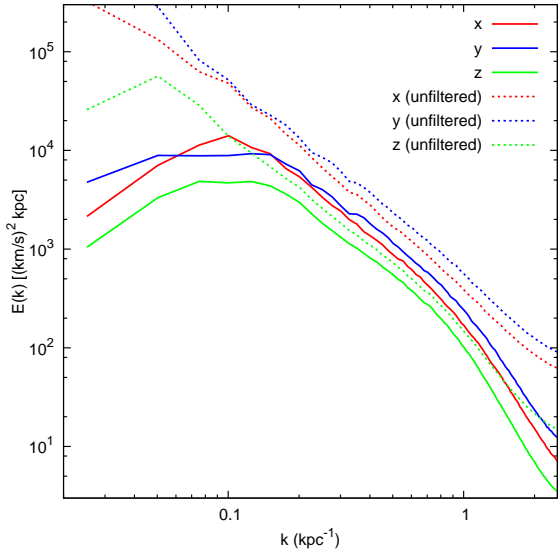}
\caption{The velocity power spectrum for the different components of velocity for the filtered (dashed lines) and unfiltered (solid lines) components of velocity. For the filtered components, the filtering box size is $\ell$ = 30~kpc. The epoch is $t$ = 3.45~Gyr.\label{fig:Pk_diff_components}}
\end{center}
\end{figure}

\begin{figure*}
\begin{center}
\plottwo{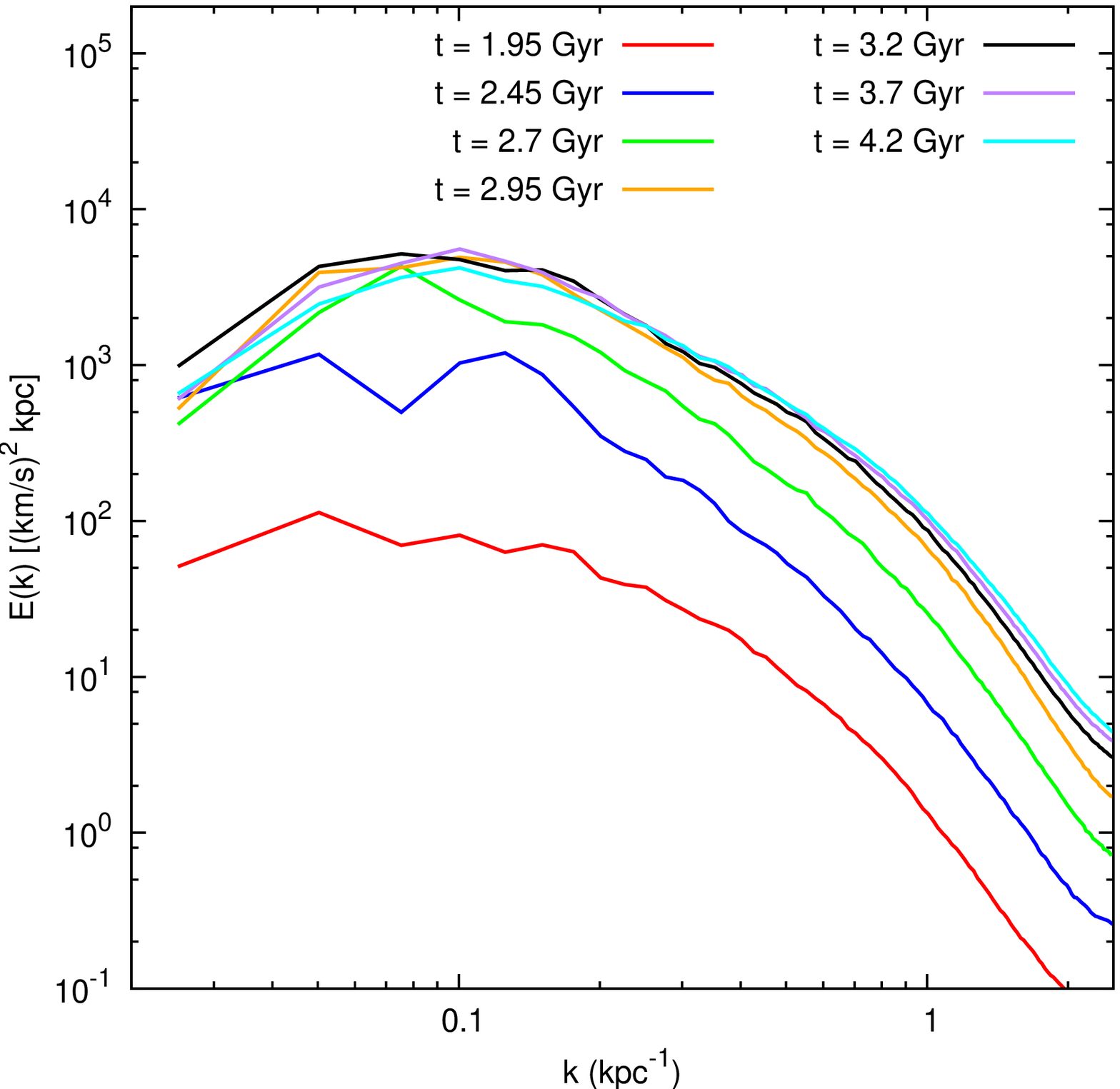}{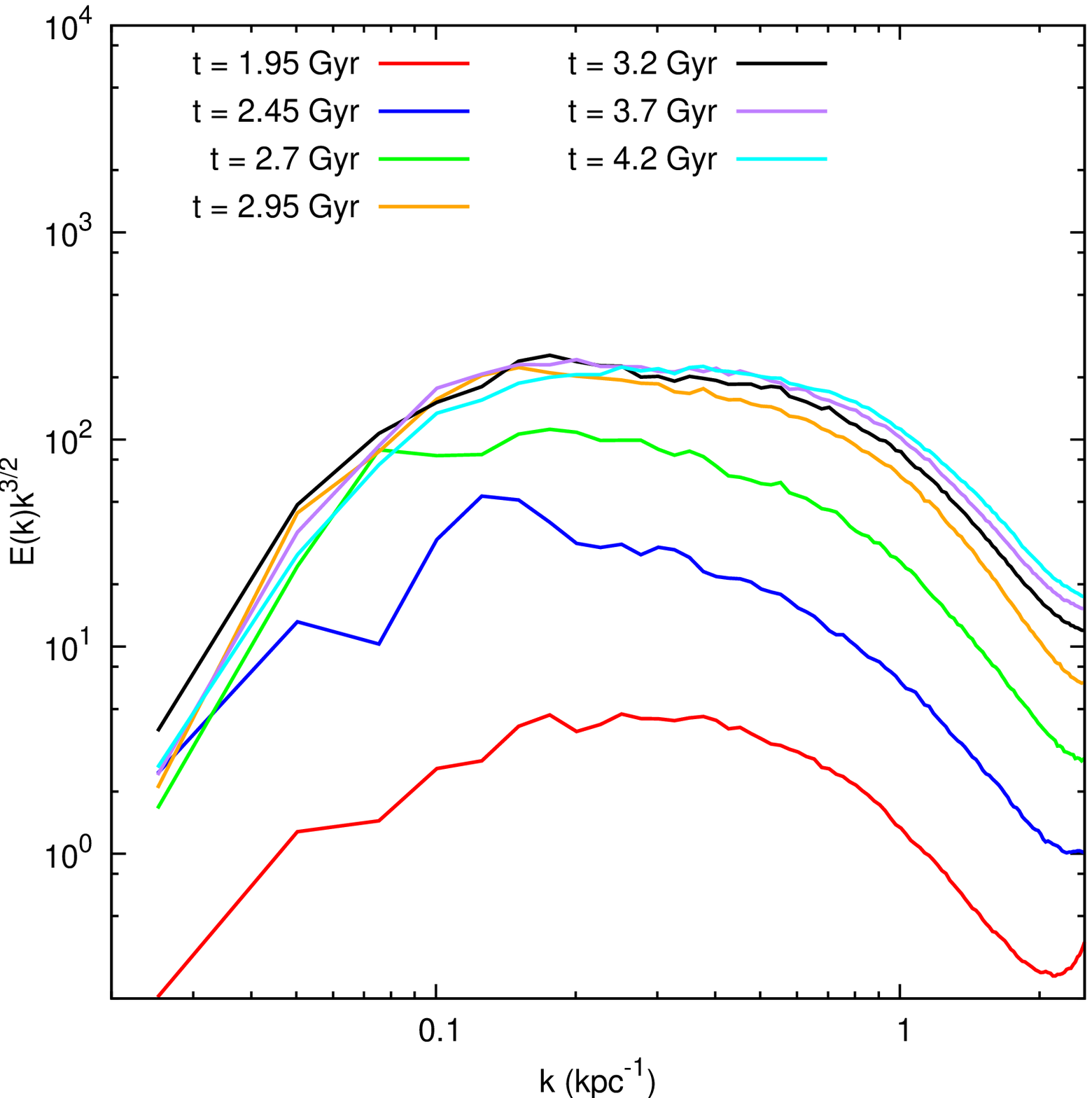}
\caption{Evolution of the $z$-component velocity power spectrum vs time. The filtering box size is $\ell = 30$~kpc. Left panel: The power spectrum $P(k)$. Right panel: The scaled power spectrum $k^{3/2}P(k)$.\label{fig:Pk_diff_epochs}}
\end{center}
\end{figure*}

It is instructive to compare the power spectra of the filtered and unfiltered velocity field, as well as the spectra of the velocity field for different filtering lengths. For this, we wish to examine the region that is dominated by the sloshing motions, so the power spectrum is taken over a box size of $L = $~250~kpc on a side, centered on the potential minimum of the cool-core cluster. The 1-D power spectrum $P(k)$ of the velocity field is defined such that 
\begin{eqnarray}\nonumber
\langle{v^2}\rangle &=& \frac{1}{V}{\displaystyle\int}|{\bf v}({\bf x})|^2d^3{\bf x} \\
\nonumber &=& \frac{1}{8\pi^3V}{\displaystyle\int}|\tilde{\bf v}({\bf k})|^2d^3{\bf k} \\
&=& \frac{1}{2\pi^2V}{\displaystyle\int}P(k)k^2dk
\label{eqn:power_spec}
\end{eqnarray}
which follows from Parseval's theorem. 

Figure \ref{fig:Pk_diff_scales} shows the total velocity power spectra for a few different values of the filtering box scale $\ell$, in comparison to the unfiltered power spectrum at the epoch $t = 3.45$~Gyr of the simulation.  The unfiltered power spectrum maintains a rough power-law shape from low to high wavenumber. This indicates that the sloshing motions contribute power at nearly all scales (as expected, given the geometry of the fronts with sharp velocity discontinuities). In contrast, the filtered power spectra are characterized by the following features: 1) an increase at low wavenumbers, followed by 2) a roughly power-law scaling at intermediate wavenumbers, and finally 3) a drop-off at high wavenumbers. The first part of the spectrum can be explained by the smoothing out of velocity structures with scales larger than the filtering box size $\ell$ due to our filtering procedure. The second part of the power spectrum corresponds to an ``inertial range'' with a power-law scaling $P(k) \propto k^{-\alpha}$. The third part of the power spectrum, a steep decrease in power beginning at wavenumbers corresponding to linear scales $\simlt 8\Delta{x}$, is due to the effect of the numerical dissipation associated with the PPM hydro method used by our simulations \citep[as noted by][]{por94,vaz09,kit09,vaz11}.\footnote{Since this is a dissipative effect on the small-scale turbulent motions and not on the sharp velocity discontinuities of the cold fronts, we do not see it in the unfiltered spectrum.} The unfortunate side effect of the filtering box is that it filters out contributions at low wavenumbers that may be part of the turbulent cascade. As $\ell$ is decreased, the normalization of the power spectrum also decreases. However, we find that that for filtering box sizes $\ell \leq 30$, the power spectrum converges at intermediate to high wavenumbers, indicating that at these scales, we have largely filtered out the sloshing motions. 

Figure \ref{fig:Pk_diff_components} shows the power spectrum in the three different velocity components, for the filtered (with a filtering box size of $\ell =$~30~kpc) and unfiltered velocity fields, at the same epoch $t$ = 3.45~Gyr. Though the curves are roughly the same shape, the power spectra of the $x$ and $y$ components of velocity are each a factor of $\sim$2-3 larger than the power spectrum of the $z$-component of velocity for all $k$, a reflection of the fact that sloshing motions which are responsible for the generation of turbulence are mostly in the $x-y$ plane. This indicates that the total kinetic energy is roughly $\sim$5-6 times higher than the kinetic energy in the $\delta{v_z}$ component of velocity alone (compared to the factor of 3 for isotropic motions). 

The left panel of Figure \ref{fig:Pk_diff_epochs} shows the power spectrum at varying epochs, with a filtering box size $\ell = $~30~kpc. At the beginning of the simulation, prior to the subcluster infall, there is a small degree of turbulence, due to the spurious velocities generated during the relaxation of the initial condition (Section \ref{sec:MHD_sims}). Once the sloshing begins, after the core passage that occurs around $t \sim$1.8~Gyr, the power spectrum normalization increases until approximately $t$ = 3~Gyr, after which the normalization is roughly constant. This is due to the fact that at these late epochs, the volume of our 250~kpc box no longer contains all of the sloshing motions, which continue to develop turbulence outside this region.  Additionally, we note that the shape of the power spectrum curves are very similar between all epochs. The right panel of Figure \ref{fig:Pk_diff_epochs} shows the power spectrum at the same epochs, multiplied by $k^{3/2}$. Within the inertial range of wavenumbers, this quantity at late epochs is fairly flat, indicating rough agreement with our assumption that the power spectrum is of Kraichnan ($P(k) \propto k^{-3/2}$) form at these wavenumbers (Equations \ref{eqn:Dpp1}-\ref{eqn:Dpp9}).\footnote{Adopting a Kolmogorov power spectrum ($P(k) \propto k^{-5/3}$) changes the numerical coefficients only slightly and does not have a substantive effect on our results.}

\subsection{Estimating Reacceleration Coefficients\label{sec:turb_coeff}}

We may use the information regarding the spectrum and spatial distribution of turbulence from our simulation to estimate the reacceleration coefficient at the positions of the tracer particles (Equations \ref{eqn:Dpp1}-\ref{eqn:Dpp9}). As mentioned above, our procedure for estimating turbulence should account for a) our use of only the $v_z$ component of velocity to minimize the influence of non-turbulent motions in the velocity field, b) the fact that only compressive MHD turbulence reaccelerates relativistic particles efficiently, and c) the turbulent cascade being artificially cut off at large scales by the size of the filtering box and at small scales by the resolution scale of the simulation. The first and third of these effects result in an underestimate in the total turbulent kinetic energy available for reacceleration, and the second results in an overestimate. The following represents an attempt to account for these effects in an averaged way.

The first consideration is the underestimate in the kinetic energy due to only taking the $v_z$ component. We have already determined that the power spectra of the three different components have roughly the same shape (see Figure \ref{fig:Pk_diff_components}), and the total turbulent kinetic energy in all three components is roughly a factor of 6 larger than the $v_z$ component alone, which we will adopt in the following estimates.\footnote{Formally, Equation \ref{eqn:Dpp1} assumes isotropic fast modes with respect to {\it the local magnetic field direction}. Still, we adopt the form of Equation 40 from \citet{bru07}, noting that checking whether this condition is completely satisfied is beyond the aim of this paper}.

\begin{figure*}
\begin{center}
\plotone{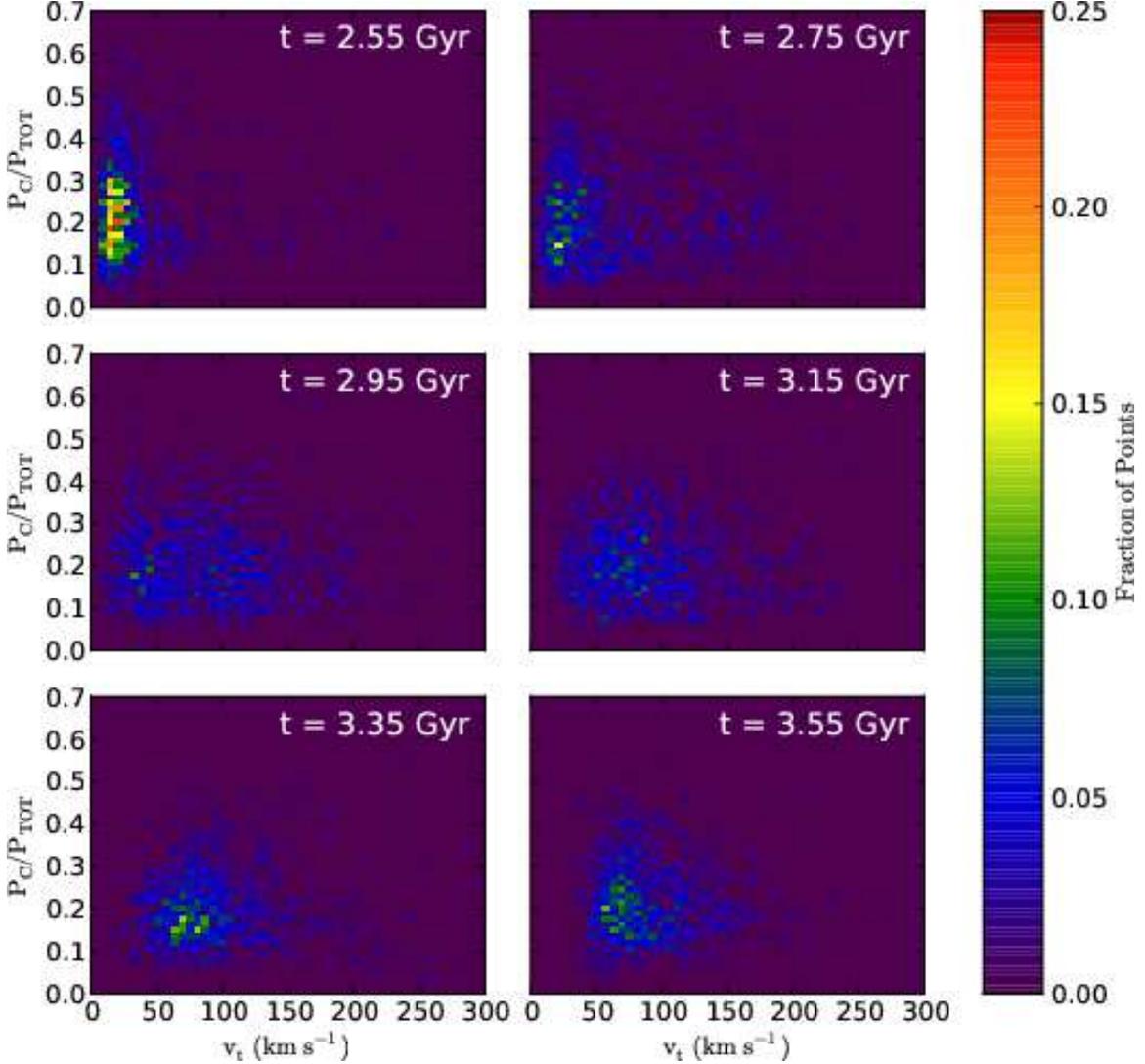}
\caption{Phase plots of turbulent velocity $v_t$ and the ratio $R^c$. As a reminder, the core passage occurs at $t = 1.8$~Gyr.\label{fig:phase_plots}}
\end{center}
\end{figure*}

The second consideration is regarding the fraction of turbulent energy in the form of compressive modes, $R^c$, that we use for the acceleration of particles in our particular model. The turbulence in our case is composed of both solenoidal and compressive components; a number of previous investigations have shown that for a purely solenoidal driving force, that the resulting turbulent cascade will be largely incompressible. For hydrodynamic turbulence, \citep{ber01} estimate a ratio $E_{\rm C}/E_{\rm S} \propto {\rm Re} \times M_t^2$, $M_t$ and ${\rm Re}$ being the turbulent Mach number and the Reynolds number in the medium (though this scaling only applies if ${\rm Re} \times M_t^2 < 10$). Using numerical simulations of MHD turbulence, \citet{cho03} showed that the drain of energy {\it from solenoidal to compressive motions} is fairly small for subsonic (and super-Alfv\'{e}nic) turbulence, consistent with (their Equation 1): 
\begin{equation}
\frac{(\delta{V})^2_S}{(\delta{V})^2_A} \simlt \left[\frac{V_A^2+c_s^2}{(\delta{V})^2_A}\frac{(\delta{V})_A}{V_A}\right]^{-1}
\end{equation}
where $V_A$ is the Alfv\'{e}n speed and $(\delta{V})_A$ is the characteristic velocity of the Alfv\'{e}nic turbulent modes. Consequently, under our subsonic, moderately super-Alfv\'{e}nic conditions\footnote{$c_s \sim 1000~{\rm km~s^{-1}}$, $V_A \sim 100~{\rm km~s^{-1}}$, $(\delta{V})_A \sim 100-200~{\rm km~s^{-1}}$}, we expect that about 10\% (or less) of the energy of incompressible motions is transferred into the compressive cascade, simply meaning that if both compressible and incompressible modes exist they generate independent cascades. 

For subsonic turbulence, $R^c$ is essentially determined by the driving force \citep[][]{fed11}. In our situation, where two dark matter cores pass by each other with a small impact parameter, both compressible and solenoidal driving is present in the region of the sloshing motions. We attempt to estimate the fraction of compressive turbulent power in our sloshing core as follows. Following other authors \citep[e.g.,][]{ryu08,fed11}, we estimate the ratio of compressive to total power $R^c$ in our simulation by applying a decomposition in the Fourier space of the velocity field into its compressive and solenoidal components. For each of these components, we may compute the power spectrum which will yield an estimate of $R^c$ for a given volume. This quantity cannot be computed at each point in space; therefore, exactly which volume to choose is an important consideration, because there is a range of turbulent velocities at any given epoch within the core, which may be associated with a range of values for $R^c$. Since our filtering procedure essentially filters out contributions to the velocity field on lengths larger than the filtering box size $\ell$, this sets a natural scale that should be sufficient to provide a "local" estimate of $R^c$ for a range of turbulent velocities within the cluster core region. 

For several epochs of the simulation, we generate uniformly gridded velocity data (at our finest resolution of $\Delta{x} = 1$~kpc) within a box size of $L^3$ = (300~kpc)$^3$ centered on the cluster potential minimum, and divide this data into 10$^3$ smaller boxes of size $\ell^3$ = (30~kpc)$^3$ each, the size of our filtering boxes. For each of these small boxes, we take the Fourier transform of the velocity components $v_i({\bf x})$, and separate the compressive and solenoidal components of the transformed velocity field $\tilde{v}_i({\bf k})$ by the following projection operations in $k$-space:
\begin{eqnarray}
\tilde{v}^C_i({\bf k}) &=& k_ik_j\tilde{v}_j({\bf k}) \\
\tilde{v}^S_i({\bf k}) &=& (\delta_{ij} - k_ik_j)\tilde{v}_j({\bf k})
\end{eqnarray}
where the indices ($i,j$) indicate the different spatial components and the Einstein summation convention over repeated indices is assumed. For each of these components a power spectrum may be computed in the usual manner (Equation \ref{eqn:power_spec}).

Figure \ref{fig:phase_plots} shows phase plots of the fraction of $V = \ell^3$ domains with a given average 3D turbulent velocity $v_t$ and ratio $R^c = P_{\rm C}/P_{\rm tot}$ over several epochs of the simulation within the central $L^3$ volume. At $t = 2.55$~Gyr (0.75~Gyr after core passage), near the beginning of the sloshing period, the turbulent velocities are mostly very low ($v_t \simlt 50$~km~s$^{-1}$--most of the sloshing motions are bulk flows), with ratios of $R^c$ clustering around $\sim$0.1-0.3, and a tail of points extending up to $R^c \sim 0.6$. As the sloshing motions expand and drive turbulence, the distribution of turbulent velocities spreads out over the range $\sim$0-250~km~s$^{-1}$, clustering mainly within the $\sim$50-150~km~s$^{-1}$ range, until the epoch $t = 3.95$~Gyr, after which the average velocity in the majority of these regions steadily decreases to the $\sim$50-150~km~s$^{-1}$ range, though a tail of strongly turbulent regions always remains. Throughout this entire time period, the value of $R^c$ for most of these regions stays within $\sim$0.1-0.3, with a tail of regions extending up to $R^c \sim 0.6$, and a typical value of $R^c \approx 0.25$ for cells with higher ($v_t \simgt 100~{\rm km~s}^{-1}$) turbulent velocities, where acceleration is important. Interestingly, the region of the phase plots with high $v_t$ and high $R^c$ is mostly devoid of points, likely related to the fact that the numerical viscosity inherent in the simulation preferentally damps compressible motions. As a consistency check, we compute the ratio $R_c$ for the entire region dominated by the sloshing motions for the same epochs, and we find that during this time $R_c \sim 0.25-0.5$, consistent with our more spatially resolved estimates. 

We conservatively adopt the value $R^c = 0.25$ as the default value for our calculations of the reacceleration and momentum-diffusion coefficients. Adopting the average value is conservative, because what is more relevant in a non-linear process such as particle reacceleration is the tail of high values of $R^c$ and $v_t$, not their average.

A caveat must be made regarding this procedure for determining $R^c$. The Fourier transform over a finite domain assumes that the field is periodic. Computing the Fourier transform of a velocity field on a non-periodic domain is equivalent to taking the transform of a field that has a sharp discontinuity such as a shock or a cold front at the boundaries of the domain. This will add spurious, unphysical power into the computation in both the compressive and solenoidal components of the velocity field, though the amount of spurious power should be small if the domain over which the FFT is taken is larger than the largest scale at which significant power is present in the velocity field. 

In order to make a rough determination of the spurious compressive power that this effect will introduce, we have performed the same analysis on a velocity field that has no compressive power {\it by construction}. To construct this field, we initialize a Gaussian random velocity field in $k$-space on a $L^3$ = (300~kpc)$^3$ domain using a power spectrum very similar to that found in our simulation, and perform a divergence-cleaning operation on this field to remove the compressible component from the field. This field is Fourier-transformed to real space, and the same analysis is performed on this field as on the velocity field from our cluster simulation. Since there is no compressive power in this field, any $R_c > 0$ in the smaller ($\ell$ = 30~kpc) boxes must arise from the aforementioned effect at the edges due to the non-periodic field in the subdomains (we have verified that over the entire domain we find $R_c$ = 0 to machine precision, since it is periodic). We find that the value of $R_c$ in these boxes ranges from 0-0.1, with a mean value of $R_c \approx 0.07$. This indicates that about this much of the compressive power that we estimate from our sloshing simulation is potentially spurious. 

The third consideration regards the inertial range of the power spectrum and its effect on the average wavenumber $\langle{k}\rangle$ as well as on the total estimated kinetic energy of turbulence. The power spectrum begins to drop off from the inertial range at high wavenumbers due to the dissipation associated with the finite resolution of the simulation. In reality, there will be a physical damping scale, and we should try to use that scale, and not the artificial numerical one, for calculating the reaccleration coefficient. 

Turbulence under conditions in cluster cool cores is mostly collisional. Collisionless damping of the magnetosonic waves with the thermal plasma becomes strong as soon as turbulence reaches the electron Coulomb mean free path $\ell_{\rm mfp}$, and it is at scales similar to this scale that we expect the turbulent cascade to end \citep{bru07}. If plasma instabilities play a role, they will make the ICM more collisional, and we may expect that the turbulent cascade would extend to much smaller scales \citep{bru11}. To be conservative, we adopt the standard picture where the turbulent cascade is cut off by collisionless damping on thermal particles. For the conditions in the core of our model cluster, the electron mean free path $\ell_{\rm mfp} \sim 0.2-0.1$~kpc. If we assume that the inertial range of the turbulent cascade extends to at least $k_{\rm cut} \approx 2\pi/\ell_{\rm mfp}$, we conservatively estimate that the kinetic energy in turbulence should be increased, over what we obtain by simply integrating the spectrum in the simulations, by a factor of $f \sim 1.5$.

Taking these considerations into account, we make the following modifications to Equation \ref{eqn:Dpp5} for the TTD coefficient:
\begin{equation}
D_{\rm pp,TTD} \approx 1.5 \times 10^{-11}{\langle{k}\rangle}\left(\frac{f}{1.5}\right)\left(\frac{v_t^2}{v_z^2}\right)\left(\frac{R^c}{0.25}\right){v_z}^2p^2
\label{eqn:Dpp_mod1}
\end{equation}
and combining Equations \ref{eqn:Dpp6}-\ref{eqn:Dpp7} for the non-resonant coefficient:
\begin{equation}
D_{\rm pp,C} \approx 1.3 \times 10^{-12}{k_{\rm mfp}}\left(\frac{f}{1.5}\right)\left(\frac{v_t^2}{v_z^2}\right)\left(\frac{R^c}{0.25}\right){v_z}^2p^2
\label{eqn:Dpp_mod2} 
\end{equation}
where $v_t^2/v_z^2 \approx 6$, $k_{\rm mfp} = 2\pi/l_{\rm mfp}$, and the approximate correction for extending the inertial range to high wavenumber is $f \sim 1.5$. The assumptions regarding the extent of the power spectrum also affect the computation of the average wavenumber $\langle{k}\rangle$. Assuming ($k_{\rm min},k_{\rm cut}$) = ($2\pi/30~{\rm kpc}^{-1}, 2\pi/0.1~{\rm kpc}^{-1}$), from Equation \ref{eqn:avg_k} we find $\langle{k}\rangle \approx 3.6~{\rm kpc}^{-1}$. For our conditions, the coefficients $D_{\rm pp,TTD}$ and $D_{\rm pp,C}$ are of a similar order of magnitude, and they are added to produce the total $D_{\rm pp}$. The various uncertainties associated with these corrections are discussed in Section \ref{sec:uncertainties}.

\section{Results: The Evolution of Relativistic Electrons, Synchrotron, and IC Emission\label{sec:rel_elec}}

\subsection{The Evolution of Electron Spectra\label{sec:electron_spectra}}

Though unobservable directly in real clusters, from our simulations we may examine the relativistic electron spectrum of the cluster as a function of time during the simulation. We construct the energy spectrum $N(\gamma)$ by binning up the relativistic particle samples into 100 equally log-spaced bins over the range $(\gamma_{\rm min}, \gamma_{\rm max}) = (10, 10^5)$. For simplicity, we include all of the relativistic particles in the cluster in the binning procedure (we will examine the spatial distribution of the relativistic electrons in the next section).

\begin{figure}
\begin{center}
\plotone{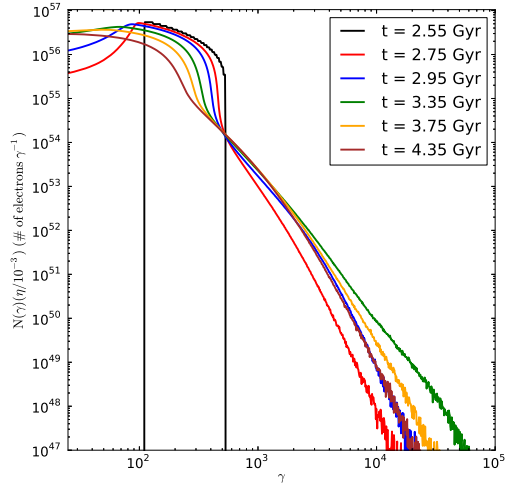}
\caption{Relativistic electron spectra for several different epochs of the simulation, beginning with the initially injected spectrum at $t$ = 2.55~Gyr.\label{fig:electron_accel}}
\end{center}
\end{figure}

\begin{figure*}
\begin{center}
\plotone{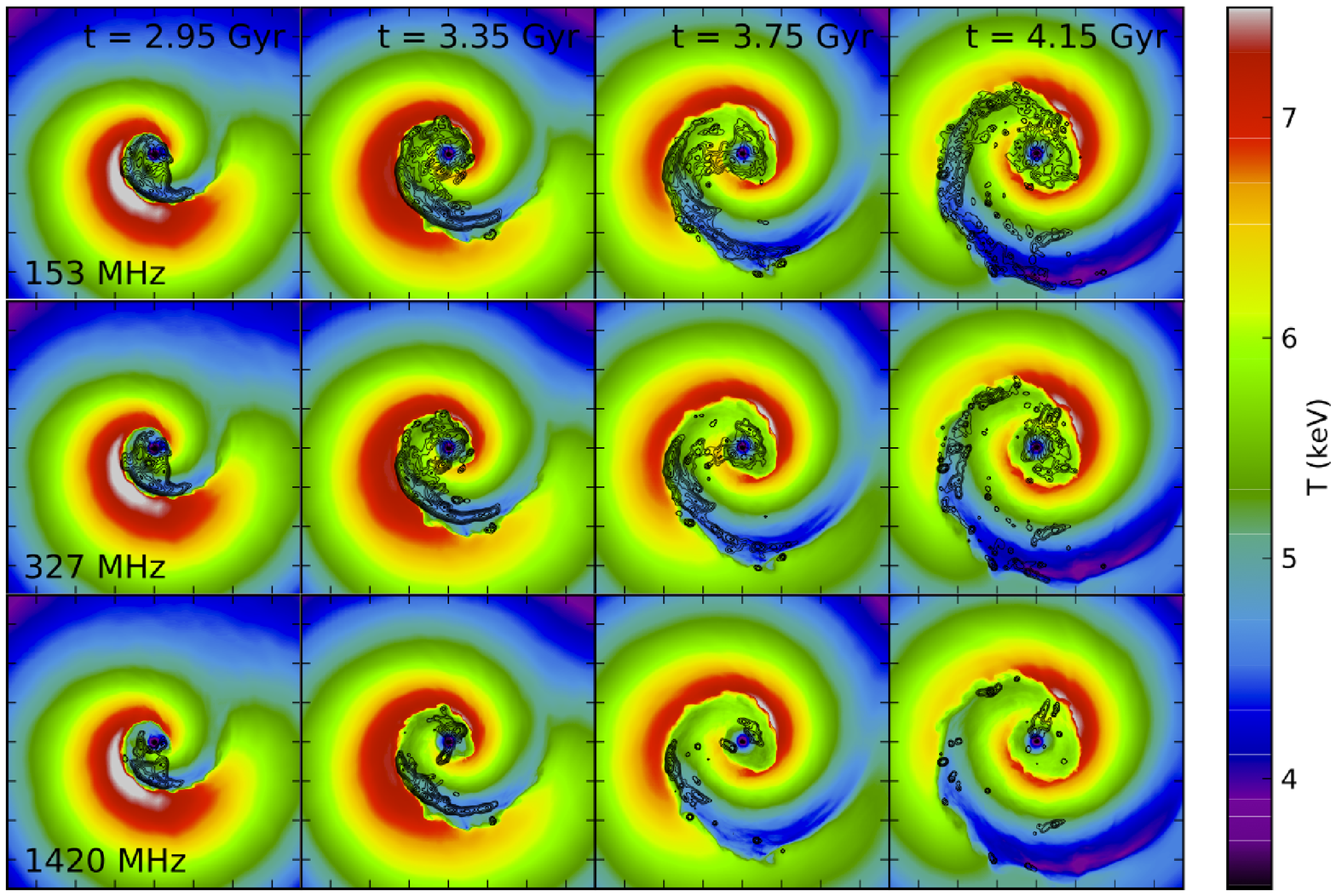}
\caption{Projected gas temperature maps with radio contours overlaid at several epochs for the frequencies 153, 327, and 1420~MHz in the $z$-projection. The colorbar is temperature in keV. Contours of radio emission at (153, 327, 1420)~MHz begin at $(1.0, 0.5, 0.125) \times 10^{-3}$~mJy arcsec$^{-2}$ and increase by a factor of 2. Each panel is 750~kpc on a side. Tick marks indicate a distance of 100~kpc.\label{fig:contours_z}}
\end{center}
\end{figure*}

\begin{figure*}
\begin{center}
\plotone{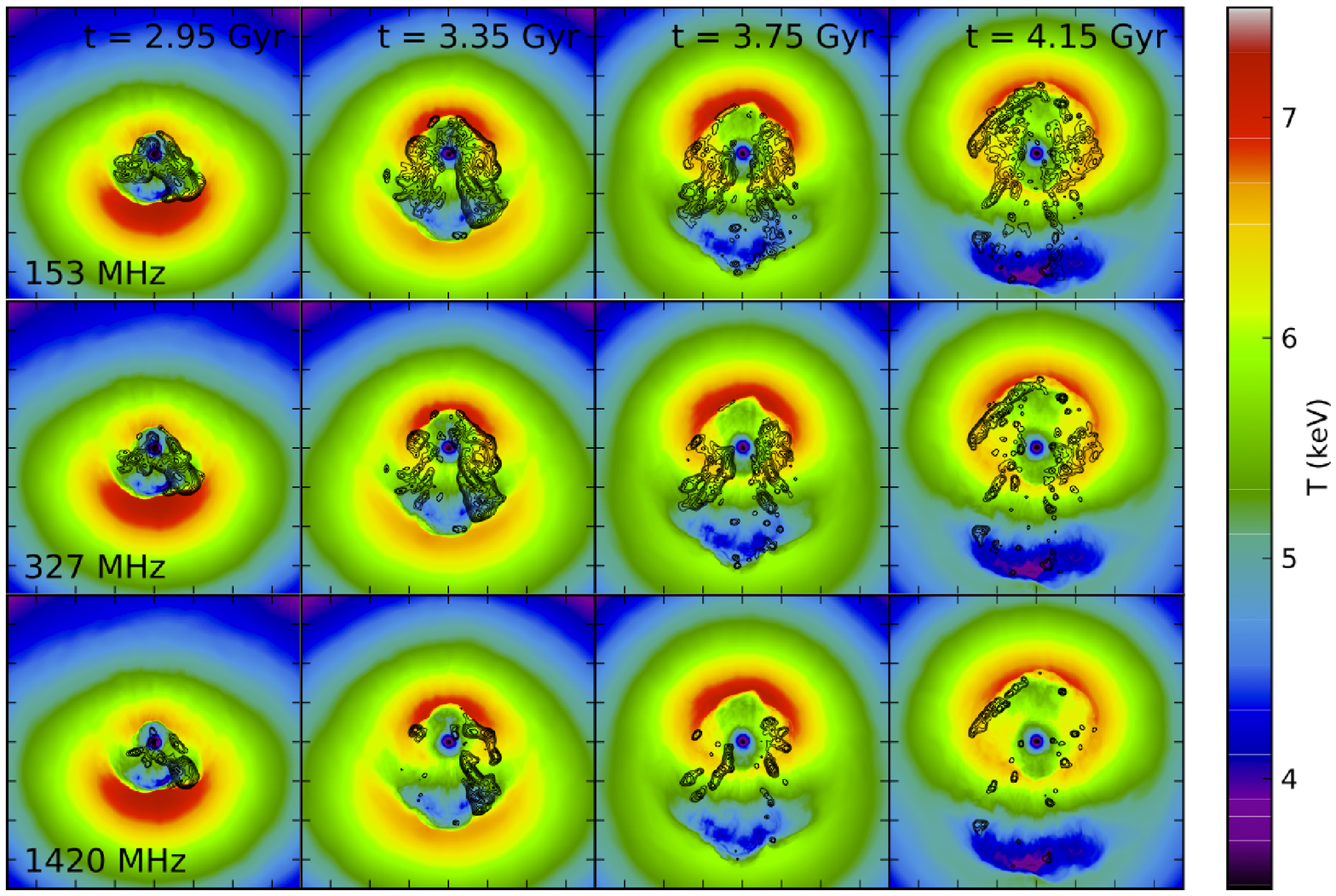}
\caption{Projected gas temperature maps with radio contours overlaid at several epochs for the frequencies 153, 327, and 1420~MHz in the $x$-projection. The colorbar is temperature in keV. Contours of radio emission at (153, 327, 1420)~MHz begin at $(1.0, 0.5, 0.125) \times 10^{-3}$~mJy arcsec$^{-2}$ and increase by a factor of 2. Each panel is 750~kpc on a side. Tick marks indicate a distance of 100~kpc.\label{fig:contours_x}}
\end{center}
\end{figure*}

\begin{figure*}
\begin{center}
\plotone{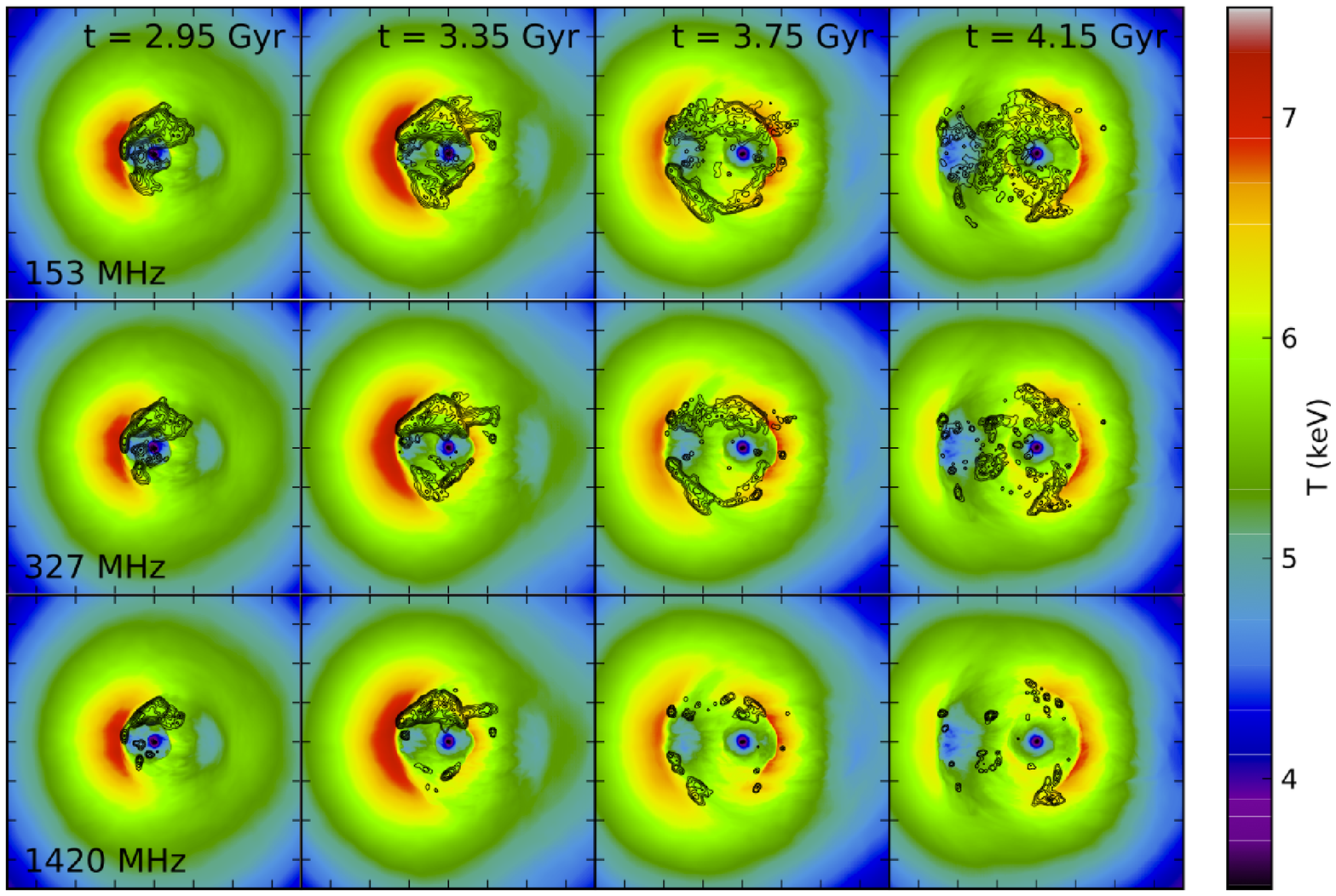}
\caption{Projected gas temperature maps with radio contours overlaid at several epochs for the frequencies 153, 327, and 1420~MHz in the $y$-projection. The colorbar is temperature in keV. Contours of radio emission at (153, 327, 1420)~MHz begin at $(1.0, 0.5, 0.125) \times 10^{-3}$~mJy arcsec$^{-2}$ and increase by a factor of 2. Each panel is 750~kpc on a side. Tick marks indicate a distance of 100~kpc.\label{fig:contours_y}}
\end{center}
\end{figure*}

Figure \ref{fig:electron_accel} shows the relativistic electron energy spectrum $N(\gamma)$ for all tracer particles for several different epochs of the simulation, with times given from the beginning of the simulation. Only 0.2~Gyr after the injection of relativistic particles, reacceleration has already generated a population of particles with energies up to $\gamma \sim 10^4$.\footnote{Since there is such a large change in the electron spectrum within the first 0.2~Gyr, we do not expect much of a dependence on the initial electron spectrum. We have experimented with a few different initial spectra to confirm this.} At later times (over the course of approximately 2~Gyr), reacceleration maintains a population of relativistic electrons up to $\gamma \approx 2 \times 10^4$. In our particular merger setup, reacceleration can no longer keep up with the cooling of the relativistic particles about 1.5~Gyr after the core passage ($t \sim 3.5$~Gyr), and the spectrum begins to steepen. 
 
\subsection{Simulated Synchrotron Radiation\label{sec:synchrotron}}

\begin{figure*}
\begin{center}
\plotone{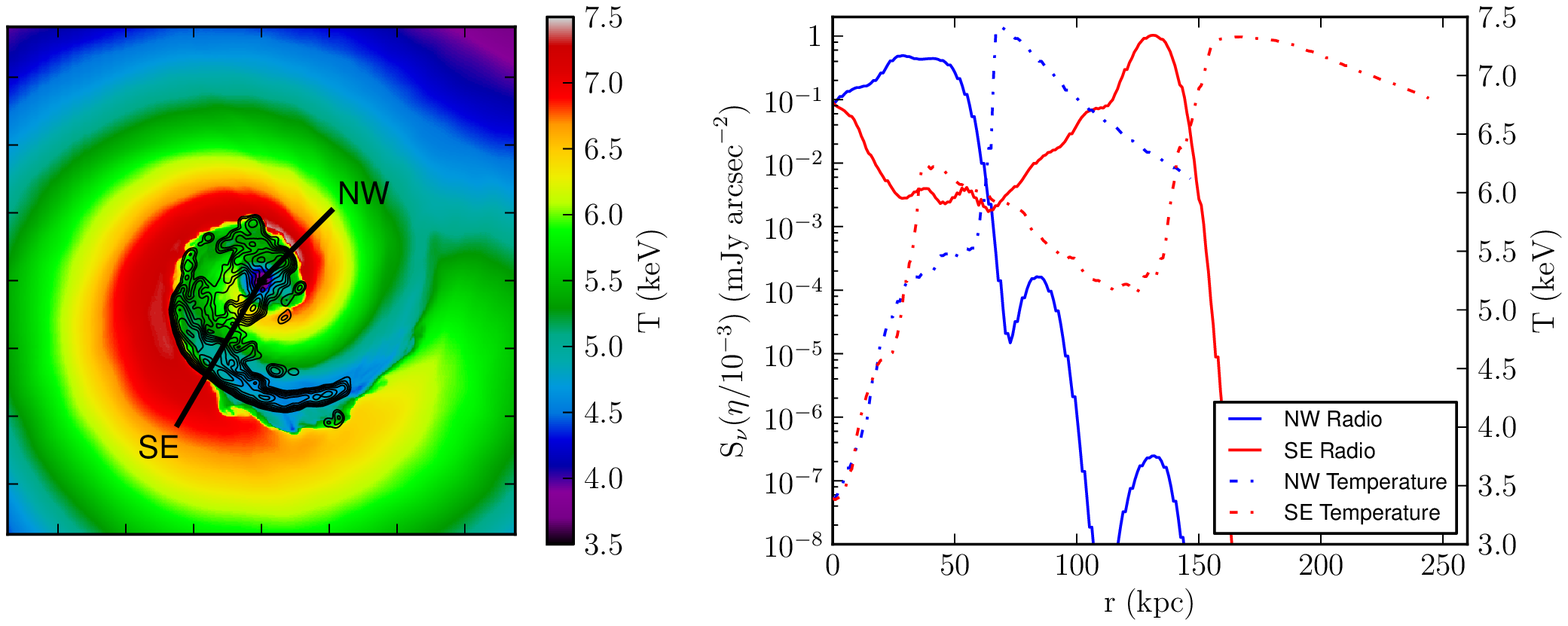}
\caption{Profiles of gas temperature and radio emission across cold front surfaces. Left: Projected gas temperature maps with radio contours overlaid at the epoch $t$ = 3.35~Gyr, in the $z$-projection. The colorbar is temperature in keV. Contours are of 327~MHz radio emission which begin at $5 \times 10^{-4}$~mJy arcsec$^{-2}$ and increase by a factor of 2. The panel is 750~kpc on a side. Tick marks indicate 100~kpc distances. Right: Profiles of projected gas temperature and radio emission from the left panel, along the lines in the left panel.\label{fig:primary_profiles}}
\end{center}
\end{figure*}

\begin{figure*}
\begin{center}
\plottwo{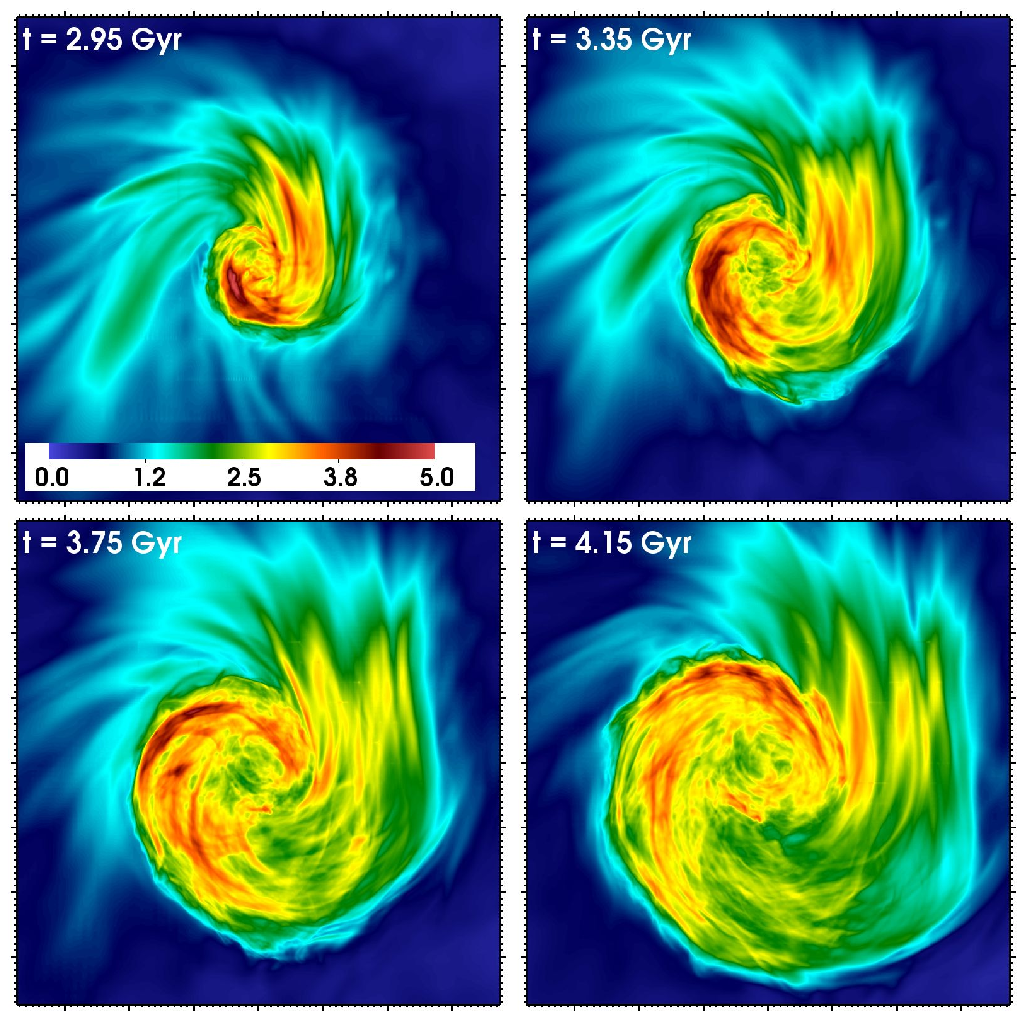}{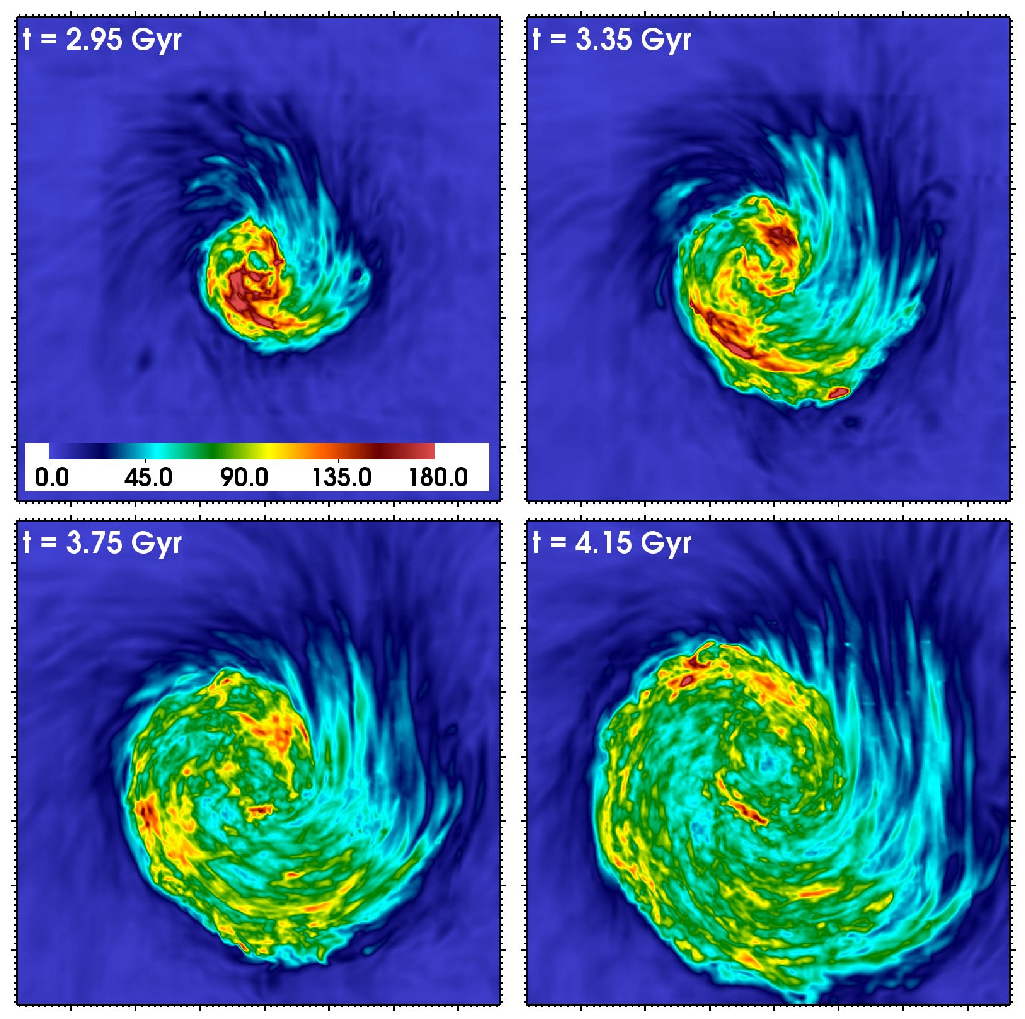}
\caption{Left: Projected (volume-weighted) magnetic field strength in the $z$-direction for the epochs $t$ = 2.95, 3.35, 3.75, and 4.15~Gyr. Right: Projected (mass-weighted) turbulent velocity (estimated using only the $v_z$ component and scaled to match the total turbulent energy) in the $z$-direction for the same epochs. Each panel is 750~kpc on a side. Major tick marks indicate 100~kpc distances.\label{fig:proj_z}}
\end{center}
\end{figure*}

\begin{figure*}
\begin{center}
\plottwo{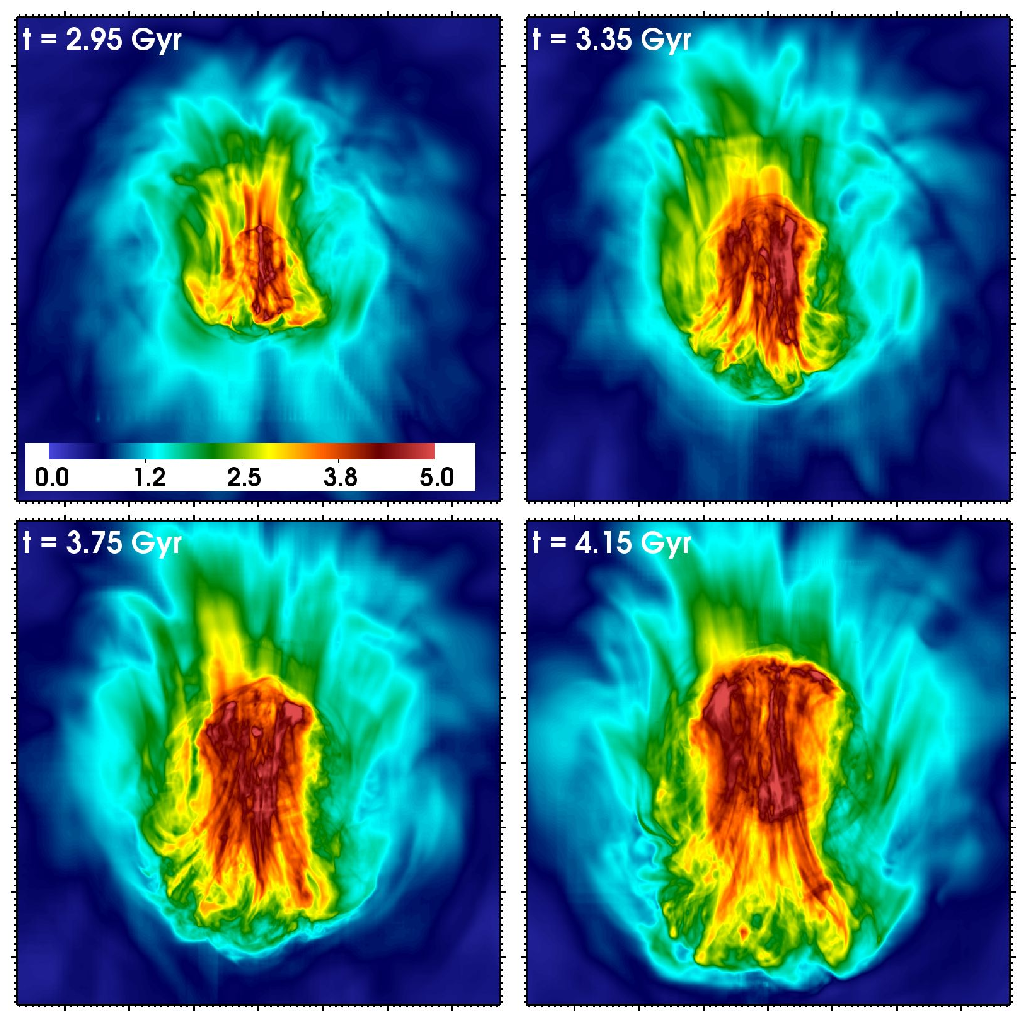}{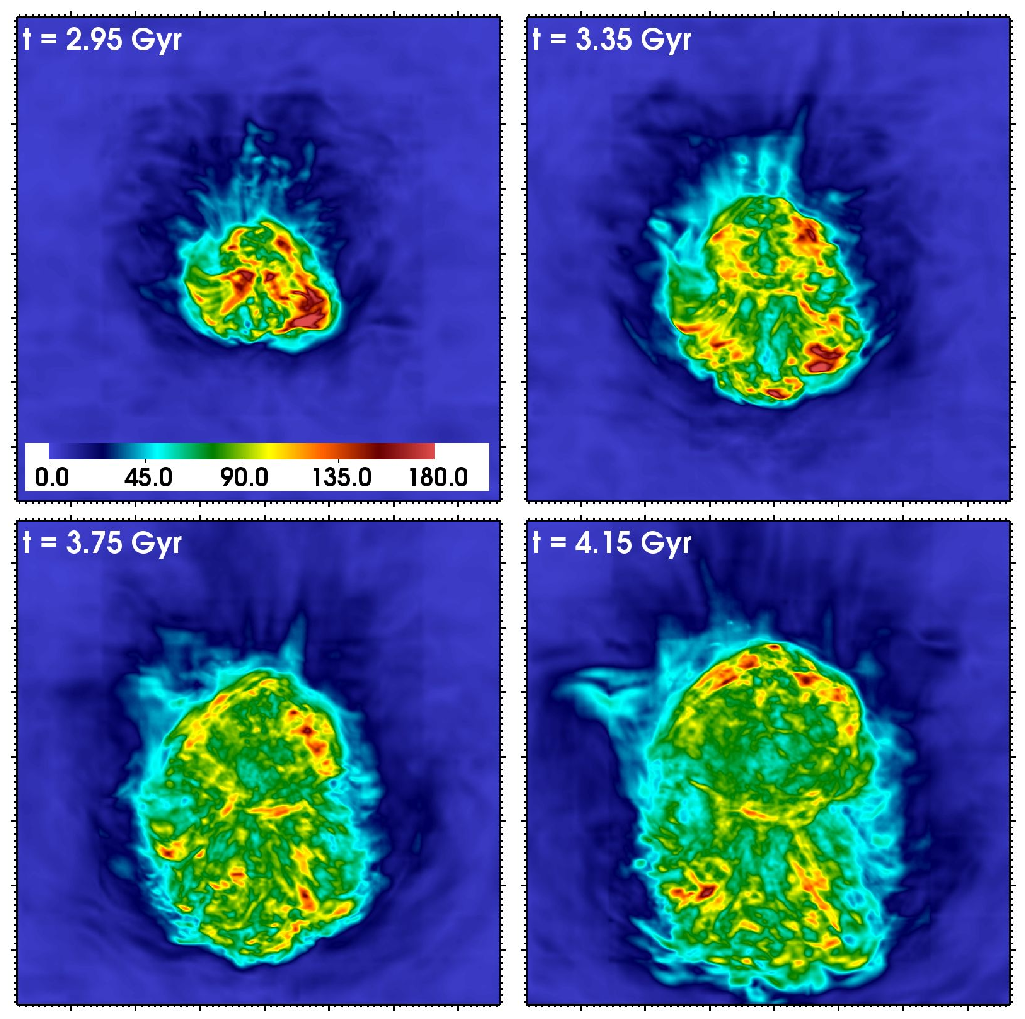}
\caption{Left: Projected (volume-weighted) magnetic field strength in the $x$-direction for the epochs $t$ = 2.95, 3.35, 3.75, and 4.15~Gyr. Right: Projected (mass-weighted) turbulent velocity (estimated using only the $v_z$ component and scaled to match the total turbulent energy) in the $x$-direction for the same epochs. Each panel is 750~kpc on a side. Major tick marks indicate 100~kpc distances.\label{fig:proj_x}}
\end{center}
\end{figure*}

\begin{figure*}
\begin{center}
\plottwo{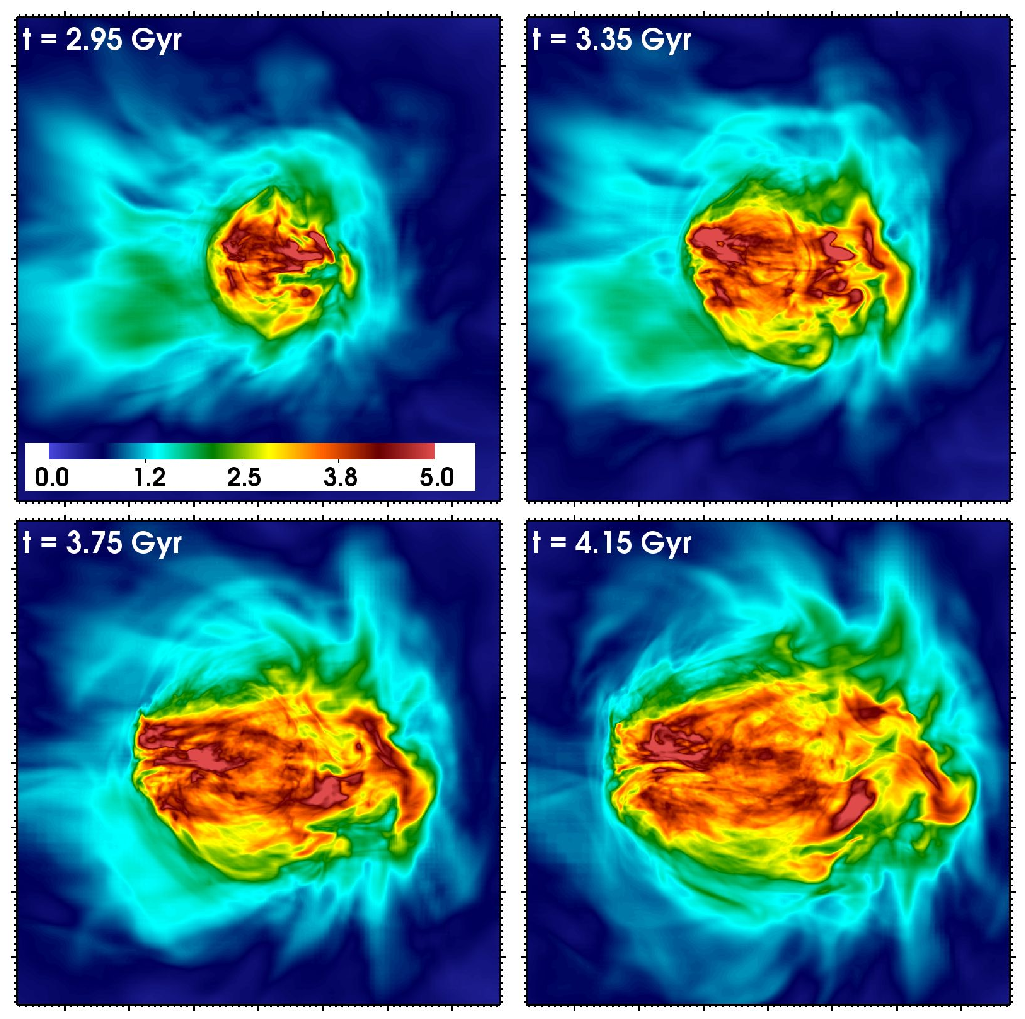}{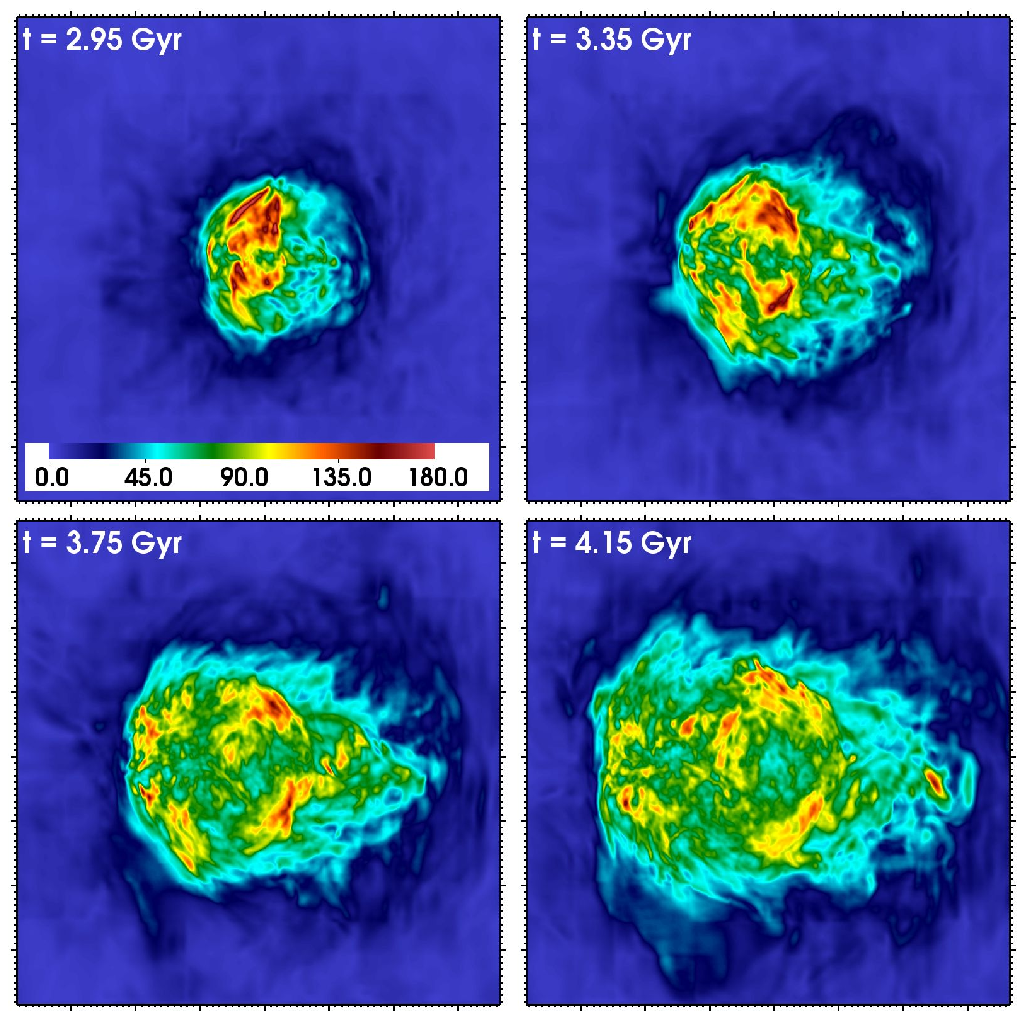}
\caption{Left: Projected (volume-weighted) magnetic field strength in the $y$-direction for the epochs $t$ = 2.95, 3.35, 3.75, and 4.15~Gyr. Right: Projected (mass-weighted) turbulent velocity (estimated using only the $v_z$ component and scaled to match the total turbulent energy) in the $y$-direction for the same epochs. Each panel is 750~kpc on a side. Major tick marks indicate 100~kpc distances.\label{fig:proj_y}}
\end{center}
\end{figure*}

Once we have the accelerated electron spectra at a given epoch for each tracer particle, we can use them to compute the synchrotron radiation they emit. The synchrotron power for a single electron as a function of frequency is \citep{ryb79}:
\begin{equation}
P(\nu,\gamma) = \frac{\sqrt{3}e^3B_{\perp}}{m_ec^2}F(x)
\end{equation}
where $F(x)$ is the synchrotron function, and $x = \nu/\nu_c$, where $\nu_c = (3/4\pi)\gamma^3eB_{\perp}/m_ec$ is the synchrotron critical frequency. The total synchrotron power at a given frequency for each tracer particle is then the sum of synchrotron powers for the electron samples associated with the tracer particle scaled by the normalization constant:
\begin{equation}
P_{\rm tot,j}(\nu) = \displaystyle\int{P(\nu,\gamma)N_j(\gamma)}d\gamma = K_j\displaystyle\sum_iP(\nu,\gamma_{i,j}).
\end{equation}
We assume an isotropic distribution of electron pitch angles. To generate maps of projected synchrotron emission, we construct a 2-D grid upon which the tracer particle luminosities are mapped according to the ``cloud-in-cell'' (CIC) prescription \citep{hoc88} and projected along the chosen line of sight. The resulting synchrotron brightness for each sky pixel is given by $I_\nu = (1+z)L_{\nu(1+z)}/4\pi{D_L^2}/\Delta{\Omega}$, where $D_L$ is the luminosity distance and $\Delta{\Omega} = \Delta{x}\Delta{y}/D_A^2$, where $D_A$ is the angular diameter distance. Our mock brightness maps are then convolved with a 2-D gaussian of FWHM 10~kpc, to simulate the effect of the PSF for a high-resolution instrument (this corresponds to 3''-10'' for $z$ = 0.05-0.2). No attempt was made to simulate the interferometric effects on the images.  

Figures \ref{fig:contours_z} through \ref{fig:contours_y} show the projected gas temperature with radio brightness contours overlaid, for the frequencies 153, 327, and 1420~MHz in the $z$-projection  (these fiducial values are selected to correspond to the GMRT and VLA frequencies). From these maps, it can be seen that the radio emission at lower frequencies persists over a long period and is bounded by the core cold fronts, apparent in the temperature maps. The minihalo emission at $\nu$ = 1.4~GHz becomes dimmer, covers a smaller area on the sky, and becomes more patchy and amorphous. The time evolution will be discussed in more detail below. The maps in the $z$-projection are particularly striking when compared to the minihalo in RXJ\,1720.1+26 from Figure \ref{fig:RXJ1720}. 

\begin{figure}
\begin{center}
\plotone{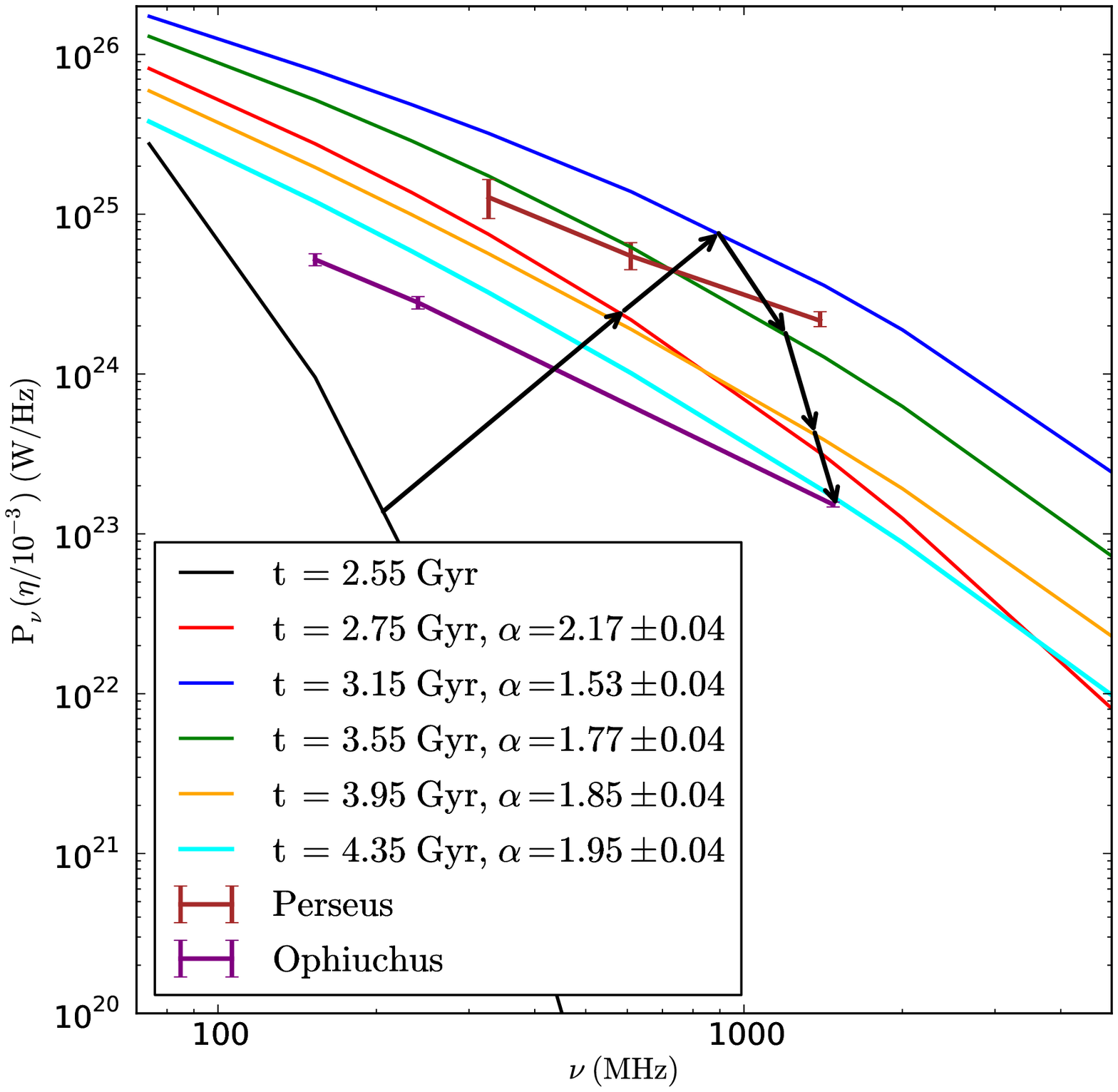}
\caption{Time dependence of the spectrum of the mini-halo within a radius of 300~kpc of the cluster center in W/Hz vs. the frequency of the emission in MHz. The effective power law slope $\alpha$ betwen the frequencies 327-1420~MHz is listed in the key, with 1$\sigma$ errors given. The arrows indicate the rise and fall of the spectrum with time. The spectra for the Perseus \citep{sij93} and Ophiuchus \citep{mur10} cluster minihalos are plotted for comparison.\label{fig:primary_spectrum}}
\end{center}
\end{figure}

In the projections that are in the orbital plane (along the $x$ and $y$-axes), it is difficult to see the cold fronts in the temperature maps, but the radio emission is still clearly bounded by them, with a radius $r \sim 100-300$~kpc that increases as the volume of the sloshing region increases with time. Figure \ref{fig:primary_profiles} shows example profiles of the 327~MHz radio emission in the $z$-projection at the epoch $t$ = 3.35~Gyr, demonstrating the lack of radio emission beyond the cold front surfaces. The spatial coincidence of the radio emission with the X-ray cold fronts in the cool core, and the steep cutoff of the mini-halo emission are in agreement with observed mini-halos \citep[e.g.,][]{maz08,gia11}.

\begin{figure}
\begin{center}
\plotone{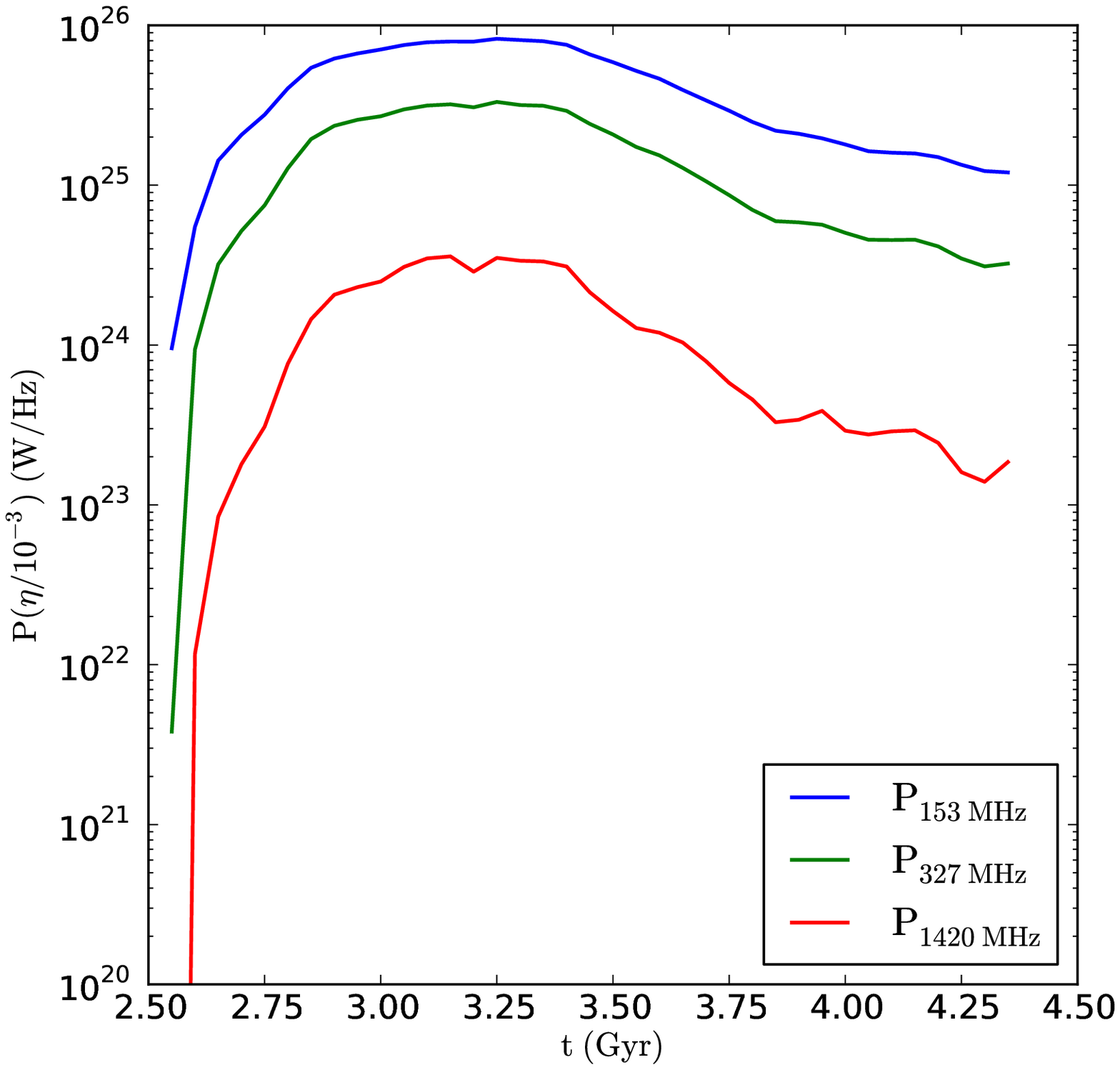}
\caption{Time dependence of the synchrotron powers of the mini-halo in W/Hz at the frequencies 153, 327, and 1420~MHz within a radius of 300~kpc of the cluster center. The initial radio power at the frequency of 1420~MHz is much smaller than shown ($P \sim 10^{10}$~W/Hz); the lower limit of the y-axis is set at a higher value for clarity.\label{fig:primary_powers}}
\end{center}
\end{figure}

It is instructive to examine the properties of the thermal plasma over the same period to compare to the features seen in the radio maps. The radio emission will be dependent upon the turbulent velocity to continuously reaccelerate electrons, and the magnetic field to produce the emission itself. Figures \ref{fig:proj_z} through \ref{fig:proj_y} show the projected magnetic field strength and turbulent velocity over the same epochs as the radio emission in the preceding figures. Within the volume of the sloshing region, the magnetic field has been significantly amplified, and the turbulence is strongest (though not necessarily in the same locations within the cool core). The fact that both of these effects are bounded within the envelopes of the cold fronts is what constrains the radio emission to these boundaries. Additionally, within the sloshing region, the amplification of the magnetic field and the strength of the turbulent velocity is far from uniform. The regions with the highest turbulent velocities ($\delta{v} \sim 100-200$~km~s$^{-1}$) span spatial scales of $\sim$50-100~kpc. Significant fluctuations in both of these quantities result in stronger emission in localized regions. In particular, it appears that the regions of brightest radio emission correspond to the regions which have the highest turbulent velocities. This is due to the fact that the reacceleration timescale, which is on the order of 0.1~Gyr, is less than the timescale of the drift of regions with high turbulent motions, which is on the order of Gyr.

One important observed characteristic of mini-halos is their steep radio spectra. This is naturally explained by the reacceleration model implemented in our simulations. The balance between reacceleration and losses sets a cutoff energy at which there is a sharp drop in the relativistic electron population, which in turn produces a steepening in the synchrotron spectrum. Since there is a range of magnetic field strengths in the cluster core, this break frequency will be different for different electrons, and the resulting spectrum will gradually steepen at higher frequencies. Assuming a power-law spectrum for the radio emission of the form $I_\nu \propto \nu^{-\alpha}$, all observed minihalos have a spectral index $\alpha \sim 1-2$ around the frequency of 1~GHz, a much steeper spectrum than the radio emission from radio galaxies \citep[$\alpha \sim 0.5-0.8$,][]{con92}. 

Figure \ref{fig:primary_spectrum} shows the evolution of the synchrotron spectrum of the emission within 300~kpc of the cluster center over the duration of the simulation. The initial spectrum at the epoch $t$ = 2.55~Gyr is not shown on the plot since it is many orders of magnitude fainter than the spectrum at later epochs. After the initial quick increase in emission on a 0.5-1~Gyr timescale, from the epoch $t \sim$ 3~Gyr onward the spectrum of the mini-halo becomes fainter and steepens at higher frequencies, as cooling begins to dominate over turbulence and fewer electrons are able to emit significant emission at these frequencies. To quantify this steepeing, we determine the spectral index of our simulated emission for these epcohs. We fit the total synchrotron spectrum to a power-law $P_\nu \propto \nu^{-\alpha}$ over the frequencies 327, 610, and 1420~MHz, assuming the errors to be 10\% of the flux at each frequency. The evolution of the spectral index is tabulated in Figure \ref{fig:primary_spectrum}. For most of the evolution of the mini-halo, the spectral index hovers around $\alpha \approx 1-2$, comparable to that of observed mini-halo sources. The spectra for the minihalos in the Perseus \citep{sij93} and Ophiuchus \citep{mur10} clusters are shown for comparison. The spectrum of simulated cluster is compatible in both shape and normalization (though note that our normalization is uncertain by an order of magnitude due to the weak constraints provided by IC measurements). 

Figure \ref{fig:primary_powers} shows the evolution of the total synchrotron power within $r$ = 300~kpc of the cluster center in the $z$-projection in W/Hz at the frequencies 153, 327, and 1420~MHz. From the beginning of the injection of the relativistic particles at $t$ = 2.55~Gyr until $t \sim$ 3.2~Gyr, the synchrotron power increases from initially low values due to the acceleration of the low-energy particles. At $t \sim$ 3.2~Gyr, the synchrotron power at each frequency reaches a maximum, in line with the evolution of the electron spectrum (see Figure \ref{fig:electron_accel}). The peak of the radio power for our simulated cluster is consistent with that of observed mini-halos, which typically fall in the range of $P_{1.4~{\rm GHz}} \sim$~a few~$\times 10^{23}$~-~a few~$\times 10^{24}$~W/Hz \citep[][again, note that our normalization is uncertain by an order of magnitude.]{cas08}. After this, the power at each frequency slowly decays. The period during which the total radio power of the simulated mini-halo is within a factor of 10 from its peak value is about 1~Gyr for 1.4~GHz, and slightly longer for lower frequencies (consistent with the evolution of the electron spectrum--the 1.4~GHz emission comes from electrons with $\gamma > (1-2) \times 10^4$). This is significant, because it indicates that if mini-halos are powered by reacceleration they must be transient sources, particularly at higher frequencies. These properties are similar to those expected in the case of giant radio halos in the turbulent reacceleration model. Based on GMRT radio survey data, \citet{cas08} concluded that mini-halos are rare, and interpreted this as support for their origin as a result of transient reacceleration events due to minor merging activity. Additionally, the longer lifetime of radio emission at lower frequencies suggests that future surveys of clusters of galaxies at lower radio frequencies with observatories such as LOFAR might reveal many more mini-halos in relaxed clusters. 

\begin{figure*}
\begin{center}
\plotone{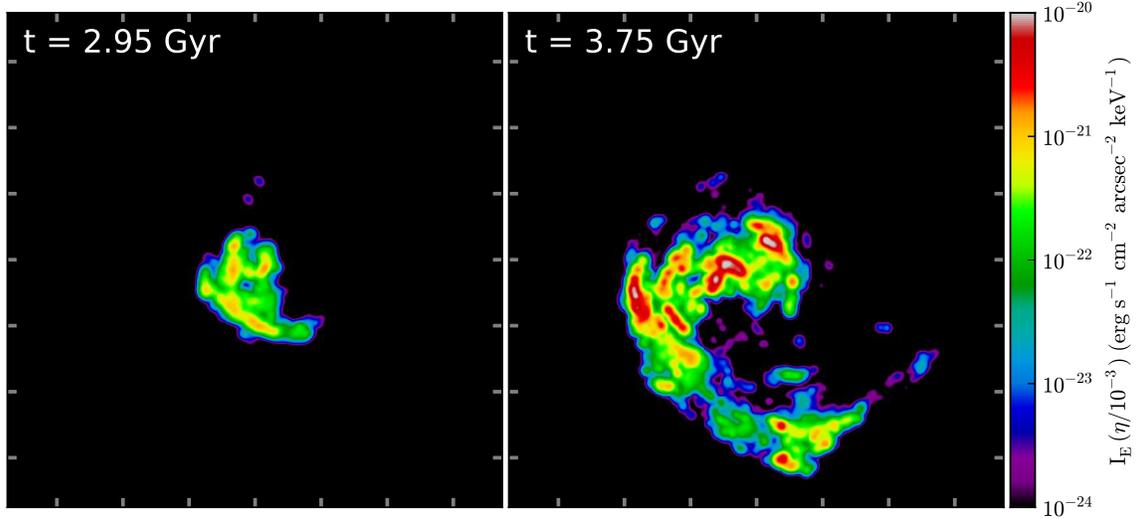}
\caption{Maps of monochromatic IC surface brightness from the relativistic electrons in the simulation for two representative epochs for 50~keV photons. Each panel is 750~kpc on a side. Tick marks indicate 100~kpc distances.\label{fig:IC_maps}}
\end{center}
\end{figure*}

\begin{figure}
\begin{center}
\plotone{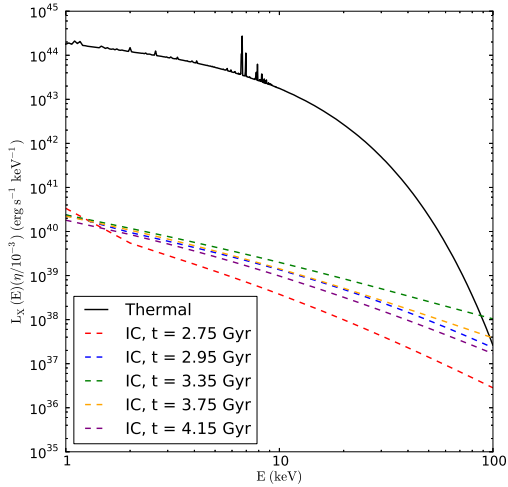}
\caption{The monochromatic X-ray luminosity for several different epochs of the simulation within a radius of 300~kpc from the cluster center. The black solid line indicates X-ray emission from the thermal gas at the epoch $t$ = 2.55~Gyr, which has been computed using the APEC model \citep{smi01} assuming a spatially uniform metallicity of $Z = 0.3~Z_\odot$. Dashed lines indicate IC emission, computed as described in the text.\label{fig:IC_spec}}
\end{center}
\end{figure}

\begin{figure}
\begin{center}
\plotone{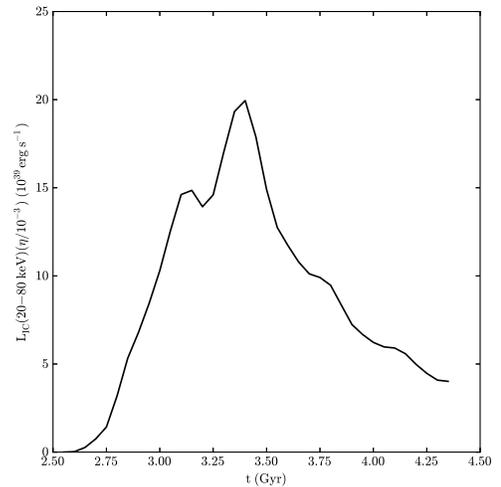}
\caption{The evolution of hard X-ray luminosity integrated over the 20-80~keV waveband for the emission within a radius of 300~kpc from the cluster center.\label{fig:IC_power}}
\end{center}
\end{figure}

\subsection{Simulated Inverse Compton Emission\label{sec:sim_IC}}

To illustrate the spatial distribution of relativistic electrons disentangled from the magnetic field distribution, we compute maps of the inverse-Compton (IC) emission for our tracer particles. The monochromatic IC power for a distribution of relativistic electrons as a function of emitted photon energy $\epsilon_1$ is \citep{ryb79}:
\begin{equation}
P(\epsilon_1) = \frac{3}{4}c\sigma_T\displaystyle\int{d\epsilon}\left(\frac{\epsilon_1}{\epsilon}\right)v(\epsilon)\displaystyle\int{d\gamma}{N(\gamma)\gamma^{-2}f\left(\frac{\epsilon_1}{4\gamma^2\epsilon}\right)},
\end{equation}
where $v(\epsilon)$ is the incident photon number density at the incident photon energy $\epsilon$, $\sigma_T$ is the Thomson cross section, and 
\begin{equation}
f(x) = 2x\ln{x} + x + 1 - 2x^2, 0 < x < 1.
\end{equation}
Given that our distribution function for each tracer particle is simply $N_j(\gamma) = K_j\sum_i\delta(\gamma-\gamma_{i,j})$, the total IC power at a given frequency for each tracer particle is then given by:
\begin{equation}
P_j(\epsilon_1) = 3c\sigma_TK_j\displaystyle\int{v(\epsilon)\left[\displaystyle\sum_i\left(\frac{\epsilon_1}{4\gamma_{i,j}^2\epsilon}\right)f\left(\frac{\epsilon_1}{4\gamma_{i,j}^2\epsilon}\right)\right]}{d\epsilon}.
\end{equation}
We assume that $v(\epsilon)$ is a blackbody spectrum with temperature $k_BT_{\rm CMB}$ (which is redshift-dependent) and integrate over all photon energies and sum over all $\gamma_{i,j}$ to obtain the emitted power at each photon energy. 

Since our main aim in this paper is to determine the properties of the radio emission of the minihalo, our examination of the properties of the IC emission will be comparatively brief. Figure \ref{fig:IC_maps} shows maps of the IC intensity at the epochs $t$ = 2.95~and~3.75~Gyr at the representative photon energy $E_{\gamma} = 50$~keV. As in the case of the radio emission, the hard X-ray emission at high energies is bounded by the cold front surfaces. The spatial distribution of the IC emission is similar to the synchrotron emission. This is expected, since, as we mentioned in the previous section, the spatial distribution of the radio emission follows the location of the turbulent regions, and both kinds of emission originate from the same populations of relativistic electrons.

Figure \ref{fig:IC_spec} shows the IC spectrum at a few different epochs of the simulation for the central 300~kpc of the cluster core, compared to the thermal spectrum at the epoch $t$ = 2.55~Gyr from the same core region (a mixture of temperatures in the range $T \sim$2-5~keV). At nearly all energies $E_\gamma \sim 1-100$~keV, the spectrum is dominated by the thermal emission. It will not be possible to detect such emission against the thermal emission from the bright cluster core with upcoming X-ray telescopes with hard X-ray detection capabilities such as {\it NuSTAR} and {\it ASTRO-H}, except possibly at energies near 100~keV.

For synchrotron and IC emission produced by the same population of relativistic electrons, $P_{\rm sync}/P_{\rm IC} = B^2/B_{\rm CMB}^2$. Taking the volume-averaged magnetic field $B_V \sim 5~\mu$G and $B_{\rm CMB} \sim 4~\mu$G (assuming $z = 0.1$), we find $P_{\rm sync}/P_{\rm IC} \approx 1.56$. Confirming this, we estimate from Figures \ref{fig:primary_spectrum} and \ref{fig:IC_spec} that the two bolometric luminosities are roughly equal with $P_{\rm sync} \sim P_{\rm IC} \sim 10^{41}$~erg~s$^{-1}$, which is orders of magnitude smaller than the thermal X-ray luminosity of $P_{\rm therm} \sim 10^{44}$~erg~s$^{-1}$. This would be the case for all clusters hosting a minihalo, as most cool-core clusters have relatively strong magnetic fields in their cores \citep[][]{tay02,tay06,tay07,bon10,vac11}. 

Figure \ref{fig:IC_power} shows the evolution of the X-ray luminosity in the energy band 20-80~keV,  for the $r \leq 300$~kpc region of the cluster core. The behavior of the hard X-ray power is similar to that of the synchrotron power in that it exhibits a fast increase (of nearly an order of magnitude) when reacceleration begins, followed by a slower dropoff. The peak of the hard X-ray power is $\sim 2 \times 10^{40}$~erg~s$^{-1}$, a few orders of magnitude smaller than the non-thermal luminosity possibly detected by {\it XMM-Newton} and {\it INTEGRAL} in Ophiuchus, $\sim~10^{43}$~erg~s$^{-1}$ \citep[][also see \citet{aje09} for upper limits from {\it Swift} and \citet{fuj08} for upper limits from {\it Suzaku}]{mur10}.

\begin{figure*}
\begin{center}
\plotone{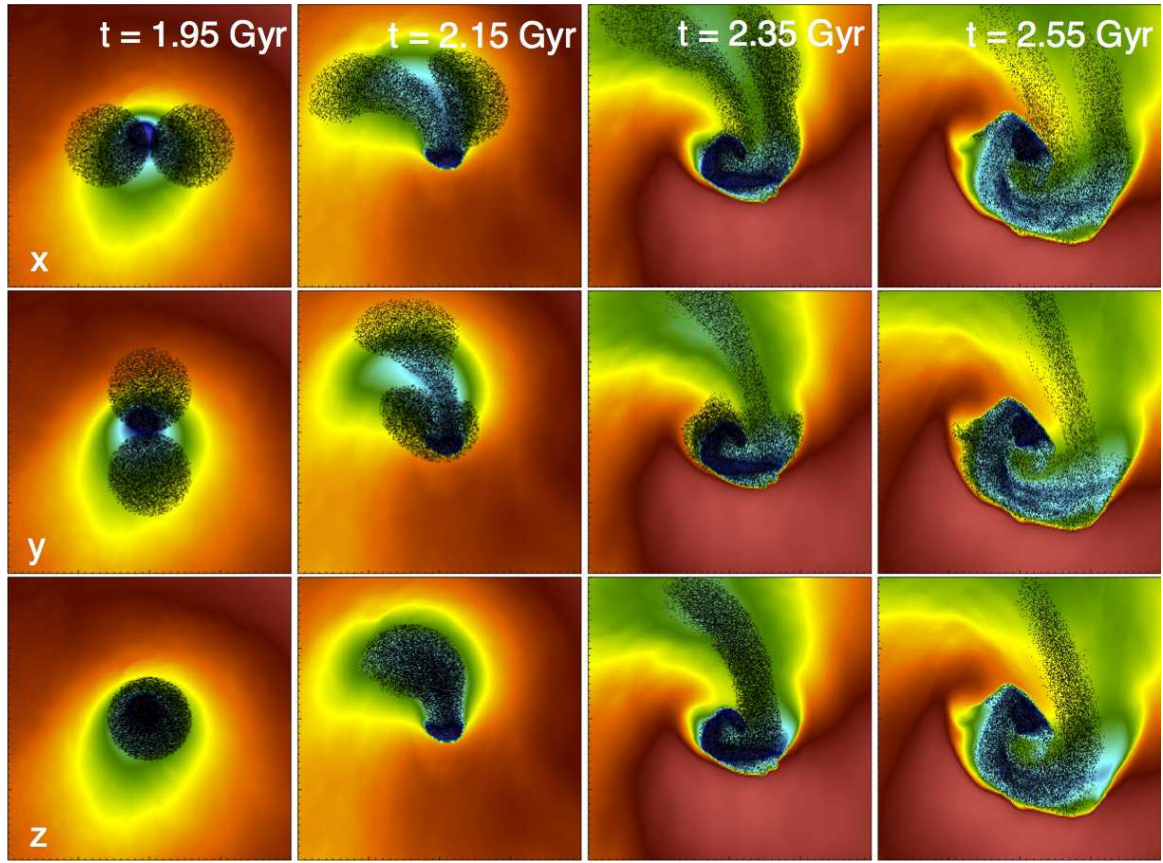}
\caption{Evolution of the positions of the gas particles between the epochs $t$ = 1.95~Gyr and 2.55~Gyr for the three different initial spatial configuations of ``bubble'' particles, overlaid on maps of projected gas temperature. Each panel is 200~kpc on a side. Major tick marks indicate 50~kpc distances.\label{fig:particle_positions}}
\end{center}
\end{figure*}

\begin{figure*}
\begin{center}
\plotone{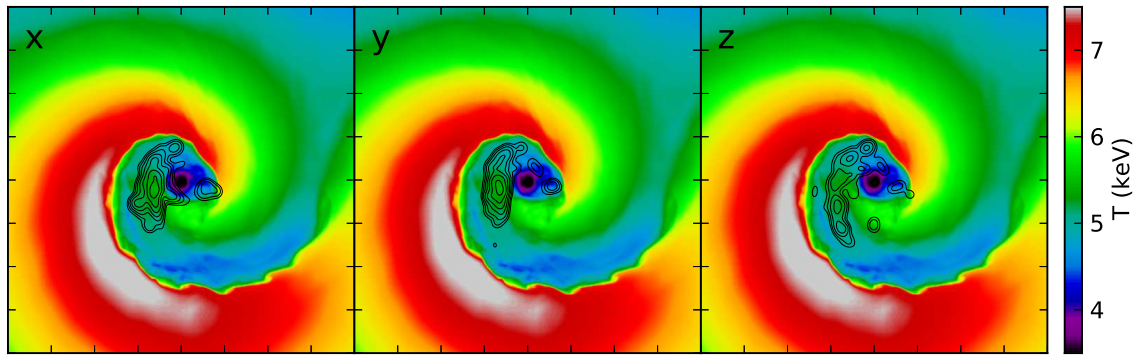}
\caption{Projected gas temperature maps with radio contours overlaid from different ``bubble'' initial conditions for the relativistic particles. The colorbar is temperature in keV. Contours are of 327~MHz radio emission which begin at $1.25 \times 10^{-4}$~mJy arcsec$^{-2}$ and increase by a factor of 2. Left: initial bubbles along x-axis. Center: initial bubbles along y-axis. Right: initial bubbles along z-axis. The epoch is $t$ = 2.95~Gyr. Each panel is 400~kpc on a side. Tick marks indicate 100~kpc distances.\label{fig:bubble_emission}}
\end{center}
\end{figure*}

\subsection{Different Initial Spatial Distributions of Relativistic Electrons\label{sec:different_space}}

Though our tracer particles are distributed throughout the simulation domain, it is not necessarily the case that in real clusters relativistic electrons are similarly distributed. Electrons are accelerated to relativistic speeds by AGN and generated as byproducts of collisions of relativistic protons with the thermal protons of the ICM. The former process will result in relativistic electrons preferentially located in AGN-blown bubbles, and the latter will produce more relativistic electrons in regions of high gas density throughout the cluster. The appearance of a mini-halo generated by sloshing may depend on the initial spatial distribution of these particles. 

Our already examined case of a large spherical distribution of relativistic particles represents the second possibility mentioned above. In order to approximately represent a population of seed electrons produced by an AGN, we perform three simulations of relativistic electrons originating in two ``bubbles''. In these cases, it order to produce radio emission that follows the X-ray cold fronts, it will not only be necessary for electrons to be reaccelerated by the turbulence, but the sloshing motions will have to adequately redistribute the seed electrons throughout this region. 

Since modeling the dynamics of the relativistic AGN bubbles is far beyond our current scope, we perform a simple exercise that tests whether or not the relativistic particles may be adequately redistributed by the sloshing motions to fill the core region. We first assume that the relativistic particles contained in the bubbles have just become mixed in with the thermal gas at the epoch $t$ = 1.95~Gyr (before the beginning of the sloshing motions), and then identify tracer particles contained within two symmetric 30~kpc ``bubbles'' centered around the cluster potential minimum (without attempting to evacuate thermal gas from these ``bubbles'', as would be the case in real clusters). Each of these particles are assumed to contain the same mass of gas $m_j$ as before, but there are far fewer of them than in our default spatial case, with approximately $\sim10^4$ tracer particles per bubble. We then follow these particles for 0.6~Gyr, and at $t$ = 2.55~Gyr, we assign them relativistic particle distributions identical to that of our default case. We perform three simulations of the evolution of the relativistic particles from these initial distributions, with the axis of the bubbles aligned along the $x$, $y$, and $z$ axes (keeping in mind that in our simulation, sloshing occurs in the $x-y$ plane. 

Figure \ref{fig:particle_positions} shows the evolution of particle positions between the epochs $t$ = 1.95~Gyr and 2.55~Gyr for our three bubble initial conditions. Regardless of the initial orientation of the bubbles, we find that the sloshing motions redistribute the gas particles so that most of them end up within the spiral shape traced out by the cold fronts. Figure \ref{fig:bubble_emission} shows the projected gas temperature in the $z$-direction of the simulation with 327~MHz radio contours overlaid at the epoch $t$ = 2.95~Gyr for two simulations where ``bubbles'' of relativistic particles have been injected. We find that faint radio emission is produced in all three cases, and it is contained within the core of the sloshing region. Although the particles have been redistributed throughout the sloshing region, only a fraction of them have encountered regions of strong turbulence, fewer than in the case where we assumed relativistic particles were originally distributed proportional to gas density everywhere within the core.

If we assume that a population of low-$\gamma$ electrons is built up over time in the core by injection from the central AGN and then mixing by sloshing motions, then we may end up with a situation similar to our default case for the spatial distribution of relativistic electrons, which succeeded in reproducing the observed mini-halo properties. This lends support to the hypothesis that mini-halos may be generated by electrons which originated from the central AGN and are reaccelerated by sloshing motions. Sloshing and AGN are both commonly found in cool-core clusters where mini-halos are found. Additionally, as pointed out by \citet{pfr04,cas08,kes10b,kes10c}, in the cool cores of galaxy clusters relativistic electrons will be efficiently produced by hadronic processes due to the high density of thermal protons. Between the AGN and hadronic interactions, the dense cores of clusters should contain adequate numbers of seed relativistic electrons to be reaccelerated by turbulence. 

\section{Discussion\label{sec:disc}}

\subsection{The Evolution of the Relativistic and Turbulent Energy Components\label{sec:energy_budget}}

\begin{figure}
\begin{center}
\plotone{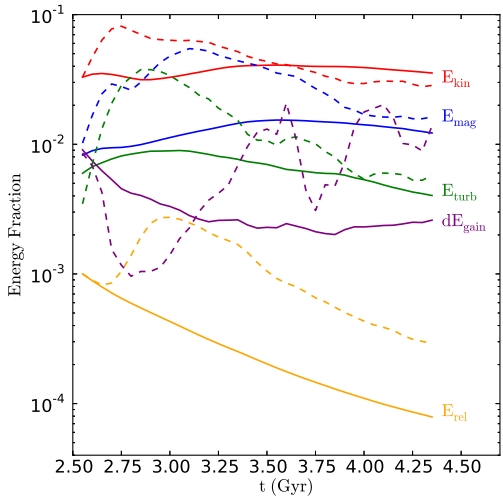}
\caption{Time dependence of the various energy components of the simulation. Solid lines indicate quantities computed for all of the tracer particles originating within a radius of $r = 100$~kpc at the epoch $t$ = 2.55~Gyr. Dashed lines indicate quantities computed for all of the tracer particles which at some point in the simulation fulfill the criterion $\bar{\gamma} > 10^3$.\label{fig:evolving_energies}}
\end{center}
\end{figure}

It is instructive to examine the energy budget of the various components of our model as they evolve during the simulation. The solid lines in Figure \ref{fig:evolving_energies} shows the evolution of the total kinetic, turbulent kinetic, magnetic, and relativistic electron energies relative to the total thermal energy of the tracer particles initially within a radius of 100~kpc from the cluster center. The largest contribution is that of the total kinetic energy, which stays around $\sim$3-4\% over time, followed by the magnetic energy at $\sim$1\% and the turbulent kinetic energy at $\sim$0.5\%. The relativistic electron energy density starts at our initial value of $\varepsilon_{rel}/\varepsilon_{th} = 10^{-3}$ and decreases as most of the electrons cool over time--the high-$\gamma$ tail of the electron distribution that is studied here does not contribute much to the total electron energy. All of the separate nonthermal energy categories combined contribute less than 10\% to the total energy budget. This is in part a consequence of the specialized initial conditions of our setup, which began with only thermal, magnetic, and a small amount of kinetic energy (due to our relaxation procedure, see Section \ref{sec:MHD_sims}). This is consistent with the cosmological simulation results of \citet{lau09}, who show that for cores of relaxed clusters, the contribution to the total pressure from turbulence, rotation, and streaming gas motions is negligible compared to the thermal pressure, as well as recent constraints placed on turbulent motions in the core of Abell 1835 by \citet{san10}, and in Abell 3112 by \citet{bul12}.

It is also interesting to examine the energies of the tracer particles that are emitting the most synchrotron radiation. These particles will be the ones that have encountered the highest turbulent velocities, as can be seen by the correspondences between the regions of brightest synchrotron emission (Figures \ref{fig:contours_z}-\ref{fig:contours_y}) and the locations with the strongest turbulence (Figures \ref{fig:proj_z}-\ref{fig:proj_x}). To isolate these regions, we select only those tracer particles that will reach an average energy $\bar{\gamma} \geq 10^3$ at any point in the simulation and emit at frequencies $\simgt$~100~MHz, at which the halos are presently observed. These particles comprise 1\% of the total number of tracer particles of our initial set; their energy ratios are shown by the dashed lines of Figure \ref{fig:evolving_energies}. The synchrotron emission indeed highlights regions with slightly higher turbulent energies and (by virtue of their selection) nearly an order of magnitude higher relativistic electron energy contribution. 

We can also determine the energy gained by relativistic particles through turbulent acceleration. Though we have not modeled self-consistently the interaction between the relativistic particles and the turbulent ICM (in particular their back-reaction on turbulence), for the reacceleration model to be a viable mechanism for generating mini-halos, the energy gain by relativistic particles should be a small fraction of the turbulent energy. To do this, we need to compare the rate of energy gained by the relativistic electrons versus the rate of energy that passes through the turbulent cascade toward the dissipation scale. The former rate will not be equal to the total change in energy in relativistic particles, as most of the energy gained (especially for the highest-energy electrons) will end up in the form of synchrotron and hard X-ray emission, but we know from Equation \ref{eqn:Dpp8} how much energy the electrons in our simulations gain. To get a rough determination of the ratio of these two quantities, we compute the former via Equation \ref{eqn:Dpp7}, and the latter via the estimate for the turbulent energy cascade rate \citep[Equation 48 from][]{bru07}:
\begin{equation}
\dot{\varepsilon}_{\rm turb} \sim \C\frac{\langle{\M}\rangle^4\langle{c_s}\rangle^3}{\ell}
\end{equation}
where $\ell$ = 30~kpc and the constant $\C \sim 5-6$. In making this estimate we have used the average turbulent Mach number $\M = v_t/c_s$ and sound speed $c_s$. The ratio of these rates for all the tracer particles and the high-$\gamma$ tracer particles is also shown in Figure \ref{fig:evolving_energies}. The rate of energy gain of the relativistic particles is typically $\sim$10$^{-2}$ of the turbulent cascade rate. 

As a consistency check, we check the ratio of the rate of the turbulent energy cascade rate to the cooling rate of the thermal gas by X-rays. We find that the rate of turbulent dissipation is at the level of $\sim$10$\%$ of the bolometric X-ray luminosity, however, this estimate is probably uncertain by an order of magnitude, since we have made an approximate estimate of the turbulent cascade rate due to our inability to measure this quantity from the simulation directly. Thus, this quantity is probably overestimated; a consequence of which is that the efficiency of reacceleration relative to the turbulent cascade rate is likely to be somewhat higher than this estimate indicates.

In general, only a fraction of the energy of turbulence is drained into relativistic particles via turbulent reacceleration \citep[see][]{bru11}. Consequently, even if we have not considered the case of cosmic ray protons that could be present in the cluster core region, the additional energy input that would be acquired by these pre-existing protons via acceleration due to sloshing-driven turbulence in our simulation is $\simlt$1\% of the thermal energy, regardless of the initial energy content of the pre-existing protons. A energy content of cosmic ray protons at percent level of the thermal gas is still consistent with present upper limits from gamma-ray observations \citep[see, e.g.][]{aha09,ack10,jel11,ale12}.

\subsection{Two Effects of Sloshing on the Relativistic Electrons\label{sec:sloshing_effects}}

From our simulations, we discern two effects of the gas sloshing on the relativistic electrons that are important for the formation of mini-halos. The first is the reacceleration itself. Without reacceleration, the electrons will simply cool down to energies below which they are incapable of emitting at observable synchrotron frequencies. With reacceleration, the electrons inside the sloshing region are maintained for a significant period of time ($\sim$1~Gyr from the onset of sloshing) at high enough $\gamma$ for radio emission to be produced (see Figure \ref{fig:electron_accel}). The second important effect of sloshing concerns the initial spatial distribution of the electrons. Sloshing motions can take rather general initial spatial distributions of relativistic electrons and redistribute them throughout the sloshing region. Thus, we anticipate that for reasonable seed populations of relativistic electrons (either resulting from hadronic interactions or injected by AGN), sloshing should be able to redistribute and reaccelerate them so that radio mini-halo emission is produced. As previously mentioned, sloshing also amplifies magnetic fields within the cluster core, an effect that is also important for hadronic/secondary models \citep{kes10b,kes10c}.

\subsection{Morphology of Mini-Halos\label{sec:morphology}}

Some mini-halos in observed galaxy clusters with attendant sloshing cold fronts have a very specific morphology--they are very similar to the shape of the sloshing region and bounded by the cold fronts as seen in X-rays \citep{maz08}. However, other mini-halos do not appear to be coincident with sloshing cold fronts \citep[][]{gov09,gia11}, and some of them appear in clusters without discernable sloshing. However, turbulence generated by sloshing may still be the origin of the radio emitting electrons, even if the corresponding cold fronts are not seen. 

Projection effects may make it difficult to associate mini-halos with cold fronts. In projections perpendicular to the plane of the mutual orbit of the main cluster and the subcluster, such as the $z$-projection of our simulations, a clear association of the radio emission with the shape of the cold front is observed. In other projections ($x$ and $y$), the spiral shape of the cold fronts is unobservable, though small portions of the cold fronts still appear. In these cases, the emission is still bounded within this region, but it is more difficult to tell from the X-ray image if sloshing is present. 

Larger cold fronts will also be more difficult to observe in X-rays due to their low surface brightness contrast. Finally, although the conditions necessary to generate the mini-halo emission may be associated with the sloshing cold fronts, it does not necessarily imply that these conditions prevail uniformly throughout the sloshing region. Figures \ref{fig:proj_z} through \ref{fig:proj_y} show the turbulent kinetic energy exhibits strong variations throughout the cluster core. Though the magnetic field strength is the strongest along the cold front surfaces, it also varies greatly inside them, also seen in Figures \ref{fig:proj_z} through \ref{fig:proj_y}. These variations may combine to produce emission maps that do not always fill the entire region bounded by the cold fronts (see Figures \ref{fig:contours_z} through \ref{fig:contours_y}), particularly since the spatial scale of these variations is comparable to the size of the core region. 

\subsection{Uncertainties in the Model\label{sec:uncertainties}}

Finally, it is important to summarize the various uncertainties involved in our model. The major sources of uncertainty arise from the assumptions of the initial energy in relativistic electrons and the assumptions involved in the estimates of the reacceleration coefficients. 

Our initial electron spectrum depends on a number of conditions as detailed in Section \ref{sec:elec_model}. We have experimented with a few different initial electron spectra under otherwise identical simulation conditions, and have found that our results do not depend significantly on the initial spectral shape. All of our results related to the relativistic electron spectra, including our synchrotron and IC emission, may be scaled by the relativistic electron energy fraction $\eta$. The hard X-ray constraints on the cluster IC emission allow values up to $\eta \sim 10^{-2}$ \citep{wik12}; a value in equipartition with the magnetic field energy would also be $\eta \sim 10^{-2}$; thus our use of $\eta = 10^{-3}$ is very conservative.

The reacceleration coefficients $D_{\rm pp,TTD}$ and $D_{\rm pp,C}$ in our model are dependent upon the characteristics of the power spectrum of turbulence, in particular the minimum and maximum wavenumbers $k_{\rm min}$ and $k_{\rm max}$, and the ratio of power in compressible motions to the total power $R^c$. While $k_{\rm min}$ is determined by the size of the sloshing structures and cannot be much different from our assumption, the value of $k_{\rm cut}$ represents an extension (albeit a very plausible and likely conservative one) by a factor of $\sim$10 above the range of wavenumbers for the inertial regime in our simulations. If our $k_{\rm cut}$ is an overestimate, we subsequently overestimate $D_{pp}$ and the reacceleration coefficient $\chi$ by a factor of $\sim$~a few. Conversely, if the turbulent cascade extends down to smaller scales in the manner of \citet{bru11}, we have underestimated $k_{\rm cut}$ by a factor of $\sim$~a few, and the reacceleration efficiency would be substantially underestimated.

We already discussed the potential overestimate in $R_c$ due to the edge effects of the non-periodic subdomains in Section \ref{sec:turb_coeff}, which showed that up to $\sim$30\% of the compressive power is potentially spurious. However, the main uncertainty in $R^c$ arises from our adoption of an average value constant in space and time. In reality, the compressive component of turbulence will vary spatially and with time. Our adoption of $R^c = 0.25$ is a conservative one, since most of the reacceleration from turbulence will arise from regions with higher $R^c$, which we found can be as high as 0.7 (Section \ref{sec:turb_coeff}). In these regions, the reacceleration due to turbulence is underestimated by a factor of a few. 

\begin{figure}
\begin{center}
\plotone{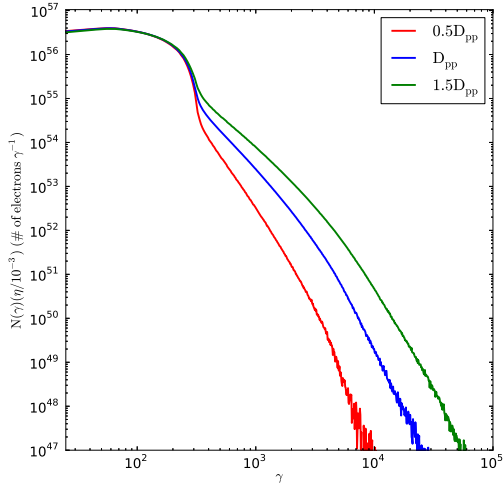}
\caption{Dependence of the total electron spectrum on the value of $D_{pp}$, with values reported relative to the default value adopted in the paper.\label{fig:different_Dpp}}
\end{center}
\end{figure}

Figure \ref{fig:different_Dpp} shows the effects of varying the coefficient $D_{pp}$ by a small factor in either direction on the total electron spectrum of the simulation. Decreasing $D_{pp}$ by a factor of two results in a decrease of $N(\gamma)$ by an order of magnitude or more at $\gamma \simgt 10^3$. Increasing $D_{pp}$ by a factor of 1.5 results in an increase of $N(\gamma)$ by nearly an order of magnitude or more at $\gamma \simgt 10^3$. This demonstrates that the resulting electron spectrum is fairly sensitive to the value of $D_{pp}$ at high $\gamma$. 

Due to the combined effects of these uncertainties, our procedure must be regarded only as an order of magnitude estimate, which is suitable for the aim of the present paper. Nevertheless, for the physically plausible choices we have made for these coefficients, our results are consistent with observations of minihalos. 

\section{Summary\label{sec:summary}}

In this work, we have performed the first simulation test of the hypothesis that radio minihalos, sometimes observed in cluster cool cores, originate from relativistic electrons reaccelerated by turbulence generated by sloshing motions. The seed low-$\gamma$ electrons are likely to remain, e.g., from past AGN activity and/or from hadronic interactions. The acceleration mechanisms that we considered are the damping of turbulence-induced magnetosonic waves on relativistic electrons and non-resonant compression, which are included in a subgrid fashion. Our high-resolution, idealized simulation consists of a cool-core galaxy cluster with magnetized gas and an infalling gasless subcluster. The interaction between the subcluster and the core of the main cluster initiates gas sloshing in the core. Two significant magnetohydrodynamic effects of the gas sloshing are the generation of turbulence and the amplification of magnetic fields, both of which occur within the region of the cool core bounded by the sloshing cold fronts. 

We have shown that in this model, faint, extended radio sources that closely resemble the spatial and spectral properties of the observed mini-halos may be generated under a variety of physically reasonable assumptions for the initial spatial distribution and input spectrum of seed relativistic electrons. The sloshing motions--bulk flows and turbulence--have two effects on the relativistic electrons, reacceleration and redistribution of seed relativistic electrons throughout the core. These two effects combine to produce radio emission that is diffuse, steep-spectrum, and that traces the spatial features of the X-ray emitting gas. The radio power generated in our model is in the range of that of observed minihalos (though our assumption of the initial energy density of the relativistic electrons is rather uncertain). Importantly, we have shown that minihalos produced by turbulent reacceleration are transient sources, particularly at high ($\sim$1~GHz) frequencies, consistent with the small number of observed mini-halos. We have also shown that the X-ray inverse-Compton emission produced by the same relativistic electrons will be difficult to observe due to the fact that the IC emission begins to dominate over the thermal emission of the bright core only at very high energies ($\sim$100~keV). 

Significant improvements could be made to bring our model more in line with observations. More sophisticated methods can be developed to better characterize the spectrum and normalization of turbulence created by sloshing \citep[such as the recent attempts in][]{vaz12}. A more accurate treatment of the evolution of the spectrum of relativistic electrons requires the use of a physical model for a time-dependent injection of these electrons. Also, in our paper we do not consider the effect of cosmic ray protons in the sloshing region. Self-consistent calculations of particle acceleration and of the resulting non-thermal emission should take into account also the proton component, since protons generate secondary electrons (that eventually may be reaccelerated) and can be important for the damping of the turbulence \citep{bru11a}. Finally, a variety of cluster initial conditions could be explored, to determine if mini-halo formation is more likely under certain merger conditions than others. We leave these considerations for future work. Our simulation data could also be used to examine the hadronic hypothesis for mini-halos, and provide a point of comparison between the two models. We will make this comparison the subject of a future paper. 

Our present work suggests that reacceleration of relativistic electrons by turbulence can produce radio mini-halo emission in cool-core clusters, though the uncertainties at various steps of our simulations are high.

\acknowledgments
JAZ thanks Pasquale Mazzotta, Uri Keshet, Dan Wik, and Eric Hallman for useful discussions and advice, and in particular Franco Vazza and Ian Parrish for guidance and advice regarding the velocity power spectra. Calculations were performed using the computational resources of the National Institute for Computational Sciences at the University of Tennessee and the NASA Advanced Supercomputing Division. Analysis of the simulation data was carried out using the AMR analysis and visualization toolset yt \citep{tur11}, which is available for download at \url{http://yt-project.org}. JAZ is supported under the NASA Postdoctoral Program. GB acknowledges partial support by grant PRIN-INAF-2009. The software used in this work was in part developed by the DOE-supported ASC / Alliances Center for Astrophysical Thermonuclear Flashes at the University of Chicago.

\appendix

\section{Test Cases for the Relativistic Integrator\label{sec:appendix}}

\begin{figure*}
\begin{center}
\plotone{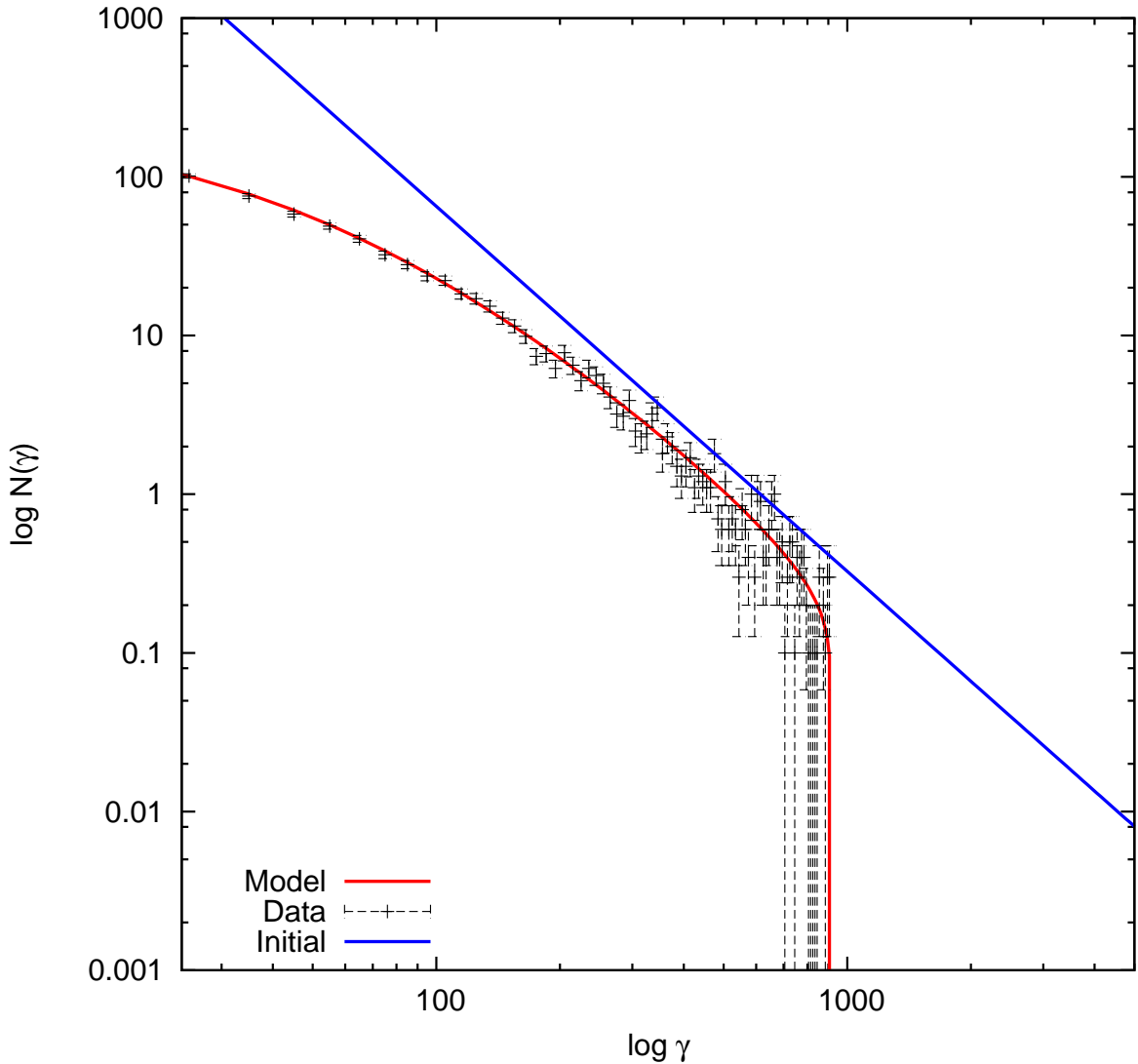}
\caption{Verification test of simulation of relativistic electrons with losses only. Parameters are $B$ = 1~$\mu$G, $n_{\rm th}$ = 10$^{-3}$, evolved from $z$ = 0.1. Blue curve: input electron spectrum. Red curve: predicted electron spectrum.\label{fig:test_losses}}
\end{center}
\end{figure*}

\begin{figure*}
\begin{center}
\plotone{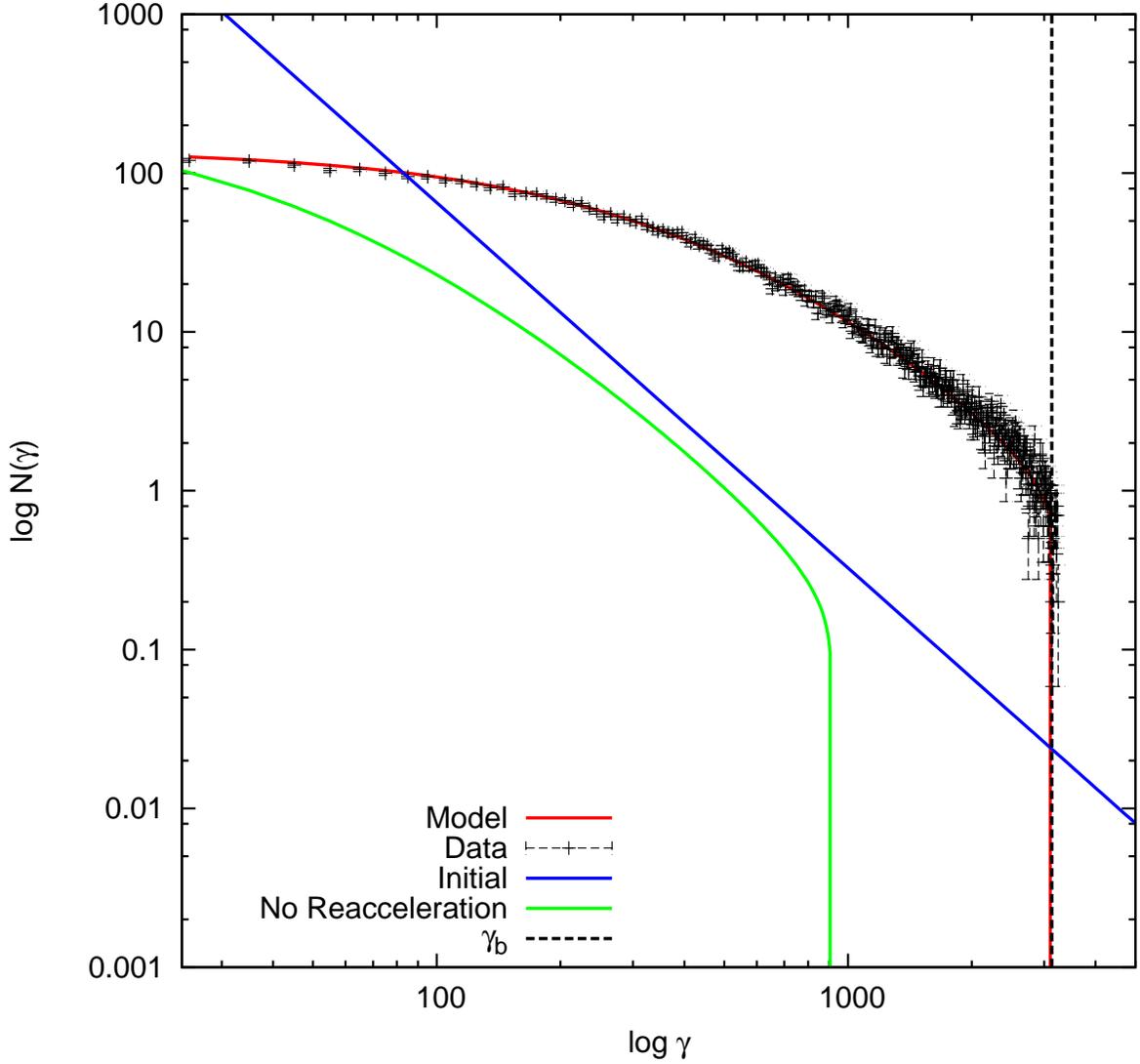}
\caption{Verification test of simulation of relativistic electrons with reacceleration. Parameters are $\delta{v} = 200$~km~s$^{-1}$, $B$ = 1~$\mu$G, $n_{\rm th}$ = 10$^{-3}$, evolved from $z$ = 0.1. Blue curve: input electron spectrum. Red curve: predicted electron spectrum. Green curve: predicted spectrum without reacceleration. The predicted ``break frequency'' $\gamma_b$ is shown by the dashed line.\label{fig:test_reaccel}}
\end{center}
\end{figure*}

\begin{figure*}
\begin{center}
\plotone{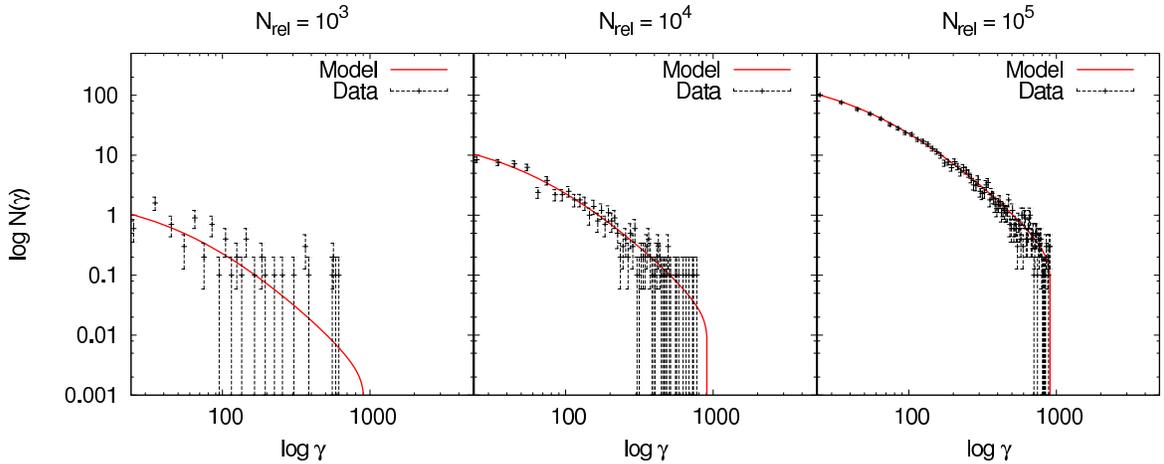}
\caption{Simulation of relativistic electrons with losses only for differing numbers of samples $N_{\rm rel}$. Parameters are $B$ = 1~$\mu$G, $n_{\rm th}$ = 10$^{-3}$, evolved from $z$ = 0.1. The red curve indicates the predicted electron spectrum. Left panel: $N_{\rm rel} = 10^3$. Center panel: $N_{\rm rel} = 10^4$. Right panel: $N_{\rm rel} = 10^5$.\label{fig:test_res}}
\end{center}
\end{figure*}

\begin{figure*}
\begin{center}
\plotone{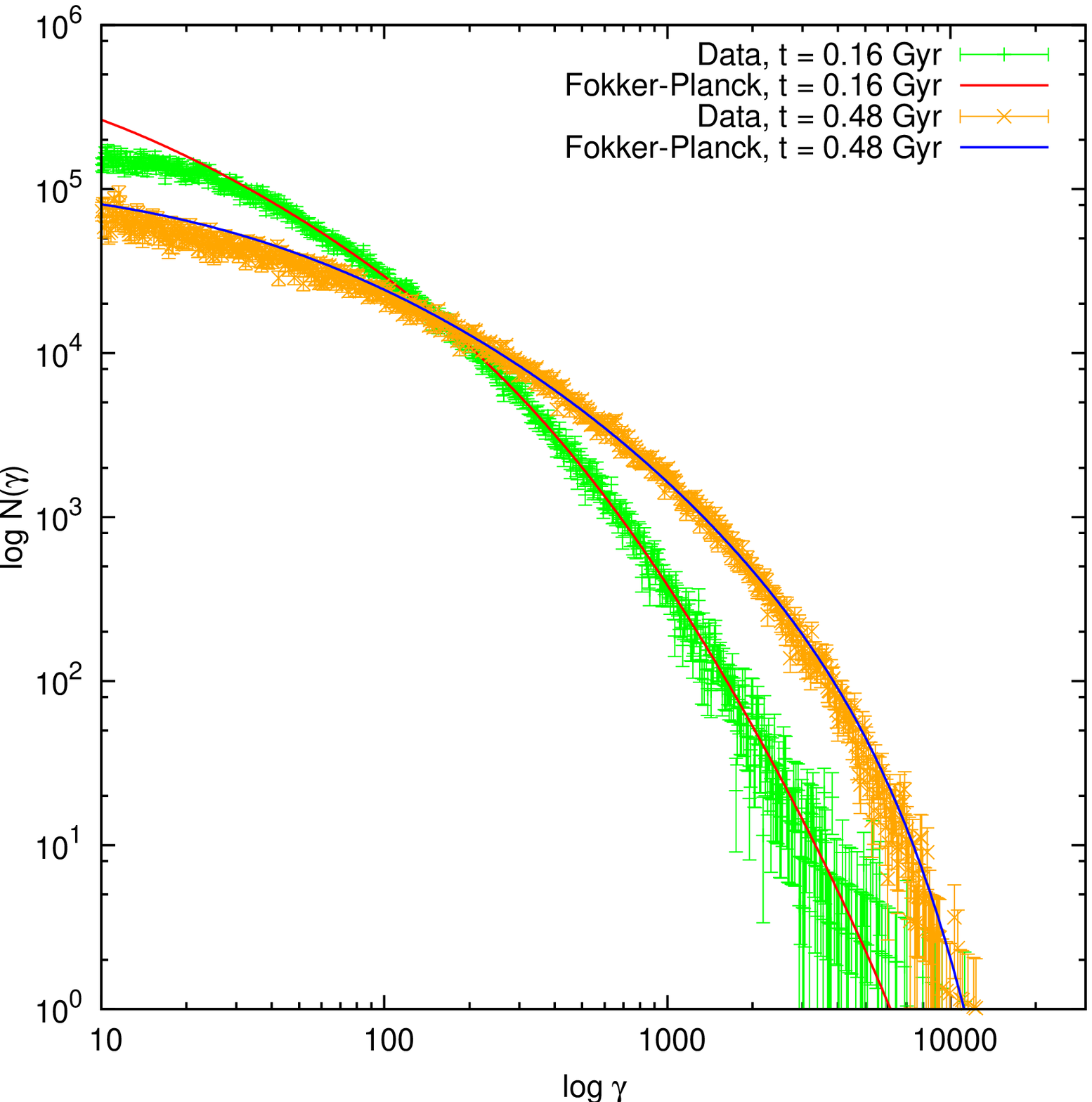}
\caption{Simulation of relativistic electrons using the SDE integrator compared with solutions from a Fokker-Planck simulation. Parameters are $B = B_{\rm CMB} = 3.25~{\rm \mu{G}}$, $n_{\rm th}$ = 10$^{-3}$, and $t_{\rm acc} = \chi^{-1} = 0.3$~Gyr.\label{fig:stochastic_comparison}}
\end{center}
\end{figure*}

In this appendix we present the results of a few verification tests of our relativistic particle integrator. We assume in these test cases time-independent conditions for the state of the gas and the magnetic field. Additionally, for timescales that are short compared to the Hubble time, the effect of the redshift dependence on the CMB energy density is small and it may be ignored. We will first perform tests of the deterministic "drift" term of Equation \ref{eqn:energy_evo} against analytic solutions, and follow with comparisons of a full solution of Equation \ref{eqn:energy_evo} with the stochastic term included with a solution generated under identical conditions by integrating a Fokker-Planck equation. 

\begin{figure*}
\begin{center}
\plotone{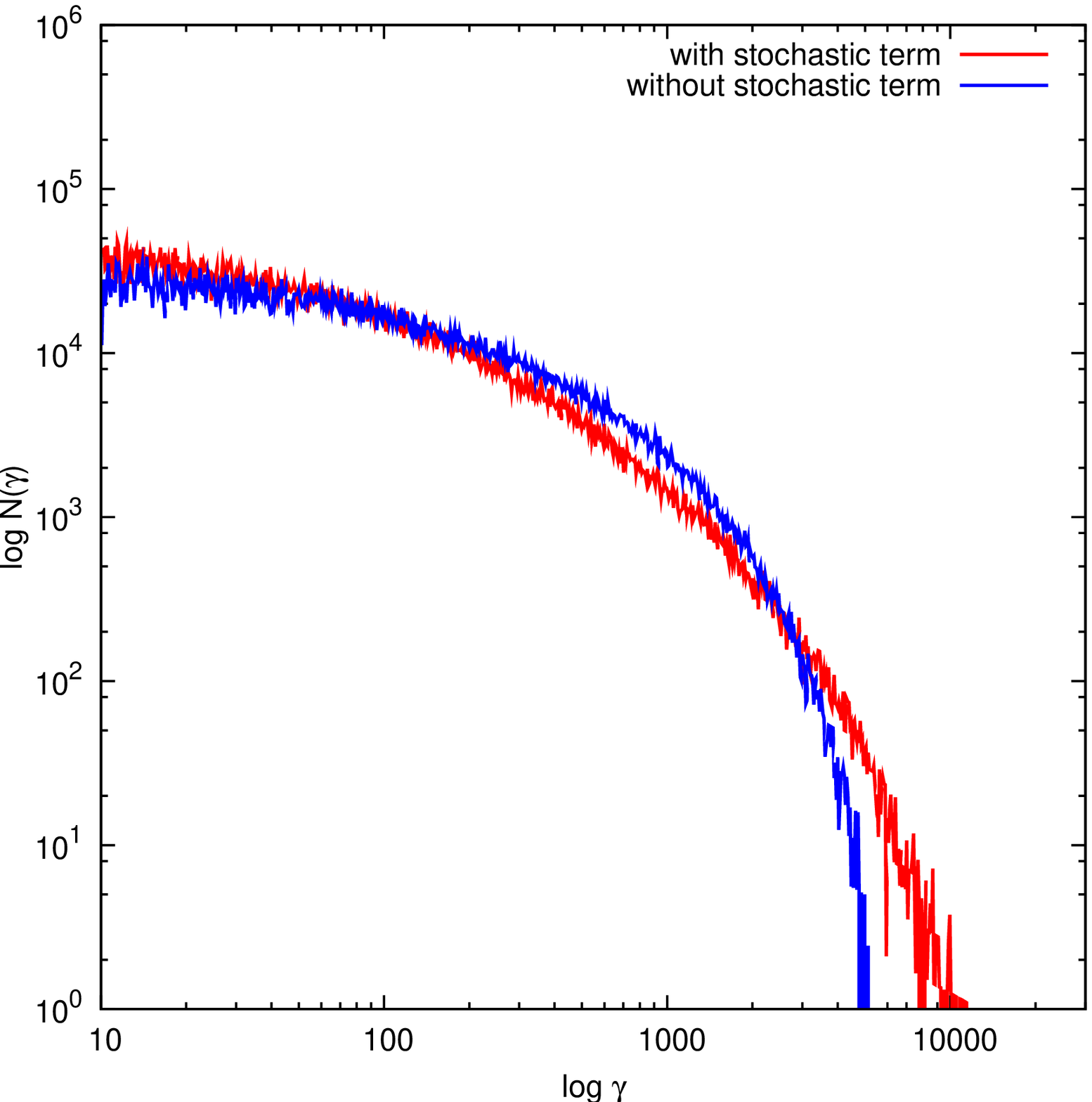}
\caption{Simulation of relativistic electrons using the SDE integrator with the stochastic term included compared with a simulation where only the systematic terms are modeled. Parameters are $B = B_{\rm CMB} = 3.25~{\rm \mu{G}}$, $n_{\rm th}$ = 10$^{-3}$, and $t_{\rm acc} = \chi^{-1} = 0.3$~Gyr.\label{fig:stochastic_off_and_on}}
\end{center}
\end{figure*}

Our tests of the deterministic evolution of particles are similar to those of \citet{sar99}, who derived the spectrum of cosmic-ray electrons in clusters of galaxies under the effects of synchrotron, IC, and Coulomb losses. For time-independent conditions the particle energy spectrum $N(\gamma, t)$ may be derived from an initial spectrum $N(\gamma_i, t_i)$ by a simple relation:
\begin{equation}
N(\gamma, t) = N[\gamma_i(\gamma, t), t_i]\frac{b[\gamma_i(\gamma, t)]}{b(\gamma)}
\label{eqn:sar99a}
\end{equation}
where $b(\gamma)$ is the rate of change of particle energy and $\gamma_i(\gamma, t)$ is given implicitly by
\begin{equation}
\displaystyle\int_{\gamma_i}^{\gamma}\frac{d\gamma'}{b(\gamma')} = (t-t_i)
\label{eqn:sar99b}
\end{equation}
 
For our tests, we adopt parameters similar to those in \citet{sar99}. We evolve an initial power-law spectrum ($N(\gamma) \propto \gamma^{-p}$) with $p$ = 2.3 from $z = 0.1$ with $B = 1~\mu$G, $n_e = 10^{-3}$~cm$^{-3}$, and a cosmology with $h$ = 0.65, $\Omega_m = 1$, and $\Omega_\Lambda$ = 0. The evolved particle distributions are binned into energy bins of $\Delta{\gamma} = 10$ to derive a particle energy spectrum $N(\gamma)$ that may be compared with analytical results derived from Equations \ref{eqn:sar99a}-\ref{eqn:sar99b}. Unless otherwise noted, the simulations used 10$^{5}$ samples for the distribution function. 

We first consider the evolution of an initial population of relativistic electrons with no reacceleration. Figure \ref{fig:test_losses} shows the resulting simulated electron spectrum $N(\gamma)$ at $z = 0$, compared to the true spectrum (red line) and the initial spectrum at $z = 0.1$ (blue line). This spectrum is very similar to the electron spectra in Figures 6-8 of \citet{sar99}. The effect of the Coulomb losses is to flatten the relativistic electron spectrum at low $\gamma$, whereas the effect of the synchrotron and IC losses is to introduce a sharp cutoff at high $\gamma$. 

Secondly, we include the effects of reacceleration. For this we use Equation \ref{eqn:Dpp5} and assume a turbulent speed $\delta{v_t} = 200$~km~s$^{-1}$ at the scale of $\ell_{\rm min} = 100$~kpc. This spectrum is shown in Figure \ref{fig:test_reaccel}, compared to the true spectrum (red line), the initial spectrum (blue line), and the spectrum without reacceleration (green line). The resulting spectrum has a similar shape to the case without reacceleration, but it has been shifted to higher energies and the normalization of the spectrum is higher. If we assume reacceleration and losses are balanced at high $\gamma$, we may derive a ``break frequency'' $\gamma_b = \chi/\beta$ from Equations \ref{eqn:Dpp9}-\ref{eqn:radloss}. For these conditions, $\gamma_b \approx 3131$, which is reproduced well by the simulated data, as seen in Figure \ref{fig:test_reaccel}.  
 
We also test the effects of varying the number of particle samples of our distribution function on the evolved spectrum. Figure \ref{fig:test_res} shows the evolved particle spectrum for three different values of $N_{\rm rel}$, assuming only particle energy losses and no reacceleration. For $N_{\rm rel} = 10^{3}$, the shape of the particle spectrum is barely discernable. For $N_{\rm rel} = 10^4$, the shape is more apparent and for $N_{\rm rel} = 10^5$ the model is well-described by the simulated data. For the simulations presented in this work, we have chosen $N_{\rm rel} = 10^4$ as the number of samples per tracer particle as a good balance between accuracy and computation time. 

Finally, we verify our integration of the full stochastic differential equation by comparing the resulting electron energy spectrum to that produced by a Fokker-Planck calculation under identical conditions. For this test, we have assumed a constant magnetic field $B = B_{\rm CMB} = 3.25~{\rm \mu{G}}$, density of thermal particles $n_{\rm th} = 10^{-3}$, and a reacceleration timescale of $t_{\rm acc} = \chi^{-1} = 0.3$~Gyr. As in the previous tests, the value of $B_{\rm CMB}$ was held fixed. The initial condition for the relativistic particle distribution is the steady-state condition with continuous injection and energy losses from \citet{sar99}, under the above conditions and a spectral index for the input spectrum of $p$ = 2.5. 

Figure \ref{fig:stochastic_comparison} shows the electron spectra produced by the two methods at the epochs of $t$ = 0.16~and 0.48~Gyr. The agreement between the two methods over a wide range of $\gamma$ is excellent, except where $\gamma \simlt$~100 (where our method underpredicts $N(\gamma)$ by $\sim$10-40\% depending on the epoch. In any case, this does not affect our results since the energies of these particles are too small to radiate at the radio frequencies of interest in this paper. 

To demonstrate the importance of including the stochastic term in our SDE, Figure \ref{fig:stochastic_off_and_on} shows the evolved electron spectrum at the epoch $t$ = 0.48~Gyr in a simulation where the stochastic effects are included with one where they are not, with identical conditions as in Figure \ref{fig:stochastic_comparison}. The effect of including the stochastic term in our model is significant for our results, as the number of electrons at the energies ($\gamma > 10^3$) required for minihalo emission at the observed frequencies is higher by orders of magnitude.


\begin{thebibliography}{}

\bibitem[Ackermann et al.(2010)]{ack10} Ackermann, M., Ajello, M., Allafort, A., et al.\ 2010, \apjl, 717, L71
\bibitem[Aharonian et al.(2009)]{aha09} Aharonian, F., Akhperjanian, A.~G., Anton, G., et al.\ 2009, \aap, 502, 437
\bibitem[Ajello et al.(2009)]{aje09} Ajello, M., Rebusco, P., Cappelluti, N., et al.\ 2009, \apj, 690, 367
\bibitem[Ascasibar \& Markevitch(2006)]{AM06} Ascasibar, Y., \& Markevitch, M. 2006, \apj, 650, 102
\bibitem[Bertoglio et al.(2001)]{ber01} Bertoglio, J.-P., Bataille, F., \& Marion, J.-D.\ 2001, Physics of Fluids, 13, 290
\bibitem[Blasi \& Colafrancesco(1999)]{bla99} Blasi, P., \& Colafrancesco, S.\ 1999, Astroparticle Physics, 12, 169
\bibitem[Bonafede et al.(2010)]{bon10} Bonafede, A., Feretti, L., Murgia, M., et al.\ 2010, \aap, 513, A30
\bibitem[Brunetti et al.(2001)]{bru01} Brunetti, G., Setti, G., Feretti, L., \& Giovannini, G.\ 2001, \mnras, 320, 365
\bibitem[Brunetti(2003)]{bru03} Brunetti, G.\ 2003, Matter and Energy in Clusters of Galaxies, 301, 349 
\bibitem[Brunetti \& Lazarian(2007)]{bru07} Brunetti, G., \& Lazarian, A.\ 2007, \mnras, 378, 245
\bibitem[Brunetti \& Lazarian(2011a)]{bru11a} Brunetti, G., \& Lazarian, A.\ 2011, \mnras, 410, 127 
\bibitem[Brunetti \& Lazarian(2011b)]{bru11} Brunetti, G., \& Lazarian, A.\ 2011, \mnras, 412, 817
\bibitem[Bulbul et al.(2012)]{bul12} Bulbul, G.~E., Smith, R.~K., Foster, A., et al.\ 2012, \apj, 747, 32 
\bibitem[Burns et al.(1992)]{bur92} Burns, J.~O., Sulkanen, M.~E., Gisler, G.~R., \& Perley, R.~A.\ 1992, \apjl, 388, L49
\bibitem[Cassano et al.(2008)]{cas08} Cassano, R., Gitti, M., \& Brunetti, G.\ 2008, \aap, 486, L31
\bibitem[Cho \& Lazarian(2003)]{cho03} Cho, J., \& Lazarian, A.\ 2003, \mnras, 345, 325
\bibitem[Cho \& Lazarian(2006)]{cho06} Cho, J., \& Lazarian, A.\ 2006, \apj, 638, 811 
\bibitem[Colella \& Woodward(1984)]{col84} Colella, P., \& Woodward, P.~R.\ 1984, Journal of Computational Physics, 54, 174
\bibitem[Condon(1992)]{con92} Condon, J.~J.\ 1992, \araa, 30, 575
\bibitem[Dolag et al.(1999)]{dol99} Dolag, K., Bartelmann, M., \& Lesch, H.\ 1999, \aap, 348, 351
\bibitem[Dolag et al.(2005)]{dol05} Dolag, K., Vazza, F., Brunetti, G., \& Tormen, G.\ 2005, \mnras, 364, 753
\bibitem[{Dubey} et~al.(2009)]{dub09} {Dubey}, A., {Antypas}, K., {Ganapathy}, M.~K., {Reid}, L.~B., {Riley}, K.~M., {Sheeler}, D., {Siegel}, A., {Weide}, K. Extensible component based architecture for FLASH, a massively parallel, multiphysics simulation code. Parallel Computing 35~(10-11), 512--522.
\bibitem[Dubois \& Teyssier(2008)]{dub08} Dubois, Y., \& Teyssier, R.\ 2008, \aap, 482, L13
\bibitem[Evans \& Hawley(1988)]{eva88} Evans, C.~R., \& Hawley, J.~F.\ 1988, \apj, 332, 659
\bibitem[Fabian \& Nulsen(1994)]{fab94} Fabian, A.~C., \& Nulsen, P.~E.~J.\ 1994, \mnras, 269, L33 
\bibitem[Federrath et al.(2011)]{fed11} Federrath, C., Chabrier, G., Schober, J., et al.\ 2011, Physical Review Letters, 107, 114504
\bibitem[Florinski \& Pogorelov(2009)]{flo09} Florinski, V., \& Pogorelov, N.~V.\ 2009, \apj, 701, 642
\bibitem[Fryxell et al.(2000)]{fry00} Fryxell, B., et al.\ 2000, \apjs, 131, 273
\bibitem[Fujita et al.(2004)]{fuj04} Fujita, Y., Matsumoto, T., \& Wada, K.\ 2004, \apjl, 612, L9
\bibitem[Fujita et al.(2008)]{fuj08} Fujita, Y., Hayashida, K., Nagai, M., et al.\ 2008, \pasj, 60, 1133
\bibitem[Giacintucci et al.(2011)]{gia11} Giacintucci, S., Markevitch, M., Brunetti, G., Cassano, R., \& Venturi, T.\ 2011, \aap, 525, L10
\bibitem[Gitti et al.(2002)]{git02} Gitti, M., Brunetti, G., \& Setti, G.\ 2002, \aap, 386, 456
\bibitem[Gitti et al.(2004)]{git04} Gitti, M., Brunetti, G., Feretti, L., \& Setti, G.\ 2004, \aap, 417, 1 
\bibitem[Gitti et al.(2007)]{git07} Gitti, M., Ferrari, C., Domainko, W., Feretti, L., \& Schindler, S.\ 2007, \aap, 470, L25
\bibitem[Govoni et al.(2009)]{gov09} Govoni, F., Murgia, M., Markevitch, M., et al.\ 2009, \aap, 499, 371
\bibitem[Hallman \& Jeltema(2011)]{hal11} Hallman, E.~J., \& Jeltema, T.~E.\ 2011, \mnras, 418, 2467
\bibitem[Hockney \& Eastwood(1988)]{hoc88} Hockney, R.~W., \& Eastwood, J.~W.\ 1988, Bristol: Hilger, 1988
\bibitem[Iapichino \& Niemeyer(2008)]{iap08} Iapichino, L., \& Niemeyer, J.~C.\ 2008, \mnras, 388, 1089
\bibitem[Iapichino et al.(2011)]{iap11} Iapichino, L., Schmidt, W., Niemeyer, J.~C., \& Merklein, J.\ 2011, \mnras, 414, 2297
\bibitem[Jeltema \& Profumo(2011)]{jel11} Jeltema, T.~E., \& Profumo, S.\ 2011, \apj, 728, 53 
\bibitem[Jones et al.(2011)]{jon11} Jones, T.~W., Porter, D.~H., Ryu, D., \& Cho, J.\ 2011, \memsai, 82, 588
\bibitem[Keshet et al.(2010)]{kes10a} Keshet, U., Markevitch, M., Birnboim, Y., \& Loeb, A.\ 2010, \apjl, 719, L74
\bibitem[Keshet \& Loeb(2010)]{kes10b} Keshet, U., \& Loeb, A.\ 2010, \apj, 722, 737
\bibitem[Keshet(2010)]{kes10c} Keshet, U.\ 2010, arXiv:1011.0729
\bibitem[Kitsionas et al.(2009)]{kit09} Kitsionas, S., Federrath, C., Klessen, R.~S., et al.\ 2009, \aap, 508, 541
\bibitem[Kloeden \& Platen(2011)]{klo11} Kloeden, P.~E., \& Platen, E.\ 2011, (Berlin: Springer)
\bibitem[Kopp et al.(2012)]{kop12} Kopp, A., B{\"u}sching, I., Strauss, R.~D., 
\& Potgieter, M.~S.\ 2012, Computer Physics Communications, 183, 530
\bibitem[Lau et al.(2009)]{lau09} Lau, E.~T., Kravtsov, A.~V., \& Nagai, D.\ 2009, \apj, 705, 1129
\bibitem[Lee \& Deane(2009)]{lee09} Lee, D., \& Deane, A.~E.\ 2009, Journal of Computational Physics, 228, 952
\bibitem[MAGIC Collaboration et al.(2011)]{ale12} MAGIC Collaboration, Aleksi{\'c}, J., Alvarez, E.~A., et al.\ 2011, arXiv:1111.5544 
\bibitem[Markevitch \& Vikhlinin(2007)]{mar07} Markevitch, M., \& Vikhlinin, A.\ 2007, \physrep, 443, 1
\bibitem[Mazzotta \& Giacintucci(2008)]{maz08} Mazzotta, P., \& Giacintucci, S.\ 2008, \apjl, 675, L9
\bibitem[Melrose(1968)]{mel68} Melrose, D.~B.\ 1968, \apss, 2, 171 
\bibitem[Mewe, Kaastra, \& Liedahl(1995)]{MKL95} Mewe, R., Kaastra, J. S., \& Liedahl, D. A. 1995, Legacy, 6, 16
\bibitem[Murgia et al.(2010)]{mur10} Murgia, M., Eckert, D., Govoni, F., et al.\ 2010, \aap, 514, A76
\bibitem[Pei et al.(2010)]{pei10} Pei, C., Bieber, J.~W., Burger, R.~A., 
\& Clem, J.\ 2010, Journal of Geophysical Research (Space Physics), 115, 12107
\bibitem[Porter \& Woodward(1994)]{por94} Porter, D.~H., \& Woodward, P.~R.\ 1994, \apjs, 93, 309
\bibitem[Pfrommer \& En{\ss}lin(2004)]{pfr04} Pfrommer, C., \& En{\ss}lin, T.~A.\ 2004, \aap, 413, 17
\bibitem[Roediger et al.(2011)]{rod11a} Roediger, E., Br{\"u}ggen, M., Simionescu, A., B{\"o}ringer, H., Churazov, E., \& Forman, W.~R.\ 2011, \mnras, 369
\bibitem[Roediger \& Zuhone(2012)]{rod12} Roediger, E., \& Zuhone, J.~A.\ 2012, \mnras, 419, 1338
\bibitem[Rybicki \& Lightman(1979)]{ryb79} Rybicki, G.~B., \& Lightman, A.~P.\ 1979, New York, Wiley-Interscience, 1979.~393 p.
\bibitem[Ryu et al.(2008)]{ryu08} Ryu, D., Kang, H., Cho, J., 
\& Das, S.\ 2008, Science, 320, 909
\bibitem[Sanders et al.(2010)]{san10} Sanders, J.~S., Fabian, A.~C., Smith, R.~K., \& Peterson, J.~R.\ 2010, \mnras, 402, L11
\bibitem[Sarazin(1999)]{sar99} Sarazin, C.~L.\ 1999, \apj, 520, 529
\bibitem[Scannapieco \& Br{\"u}ggen(2008)]{sca08} Scannapieco, E., \& Br{\"u}ggen, M.\ 2008, \apj, 686, 927
\bibitem[Sijbring(1993)]{sij93} Sijbring, D. 1993, Ph.D. Thesis, Groningen
\bibitem[Smith et al.(2001)]{smi01} Smith, R.~K., Brickhouse, N.~S., Liedahl, D.~A., \& Raymond, J.~C.\ 2001, \apjl, 556, L91
\bibitem[Sunyaev et al.(2003)]{sun03} Sunyaev, R.~A., Norman, M.~L., \& Bryan, G.~L.\ 2003, Astronomy Letters, 29, 783
\bibitem[Strauss et al.(2011)]{str11} Strauss, R.~D., Potgieter, M.~S., B{\"u}sching, I., \& Kopp, A.\ 2011, \apj, 735, 83 
\bibitem[Taylor et al.(2002)]{tay02} Taylor, G.~B., Fabian, A.~C., \& Allen, S.~W.\ 2002, \mnras, 334, 769
\bibitem[Taylor et al.(2006)]{tay06} Taylor, G.~B., Gugliucci, N.~E., Fabian, A.~C., Sanders, J.~S., Gentile, G., \& Allen, S.~W.\ 2006, \mnras, 368, 1500
\bibitem[Taylor et al.(2007)]{tay07} Taylor, G.~B., Fabian, A.~C., Gentile, G., Allen, S.~W., Crawford, C., \& Sanders, J.~S.\ 2007, \mnras, 382, 67
\bibitem[Turk et al.(2011)]{tur11} Turk, M.~J., Smith, B.~D., Oishi, J.~S., Skory, S., Skillman, S.~W., Abel, T., \& Norman, M.~L.\ 2011, \apjs, 192, 9
\bibitem[Vacca et al.(2011)]{vac11} Vacca, V., Murgia, M., Govoni, F., et al.\ 2011, \memsai, 82, 658
\bibitem[Vazza et al.(2006)]{vaz06} Vazza, F., Tormen, G., Cassano, R., Brunetti, G., \& Dolag, K.\ 2006, \mnras, 369, L14
\bibitem[Vazza et al.(2009)]{vaz09} Vazza, F., Brunetti, G., Kritsuk, A., Wagner, R., Gheller, C., \& Norman, M.\ 2009, \aap, 504, 33
\bibitem[Vazza et al.(2010)]{vaz10} Vazza, F., Gheller, C., \& Brunetti, G.\ 2010, \aap, 513, A32 
\bibitem[Vazza et al.(2011)]{vaz11} Vazza, F., Brunetti, G., Gheller, C., Brunino, R., \& Br{\"u}ggen, M.\ 2011, \aap, 529, A17
\bibitem[Vazza et al.(2012)]{vaz12} Vazza, F., Roediger, E., \& Brueggen, M.\ 2012, arXiv:1202.5882
\bibitem[Wik et al.(2012)]{wik12} Wik, D.~R., Sarazin, C.~L., Zhang, Y.-Y., et al.\ 2012, \apj, 748, 67
\bibitem[Zhang(1999)]{zha99} Zhang, M.\ 1999, \apj, 513, 409
\bibitem[ZuHone et al.(2010)]{zuh10} ZuHone, J.~A., Markevitch, M., \& Johnson, R.~E.\ 2010, \apj, 717, 908 (ZMJ10)
\bibitem[ZuHone et al.(2011)]{zuh11} ZuHone, J.~A., Markevitch, M., \& Lee, D.\ 2011, \apj, 743, 16
\end{thebibliography}
\end{document}